\documentclass[conference,compsoc]{IEEEtran}
\usepackage{xspace}
\usepackage{xparse}
\usepackage{amsfonts,amsmath}
\usepackage{stackrel}
\usepackage{mathtools}
\usepackage{amsthm}
\usepackage[english]{babel}
\usepackage{amssymb}
\usepackage{txfonts}
\usepackage{enumitem}
\usepackage{xcolor}
\usepackage{array}
\usepackage{tikz}
\usetikzlibrary{arrows}
\usepackage{pgfplots, pgfplotstable}
\pgfplotsset{compat=1.9}
\usepgfplotslibrary{groupplots}
\usetikzlibrary{patterns}
\usepackage{pstricks}
\usepackage{subcaption}
\usepackage{wrapfig}
\usepackage[utf8]{inputenc}
\usepackage{booktabs}
\usepackage{expl3}
\usepackage{collcell}
\usepackage{url}
\usepackage{hyperref}
\usepackage{graphicx}
\usepackage{multirow}
\usepackage{framed}
\usepackage{verbatim}
\usepackage[square,sort,comma,numbers]{natbib}
\usepackage{mdframed}
\usepackage{algorithm}
\usepackage{etoolbox}
\usepackage{titlesec}
\titlespacing*{\subsection}{0pt}{0.5\baselineskip}{0.5\baselineskip}
\mdfsetup{frametitlealignment=\center}
\usepackage[utf8]{inputenc}

\newcounter{note}[section]
\renewcommand{\thenote}{\thesection.\arabic{note}}
\newcommand{\fixmecolor}{red}

\newcommand{\mkr}[1]{\refstepcounter{note}{\bf
\textcolor{\fixmecolor}{$\ll$MKR~\thenote: {\sf #1}$\gg$}}}

\newcommand{\assign}{\ensuremath{\gets}\xspace}

\newcommand{\secref}[1]{\mbox{Sec.~\ref{#1}}\xspace}

\newcommand{\figref}[1]{\mbox{Fig.~\ref{#1}}\xspace}

\newcommand{\tblref}[1]{\mbox{Table~\ref{#1}}\xspace}

\newcommand{\appref}[1]{\mbox{App.~\ref{#1}}\xspace}
\newcommand{\eqnref}[1]{\mbox{(\ref{#1})}\xspace}
\newcommand{\eqnsref}[2]{\mbox{(\ref{#1})--(\ref{#2})}\xspace}
\newcommand{\propref}[1]{\mbox{Prop.~\ref{#1}}\xspace}

\newcommand{\lemmaref}[1]{\mbox{Lemma~\ref{#1}}\xspace}
\newcommand{\thmref}[1]{\mbox{Theorem~\ref{#1}}\xspace}
\newcommand{\stepref}[1]{\mbox{Step~\ref{#1}}\xspace}
\newcommand{\stepsref}[2]{\mbox{Steps~\ref{#1}--\ref{#2}}\xspace}

\newcommand{\myparagraph}[1]{\smallskip\noindent\textbf{#1}:\xspace}

\newtheorem{defn}{Definition}
\newtheorem{theorem}{Theorem}[section]
\newtheorem*{theorem*}{Theorem}
\newtheorem{lemma}[theorem]{Lemma}
\newtheorem{prop}[theorem]{Proposition}
\newtheorem*{prop*}{Proposition}

{\hfil\hskip2em\penalty250\parfillskip=0pt\finalhyphendemerits=0$\qed$%
	\par\smallskip\par}

\newcommand{\symmUR}{\ensuremath{\mathsf{SUR}}\xspace}
\newcommand{\symmURDist}[2]{\ensuremath{\vectorNorm{\symmUR}{{#1}-{#2}}}\xspace}
\newcommand{\symmURDefn}{\ensuremath{\vectorNorm{\symmUR}{\cdot}}\xspace}

\DeclareMathOperator{\symmDiff}{\triangle}

\NewDocumentCommand{\distance}{o}%
{%
	\IfNoValueTF{#1}%
	{\ensuremath{\fnNotation{d}}\xspace}
	{\ensuremath{\fnNotation{d}_{#1}}\xspace}
}

\NewDocumentCommand{\distanceNotation}{o o o}
{
\IfNoValueTF{#2}
{\ensuremath{\left\Vert . \right\Vert_{\mathsf{#1}}}\xspace}
{
\IfNoValueTF{#3}
{\ensuremath{\left\Vert #2 \right\Vert_{#1}}\xspace}
{\ensuremath{\left\Vert #2 - #3 \right\Vert_{#1}}\xspace}
}
}

\NewDocumentCommand{\hammingDist}{ g g }{\distanceNotation[H][#1][#2]}

\NewDocumentCommand{\embed}{o}%
{%
  \IfNoValueTF{#1}%
              {\ensuremath{\fnNotation{f}}\xspace}
              {\ensuremath{\fnNotation{f}_{#1}}\xspace}
}

\newcommand{\embedESP}{\ensuremath{\embed[\embedKey]}\xspace}
\newcommand{\embedSim}{\embedESP}

\NewDocumentCommand{\setToVecMap}{o}
{\IfNoValueTF{#1}
{\ensuremath{\mathsf{SetToVec}}\xspace}
{\ensuremath{\mathsf{SetToVec}_{#1}}\xspace}
}
\newcommand{\setOfEvalPts}{\setNotation{X}\xspace}
\newcommand{\ESPinput}{\ensuremath{\valNotation{s}}\xspace}
\NewDocumentCommand{\ESPNodeVec}{ g o
}{\ensuremath{\textsc{esp}({#1}).\mathsf{nvec}\IfNoValueF{#2}{_{#2}}}\xspace}
\NewDocumentCommand{\ESPCharVec}{ g o
}{\ensuremath{\textsc{esp}({#1}).\mathsf{cvec}\IfNoValueF{#2}{_{#2}}}\xspace}
\NewDocumentCommand{\ESPCharSet}{ g o
}{\ensuremath{\textsc{esp}({#1}).\mathsf{cset}\IfNoValueF{#2}{\mathopen{}({#2}
)\mathclose{}}}\xspace}
\newcommand{\ESPInternalNode}[1]{\ensuremath{X_{#1}}\xspace}

\NewDocumentCommand{\fileBlock}{o}{\ensuremath{\cInput\IfNoValueF{#1}{_{#1}}
}\xspace}
\NewDocumentCommand{\fileBlockAlt}{o}{\ensuremath{\cInputAlt\IfNoValueF{#1}{_{
#1}}}\xspace}
\NewDocumentCommand{\fileBlockVecAlt}{o}
{\IfNoValueTF{#1}
	{\ensuremath{\vecNotation{\fileBlockAlt}}\xspace}
	{\ensuremath{\fileBlockAlt_{#1}}\xspace}
}
\newcommand{\maxFileLen}{\ensuremath{\{0,1\}^{\ast}}\xspace}

\NewDocumentCommand{\cInputVec}{o}
{\IfNoValueTF{#1}
	{\ensuremath{\vecNotation{\cInput}}\xspace}
	{\ensuremath{\cInput_{#1}}\xspace}
}

\NewDocumentCommand{\cInputVecAlt}{o}
{\IfNoValueTF{#1}
	{\ensuremath{\vecNotation{\cInput}'}\xspace}
	{\ensuremath{\cInput'_{#1}}\xspace}
}

\NewDocumentCommand{\simHash}{o}
{
	\IfNoValueTF{#1}
	{\ensuremath{\mathsf{SimHash}}\xspace}
	{\ensuremath{\mathsf{SimHash}_{#1}}\xspace}
}
\NewDocumentCommand{\sdHash}{o}
{
	\IfNoValueTF{#1}
	{\ensuremath{\mathsf{sdHash}}\xspace}
	{\ensuremath{\mathsf{sdHash}_{#1}}\xspace}
}

\newcommand{\numberOfPointsSim}{\ensuremath{\theta}\xspace}
\newcommand{\numberOfPointsSimDefn}{\ensuremath{2\degreeOfKey{1} -
\frac{\threshold}{2} + 2}\xspace}
\newcommand{\metricSpaceESP}{\ensuremath{\finiteField^{\numberOfPointsSim}
}\xspace}

\newcommand{\metricSpaceGen}{\ensuremath{\finiteField^{\metricSpaceDimGen}
}\xspace}
\newcommand{\metricSpaceDimGen}{\ensuremath{\theta}\xspace}

\makeatletter
\newcommand{\vast}{\bBigg@{4}}
\newcommand{\Vast}{\bBigg@{5}}
\makeatother

\newcommand{\rangePRF}{\ensuremath{\{0,1\}^{\secParam}}\xspace}
\newcommand{\domainHash}{\ensuremath{\{0,1\}^{\ast}}\xspace}

\newcommand{\oleFunc}{\ensuremath{\mathcal{F}_{\mathit{ole}}}\xspace}
\newcommand{\otFunc}{\ensuremath{\mathcal{F}_{\mathit{GC}}}\xspace}
\newcommand{\oprfFunc}{\ensuremath{\mathcal{F}_{\mathit{oprf}}}\xspace}
\newcommand{\npaFunc}{\ensuremath{\mathcal{F}_{\mathit{npa}}}\xspace}

\NewDocumentCommand{\hash}{o o}%
{%
	\IfNoValueTF{#1}%
	{\ensuremath{\mathsf{H}\IfNoValueF{#2}{_{#2}}}\xspace}
	{\ensuremath{\mathsf{H{#1}}\IfNoValueF{#2}{_{#2}}}\xspace}
}

\NewDocumentCommand{\Enc}{o}%
{
  \IfNoValueTF{#1}%
	{\ensuremath{\constNotation{AuthEnc}}\xspace}
	{\ensuremath{\constNotation{AuthEnc}_{#1}}\xspace}
}
\NewDocumentCommand{\PRP}{o}%
{%
	\IfNoValueTF{#1}%
	{\ensuremath{\mathsf{P}}\xspace}
	{\ensuremath{\mathsf{P}_{#1}}\xspace}
}

\NewDocumentCommand{\PRF}{o o}%
{%
	\IfNoValueTF{#1}%
	{\ensuremath{\mathsf{OPRF}\IfNoValueF{#2}{_{#2}}}\xspace}%
{\ensuremath{\ifblank{#1}{\mathsf{OPRF}\IfNoValueF{#2}{_{#2}}}{\mathsf{PRF
}\IfNoValueF{#2}{_{#2}}}}\xspace}%
}

\NewDocumentCommand{\aPAKEOPRF}{o}
{
\IfNoValueTF{#1}
{\ensuremath{\mathsf{PRF'}}\xspace}
{\ensuremath{\mathsf{PRF'_{#1}}}\xspace}
}

\NewDocumentCommand{\genericFnFamily}{ o
}{\ensuremath{\fnNotation{G}\IfNoValueF{#1}{_{#1}}}\xspace}
\newcommand{\genericFnKeyspace}{\ensuremath{\setNotation{K}}\xspace}
\newcommand{\genericFnKey}{\ensuremath{k}\xspace}
\newcommand{\genericFnDomain}{\ensuremath{\setNotation{D}}\xspace}
\newcommand{\genericFnDomainElmt}{\ensuremath{d}\xspace}
\newcommand{\genericFnRange}{\ensuremath{\setNotation{R}}\xspace}

\newcommand{\rootIdx}{\ensuremath{j}\xspace}
\newcommand{\ptIdx}{\ensuremath{k}\xspace}
\newcommand{\setIdx}{\ensuremath{i}\xspace}
\newcommand{\vecIdx}{\ensuremath{i}\xspace}

\NewDocumentCommand{\nats}{o o}%
{%
	\IfNoValueTF{#1}%
	{\ensuremath{\mathbb{N}}\xspace}%
	{\IfNoValueTF{#2}%
		{\ensuremath{[{#1}]}\xspace}%
		{\ensuremath{[{#1},{#2}]}\xspace}%
	}%
}

\newcommand{\cInputPRPOutAlt}{\ensuremath{y'}\xspace}

\NewDocumentCommand{\cInput}{o}%
{%
  \IfNoValueTF{#1}%
              {\ensuremath{\valNotation{w}}\xspace}
              {\ensuremath{\valNotation{w}_{#1}}\xspace}
}

\NewDocumentCommand{\cInputAlt}{o}%
{%
  \IfNoValueTF{#1}%
              {\ensuremath{\valNotation{w}'}\xspace}
              {\ensuremath{\valNotation{w}'_{#1}}\xspace}
}

\newcommand{\cInputEmbedding}{\ensuremath{z}\xspace}

\newcommand{\authToken}{\ensuremath{\vec{y}}\xspace}
\newcommand{\authTokenAlt}{\ensuremath{\vec{y}'}\xspace}
\newcommand{\authTokenDblAlt}{\ensuremath{\vec{y''}}\xspace}
\newcommand{\authTokenComp}[1]{\ensuremath{y_{#1}}\xspace}
\newcommand{\authTokenAltComp}[1]{\ensuremath{y'_{#1}}\xspace}

\NewDocumentCommand{\PRPKey}{o}
{
\IfNoValueTF{#1}
{\ensuremath{\valNotation{k}_{\PRP}}\xspace}
{\ensuremath{\valNotation{k#1}_{\PRP}}\xspace}
}

\newcommand{\PRPKeySpace}{\ensuremath{\setNotation{K}_{\PRP}}\xspace}
\NewDocumentCommand{\PRFKeySpace}{o}{\ensuremath{\setNotation{K}_{\PRF[{#1}]}
}\xspace}
\NewDocumentCommand{\PRFKey}{o}{\ensuremath{\valNotation{k}_{\PRF[{#1}]}
}\xspace}
\NewDocumentCommand{\PRFKeyAlt}{o}{\ensuremath{\valNotation{k}'_{\PRF[{#1}]}
}\xspace}

\newcommand{\embedKeySpace}{\ensuremath{\setNotation{K}_{\embed}}\xspace}
\newcommand{\embedKey}{\ensuremath{\valNotation{k}_{\embed}}\xspace}
\newcommand{\embedKeyAlt}{\ensuremath{\valNotation{k}_{\embed}'}\xspace}

\newcommand{\degreeOfKey}[1]{
\ifstrequal{#1}{1}
{\ensuremath{\delta}\xspace}
{\ensuremath{2\delta}\xspace}
}

\NewDocumentCommand{\PRPKeyVec}{o o}
{
	\ifstrequal{#1}{1}
	{
		\IfNoValueTF{#2}
		{\ensuremath{\vecNotation{a}}\xspace}
		{\ifblank{#2}{\ensuremath{a}\xspace}{\ensuremath{a_{#2}}\xspace} }
	}
	{
		\IfNoValueTF{#2}
		{\ensuremath{\vecNotation{b}}\xspace}
		{\ifblank{#2}{\ensuremath{b}\xspace}{\ensuremath{b_{#2}}\xspace} }
	}
}

\NewDocumentCommand{\PRPKeyEmbeddingVec}{o o}
{
	\ifstrequal{#1}{1}
	{
		\IfNoValueTF{#2}
		{\ensuremath{\vecNotation{a}}\xspace}
		{\ifblank{#2}{\ensuremath{a}\xspace}{\ensuremath{a_{#2}}\xspace} }
	}
	{
		\IfNoValueTF{#2}
		{\ensuremath{\vecNotation{b}}\xspace}
		{\ifblank{#2}{\ensuremath{b}\xspace}{\ensuremath{b_{#2}}\xspace} }
	}
}

\NewDocumentCommand{\PRPKeyHashVec}{o o}
{
	\ifstrequal{#1}{1}
	{
		\IfNoValueTF{#2}
		{\ensuremath{\vecNotation{a'}}\xspace}
		{\ifblank{#2}{\ensuremath{a'}\xspace}{\ensuremath{a'_{#2}}\xspace} }
	}
	{
		\IfNoValueTF{#2}
		{\ensuremath{\vecNotation{b'}}\xspace}
		{\ifblank{#2}{\ensuremath{b'}\xspace}{\ensuremath{b'_{#2}}\xspace} }
	}
}

\NewDocumentCommand{\aPAKEOPRFKeySpace}{}{\ensuremath{\setNotation{K}_{\aPAKEOPRF}}\xspace}
\NewDocumentCommand{\aPAKEOPRFKey}{}{\ensuremath{\valNotation{k}_{\aPAKEOPRF}}\xspace}

\newcommand{\bigO}[1]{\ensuremath{\mathsf{O}(#1)}\xspace}

\newcommand{\reals}{\ensuremath{\mathbb{R}}\xspace}
\newcommand{\nonnegativeReals}{\ensuremath{\reals_{\ge 0}}\xspace}

\newcommand{\setSize}[1]{\ensuremath{\left|{#1}\right|}\xspace}

\newcommand{\valNotation}[1]{\ensuremath{\expandafter\MakeLowercase{#1}}\xspace}
\newcommand{\setNotation}[1]{\ensuremath{\expandafter\mathcal{#1}}\xspace}
\newcommand{\rvNotation}[1]{\ensuremath{\expandafter\varmathbb{#1}}\xspace}
\newcommand{\fnNotation}[1]{\ensuremath{\expandafter\MakeLowercase{#1}}\xspace}
\newcommand{\algNotation}[1]{\ensuremath{\expandafter\mathtt{\MakeUppercase{#1}
}}\xspace}
\newcommand{\constNotation}[1]{\ensuremath{\expandafter\mathsf{#1}}\xspace}
\newcommand{\polyNotation}[1]{\ensuremath{\expandafter\MakeLowercase{#1}
}\xspace}
\newcommand{\vecNotation}[1]{\ensuremath{\expandafter\vec{#1}}\xspace}
\newcommand{\getsr}{\ensuremath{\stackrel{\scriptsize\$}{\leftarrow}}\xspace}
\newcommand{\defeq}{\ensuremath{\mathbin{\stackrel{\scriptsize\mbox{def}}{=}}
}\xspace}
\newcommand{\tru}{\ensuremath{\mathsf{TRUE}}\xspace}
\newcommand{\fals}{\ensuremath{\mathsf{FALSE}}\xspace}
\newcommand{\equals}{\stackrel{?}{=}}
\newcommand{\hadamardProd}{\ensuremath{\times}\xspace}
\newcommand{\starNot}{*\xspace}

\newcommand{\OLEAliceInput}{\ensuremath{x}\xspace}
\newcommand{\OLEBobInputOne}{\ensuremath{u}\xspace}
\newcommand{\OLEBobInputTwo}{\ensuremath{v}\xspace}

\NewDocumentCommand{\genericVecComp}{o o}
{
	\IfNoValueTF{#2}
	{\ensuremath{v_{#1}}\xspace}
	{\ensuremath{v_{#1, #2}}\xspace}
}

\NewDocumentCommand{\genericVecAltComp}{o o}
{
	\IfNoValueTF{#2}
	{\ensuremath{v'_{#1}}\xspace}
	{\ensuremath{v'_{#1, #2}}\xspace}
}

\newcommand{\linearComb}{\ensuremath{+}\xspace}

\newcommand{\sessionID}{\ensuremath{ssid}}

\newcommand{\prpOutputEmbedding}{\ensuremath{\valNotation{p_1}}\xspace}
\newcommand{\prpOutputEmbeddingVec}{\ensuremath{\vecNotation{p_1}}\xspace}
\newcommand{\prpOutputHash}{\ensuremath{\valNotation{p_2}}\xspace}
\newcommand{\prpOutputEmbeddingAlt}{\ensuremath{\valNotation{p'_1}}\xspace}
\newcommand{\prpOutputHashAlt}{\ensuremath{\valNotation{p'_2}}\xspace}
\newcommand{\prpOutputHashVec}{\ensuremath{\vecNotation{p_2}}\xspace}

\newcommand{\prfOutput}{\ensuremath{\valNotation{\gamma}}\xspace}
\newcommand{\prfOutputAlt}{\ensuremath{\prfOutput'}\xspace}
\newcommand{\prpOutputEmbeddingComp}[1]{\ensuremath{\valNotation{p_{1,#1}}
}\xspace}

\newcommand{\prpOutputHashComp}[1]{\ensuremath{\valNotation{p_{2,#1}}}\xspace}

\newcommand{\mapFn}{\ensuremath{\fnNotation{M}}\xspace}
\NewDocumentCommand{\sInputPW}{o}{\ensuremath{\lanElmt\IfNoValueF{#1}{_{#1}}
}\xspace}
\newcommand{\vectorNorm}[2]{\ensuremath{\left\Vert {#2}
\right\Vert_{#1}}\xspace}
\newcommand{\credKey}{\ensuremath{\mathsf{rw}}\xspace}
\newcommand{\clientID}{\ensuremath{cid}\xspace}
\newcommand{\pakeClientPubKey}{\ensuremath{P_u}\xspace}
\newcommand{\pakeClientPrivKey}{\ensuremath{p_u}\xspace}
\newcommand{\pakeServerPubKey}{\ensuremath{P_s}\xspace}
\newcommand{\pakeServerPrivKey}{\ensuremath{p_s}\xspace}
\NewDocumentCommand{\pw}{o}{\ensuremath{\cInput\IfNoValueF{#1}{_{#1}}}\xspace}
\NewDocumentCommand{\pwAlt}{o}{\ensuremath{\cInputAlt\IfNoValueF{#1}{_{#1}}
}\xspace}

\newcommand{\blockedElmts}{\ensuremath{\universe_{\mathsf{blk}}}\xspace}
\newcommand{\policyList}{\ensuremath{\mathcal{E}_{\mathsf{blk}}}\xspace}
\newcommand{\PSIList}{\ensuremath{\mathcal{S}_{\mathsf{blk}}}\xspace}
\newcommand{\policyLan}{\ensuremath{\setNotation{L}}\xspace}
\newcommand{\LanPolicy}{\policyLan}
\newcommand{\blockList}{\policyLan}
\newcommand{\blockListSize}{\ensuremath{n}}
\newcommand{\lanElmt}{\ensuremath{\valNotation{l}}\xspace}
\newcommand{\listElmt}{\ensuremath{\valNotation{u}}\xspace}

\NewDocumentCommand{\lanElmtComp}{g
g}{\ensuremath{\valNotation{l}_{#1\IfValueT{#2}{,#2}}}\xspace}

\NewDocumentCommand{\lElmtVecComp}{g
g}{\ensuremath{\vecNotation{l}_{#1\IfValueT{#2}{,#2}}}\xspace}

\NewDocumentCommand{\lanElmtVec}{g}{\ensuremath{\vecNotation{\lanElmt\IfValueF{
#1}{_{#1}}}\ }\xspace}

\newcommand{\threshold}{\ensuremath{T}\xspace}
\newcommand{\secParam}{\ensuremath{\lambda}\xspace}

\newcommand{\universe}{\ensuremath{\mathcal{U}}\xspace}

\NewDocumentCommand{\commitVec}{o}
{\IfNoValueTF{#1}
	{\ensuremath{\vecNotation{C}}\xspace}
	{\ensuremath{\vecNotation{c}_{#1}}\xspace}
}

\newcommand{\finiteField}{\ensuremath{\mathbb{F}}\xspace}
\newcommand{\fieldOrder}{\ensuremath{q}\xspace}
\newcommand{\polyField}{\ensuremath{\mathbb{F}[x]}\xspace}

\NewDocumentCommand{\genericSetVar}{o}%
{%
	\IfNoValueTF{#1}%
	{\ensuremath{\setNotation{S}}\xspace}
	{\ensuremath{\setNotation{S}_{#1}}\xspace}
}

\NewDocumentCommand{\genericVec}{o}%
{%
	\IfNoValueTF{#1}%
	{\ensuremath{\vecNotation{v}}\xspace}
	{\ensuremath{\vecNotation{v}_{#1}}\xspace}
}

\NewDocumentCommand{\genericVecAlt}{o}%
{%
	\IfNoValueTF{#1}%
	{\ensuremath{\vecNotation{v}'}\xspace}
	{\ensuremath{\vecNotation{v}_{#1}'}\xspace}
}

\newcommand{\genericSetElmt}{\ensuremath{s}\xspace}

\newcommand{\genericFieldElmt}{\ensuremath{x}}
\newcommand{\genericFieldElmtAlt}{\ensuremath{y}}

\newcommand{\numberOfPoints}{\ensuremath{2\delta + 1}\xspace}
\newcommand{\numberOfMasks}{\ensuremath{\degreeOfKey{1}}\xspace}
\newcommand{\distInMaskedSets}{\ensuremath{t}\xspace}
\newcommand{\distInUniverse}{\ensuremath{t}\xspace}
\newcommand{\tmpVar}{\ensuremath{t}\xspace}

\NewDocumentCommand{\tmpVarVec}{o o}
{
\IfNoValueTF{#2}
{\ensuremath{\vecNotation{t_{#1}}}\xspace}
{\ensuremath{t_{#1, #2}}\xspace}
}

\NewDocumentCommand{\tmpVarAltVec}{o o}
{
\IfNoValueTF{#2}
{\ensuremath{\vecNotation{t'_{#1}}}\xspace}
{\ensuremath{t'_{#1, #2}}\xspace}
}

\NewDocumentCommand{\bitMask}{o}
{
	\IfNoValueTF{#1}{\ensuremath{M}\xspace}{\ensuremath{M_{#1}}\xspace}
}

\newcommand{\vecMap}{\ensuremath{M_{\setOfEvalPts}}\xspace}
\newcommand{\secretShare}[1]{\ensuremath{ss_{#1}}\xspace}
\newcommand{\diffInComps}{\ensuremath{d}\xspace}

\newcommand{\POPRFtag}{\ensuremath{t}\xspace}

\NewDocumentCommand{\client}{o}%
{%
	\IfNoValueTF{#1}%
	{\ensuremath{\mathsf{C}}\xspace}
	{\ensuremath{\mathsf{C}_{#1}}\xspace}
}
\newcommand{\server}{\ensuremath{\mathsf{S}}\xspace}
\newcommand{\policyServer}{\ensuremath{\mathsf{PS}}\xspace}
\newcommand{\storageServer}{\ensuremath{\mathsf{SS}}\xspace}

\newcommand{\degreeOfPoly}{\ensuremath{\mathsf{Deg}}\xspace}
\newcommand{\uniqueRoots}[1]{\ensuremath{\setNotation{R}_{#1}}\xspace}
\newcommand{\polyFromVec}[1]{\ensuremath{\polyNotation{p_{#1}}(x)}\xspace}

\NewDocumentCommand{\polyFromEmbedding}{o o}
{
\IfNoValueTF{#2}
{\ensuremath{\polyNotation{e_{#1}}(x)} \xspace}
{\ensuremath{\polyNotation{e_{#1}}(#2)} \xspace}
}

\NewDocumentCommand{\clientsBlindingPolyVOLENoSub}{o}%
{%
	\IfNoValueTF{#1}%
	{\ensuremath{\polyNotation{R'}(x)}\xspace}
	{\ensuremath{\polyNotation{R'}(#1)}\xspace}
}

\NewDocumentCommand{\clientsBlindingPolyVOLE}{o o}%
{%
	\IfNoValueTF{#2}%
		{\IfNoValueTF{#1}%
			{\ensuremath{\polyNotation{R'}(x)}\xspace}
			{\ensuremath{\polyNotation{R'}_{#1}(x)}\xspace}
		}
	{
		{\ensuremath{\polyNotation{R'}_{#1}(#2)}\xspace}
	}
}

\NewDocumentCommand{\genericRand}{o}
{\IfNoValueTF{#1}
	{\ensuremath{r}\xspace}
	{\ensuremath{r_{#1}}\xspace}
}

\NewDocumentCommand{\PRPKeyPolyHash}{o o}
{
	\ifstrequal{#1}{1}
	{
		\IfNoValueTF{#2}
		{\ensuremath{\polyNotation{A'}(x)}\xspace}
		{\ensuremath{\polyNotation{A'}(#2)}\xspace}
	}
	{
		\IfNoValueTF{#2}
		{\ensuremath{\polyNotation{B'}(x)}\xspace}
		{\ensuremath{\polyNotation{B'}(#2)}\xspace}
	}
}

\NewDocumentCommand{\clientsRandomPoly}{o o}
{
	\IfNoValueTF{#2}
	{\ensuremath{\polyNotation{R}_{#1}(x)}\xspace}
	{\ensuremath{\polyNotation{R}_{#1}(#2)}\xspace}
}

\NewDocumentCommand{\serversRandomPoly}{o o}
{
	\IfNoValueTF{#2}
	{\ensuremath{\polyNotation{R}_{#1}'(x)}\xspace}
	{\ensuremath{\polyNotation{R}_{#1}'(#2)}\xspace}
}

\NewDocumentCommand{\interpolatingPoly}{o}
{
	\IfNoValueTF{#1}
	{\ensuremath{\polyNotation{G}(x)}\xspace}
	{\ensuremath{\polyNotation{G}(#1)}\xspace}
}

\NewDocumentCommand{\evalPt}{o}
{
	\IfNoValueTF{#1}
	{\ensuremath{\genericFieldElmt}\xspace}
	{\ensuremath{\genericFieldElmt_{#1}}\xspace}
}

\NewDocumentCommand{\genericPoly}{o}%
{%
	\IfNoValueTF{#1}%
	{\ensuremath{\polyNotation{p}(x)}\xspace}
	{\ensuremath{\polyNotation{p}_{#1}(x)}\xspace}
}

\NewDocumentCommand{\genericPolyEval}{o o}%
{%
	\IfNoValueTF{#2}%
	{\ensuremath{\polyNotation{p}(#1)}\xspace}
	{\ensuremath{\polyNotation{p}_{#2}(#1)}\xspace}
}

\NewDocumentCommand{\genericPolyAlt}{o}%
{%
	\IfNoValueTF{#1}%
	{\ensuremath{\polyNotation{p'}(x)}\xspace}
	{\ensuremath{\polyNotation{p'}_{#1}(x)}\xspace}
}

\NewDocumentCommand{\genericPolyAltEval}{o o}%
{%
	\IfNoValueTF{#2}%
	{\ensuremath{\polyNotation{p'}(#1)}\xspace}
	{\ensuremath{\polyNotation{p'}_{#2}(#1)}\xspace}
}

\newcommand{\hashComp}[1]{\ensuremath{h_{#1}}\xspace}

\NewDocumentCommand{\randomPoly}{o}%
{%
	\IfNoValueTF{#1}%
	{\ensuremath{\polyNotation{r}(x)}\xspace}
	{\ensuremath{\polyNotation{r}_{#1}(x)}\xspace}
}

\NewDocumentCommand{\randomPolyComb}{o}%
{%
	\IfNoValueTF{#1}%
	{\ensuremath{\polyNotation{r_c}(x)}\xspace}
	{\ensuremath{\polyNotation{r_c}(#1)}\xspace}
}

\NewDocumentCommand{\randomPolyEval}{o o}%
{%
	\IfNoValueTF{#2}%
	{\ensuremath{\polyNotation{r}(#1)}\xspace}
	{\ensuremath{\polyNotation{r}_{#2}(#1)}\xspace}
}

\NewDocumentCommand{\randomVec}{o}{\ensuremath{\vecNotation{r\IfValueT{#1}{_{#1
}}}}\xspace}

\newcommand{\serversRandomPolyEvalVecDefn}[1]{\ensuremath
{\serversRandomPolyEvalVec[#1] \gets \langle
\serversRandomPoly[\setIdx]\rangle_{\evalPt \in \setOfEvalPts}}\xspace}
\newcommand{\clientsRandomPolyEvalVecDefn}[1]{\ensuremath
{\clientsRandomPolyEvalVec[#1] \gets \langle
\clientsRandomPoly[\setIdx]\rangle_{\evalPt \in \setOfEvalPts}}\xspace}
\newcommand{\randomPolyCombEvalVec}{\vecNotation{r_{c}}\xspace}

\NewDocumentCommand{\clientsESPPolyEvalVec}{o}
{\IfNoValueTF{#1}
	{\ensuremath{\vecNotation{W}}\xspace}
	{\ensuremath{w_{#1}}\xspace}
}

\NewDocumentCommand{\clientsSimPolyEvalVec}{o}
{\IfNoValueTF{#1}
	{\ensuremath{\vecNotation{W}}\xspace}
	{\ensuremath{w_{#1}}\xspace}
}

\NewDocumentCommand{\clientsInputPolyEvalVec}{o}
{\IfNoValueTF{#1}
	{\ensuremath{\vecNotation{p}}\xspace}
	{\ensuremath{p_{#1}}\xspace}
}

\NewDocumentCommand{\clientsRandomPolyEvalVec}{o o}
{\IfNoValueTF{#2}
	{\ensuremath{\vecNotation{r}_{#1}}\xspace}
	{\ensuremath{r_{#1, #2}}\xspace}
}

\NewDocumentCommand{\serversRandomPolyEvalVec}{o o}
{\IfNoValueTF{#2}
	{\ensuremath{\vecNotation{r}_{#1}'}\xspace}
	{\ensuremath{r_{#1, #2}'}\xspace}
}

\NewDocumentCommand{\serversInputPolyEvalVec}{o o}
{\IfNoValueTF{#2}
	{\ensuremath{\vecNotation{Q}_{#1}}\xspace}
	{\ensuremath{q_{#1, #2}}\xspace}
}

\NewDocumentCommand{\simRandomPoly}{o}
{
\IfNoValueTF{#1}
{\ensuremath{\overline{\polyNotation{r}}(x)}\xspace}
{\ensuremath{\overline{\polyNotation{r}}_{#1}(x)}\xspace}
}

\NewDocumentCommand{\simRandomPolyAlt}{o}
{
	\IfNoValueTF{#1}
	{\ensuremath{\overline{\polyNotation{r'}}(x)}\xspace}
	{\ensuremath{\overline{\polyNotation{r'}}_{#1}(x)}\xspace}
}

\NewDocumentCommand{\simRandomPolyEval}{o o}
{
\IfNoValueTF{#2}{\ensuremath{\overline{\polyNotation{r}}(#1)}\xspace}
{\ensuremath{\overline{\polyNotation{r}}_{#2}(#1)}\xspace}
}

\NewDocumentCommand{\simRandomPolyAltEval}{o o}
{
\IfNoValueTF{#2}{\ensuremath{\overline{\polyNotation{r'}}(#1)}\xspace}
{\ensuremath{\overline{\polyNotation{r'}}_{#2}(#1)}\xspace}
}

\newcommand{\combVector}{\ensuremath{\vecNotation{c}}\xspace}
\newcommand{\combPoly}{\ensuremath{\polyNotation{c}(x)}\xspace}

\newcommand{\prob}[1]{\ensuremath{\mathbb{P}\mathopen{}\left({#1}\right
)\mathclose{}}\xspace}
\newcommand{\cprob}[3]{\ensuremath{\mathbb{P}\mathopen{}{#1(}{#2} \; {#1|} \;
{#3}{#1)}\mathclose{}}\xspace}

\newcommand{\falseRejectRate}[3]{\ensuremath{\mathsf{FRR}^{#1}_{{#2},{#3}}
}\xspace}
\newcommand{\falseAcceptRate}[3]{\ensuremath{\mathsf{FAR}^{#1}_{{#2},{#3}}
}\xspace}
\newcommand{\trueRejectRate}[3]{\ensuremath{\mathsf{TRR}^{#1}_{{#2},{#3}}
}\xspace}
\newcommand{\trueAcceptRate}[3]{\ensuremath{\mathsf{TAR}^{#1}_{{#2},{#3}}
}\xspace}

\newcommand{\Adv}[1]{\ensuremath{\algNotation{A}_{#1}}\xspace}
\newcommand{\AdvS}[1]{\ensuremath{\algNotation{S}_{#1}}\xspace}
\newcommand{\genericAdv}{\ensuremath{\mathcal{A}}\xspace}

\newcommand{\semiHonest}{\text{SH}\xspace}
\newcommand{\malicious}{\text{M}\xspace}
\newcommand{\maliciousC}{\text{M$^{\dagger}$}\xspace}

\NewDocumentCommand{\metricSpace}{o}%
{%
	\IfNoValueTF{#1}%
	{\ensuremath{\finiteField^{\numberOfPointsSim}}\xspace}%
	{\ensuremath{\setNotation{M}_{#1}}\xspace}
}

\newcommand{\PRPsecdef}{\textsf{\scriptsize opu}\xspace}

\newcommand{\WPRsecdef}{\textsf{\scriptsize wpr}\xspace}
\newcommand{\PRFINDsecdef}{\textsf{\scriptsize wpr}\xspace}
\newcommand{\Ssecdef}{\textsf{\scriptsize S}\xspace}
\newcommand{\DCOPRFsecdef}{\scriptsize\textsf{\exptAcro}\xspace}

\NewDocumentCommand{\ExptCC}{o}%
{%
  \IfNoValueTF{#1}%
{\ensuremath{\mathbf{Expt}^{\mbox{\scriptsize\textsf{\exptAcro}}}}\xspace}%
{\ensuremath{\mathbf{Expt}^{\mbox{\scriptsize\textsf{\exptAcro}}}_{#1}}\xspace}%
}

\NewDocumentCommand{\ExptS}{o}%
{%
  \IfNoValueTF{#1}%
{\ensuremath{\mathbf{Expt}^{\mbox{\scriptsize\textsf{\algNotation{S}}}}}\xspace
}%
{\ensuremath{\mathbf{Expt}^{\mbox{\scriptsize\textsf{\algNotation{S}}}}_{#1}
}\xspace}%
}

\NewDocumentCommand{\blocked}{g g g}%
{%
	\IfNoValueTF{#2}
{\ensuremath{\fnNotation{\mathit{blocked}}^{{#1}}}\xspace}
{
\IfNoValueTF{#3}%
	{\ensuremath{\fnNotation{\mathit{blocked}}^{{#1},{#2}}}\xspace}
	{\ensuremath{\fnNotation{\mathit{blocked}}_{#1}^{{#2},{#3}}}\xspace}
}
}

\newcommand{\ExptWPR}[1]{\ensuremath{\mathbf{Expt}^{\mbox{\textsf{\WPRsecdef}}}
_{#1}}\xspace}
\NewDocumentCommand{\ExptPRP}{o}%
{%
  \IfNoValueTF{#1}%
{\ensuremath{\mathbf{Expt}^{\mbox{\textsf{\PRPsecdef}}}_{\PRP}}\xspace}
{\ensuremath{\mathbf{Expt}^{\mbox{\textsf{\PRPsecdef}-{#1}}}_{\PRP}}\xspace}
}
\NewDocumentCommand{\ExptPRFIND}{o}%
{%
  \IfNoValueTF{#1}%
{\ensuremath{\mathbf{Expt}^{\mbox{\textsf{\PRFINDsecdef}}}_{\PRF[1]}}\xspace}
{\ensuremath{\mathbf{Expt}^{\mbox{\textsf{\PRFINDsecdef}-{#1}}}_{\PRF[1]}}\xspace}
}

\newcommand{\Advstate}[1]{\ensuremath{\phi_{#1}}\xspace}

\newcommand{\Advantage}[3]{\ensuremath{\mathsf{Adv}_{#2}^{\mbox{\textsf{#1}}
}\mathopen{}\left({#3}\right)\mathclose{}}\xspace}

\newcommand{\exptAcro}{CC}
\newcommand{\timeBound}{\ensuremath{t}\xspace}
\newcommand{\timeBoundAlt}{\ensuremath{t'}\xspace}
\newcommand{\forwardOracleQueries}{\ensuremath{q}\xspace}

\newcommand{\hashOracleQueries}{\ensuremath{\forwardOracleQueries_{\mathsf{H}}
}\xspace}

\newcommand{\cInputguess}{\ensuremath{\hat{\cInput}}\xspace}
\newcommand{\simOp}[1]{\ensuremath{\overline{#1}}\xspace}

\newcommand{\primitiveName}{\textit{blocklisted oblivious PRF}\xspace}
\newcommand{\PrimitiveName}{\textit{Blocklisted oblivious PRF}\xspace}
\newcommand{\primitiveNameShort}{B-OPRF\xspace}
\newcommand{\implicitCheckPhase}{implicit check\xspace}
\newcommand{\explicitCheckPhase}{explicit check\xspace}
\newcommand{\ImplicitCheckPhase}{Implicit check\xspace}
\newcommand{\ExplicitCheckPhase}{Explicit check\xspace}
\newcommand{\ImplicitCheckPhaseHeading}{Implicit Check\xspace}
\newcommand{\ExplicitCheckPhaseHeading}{Explicit Check\xspace}
\newcommand{\baselineNoEmbed}{PSI-C\xspace}
\newcommand{\baselineEmbed}{GC-ET-C\xspace}
\newcommand{\serverState}{\ensuremath{\phi}\xspace}

\newcommand{\genericProtocol}{\ensuremath{\Pi}\xspace}
\newcommand{\protocolEmbedESP}{\ensuremath{\genericProtocol_
{\mathit{EM}}}\xspace}
\newcommand{\protocolCommitESP}{\ensuremath{\genericProtocol_
{\mathit{TC}}}\xspace}
\newcommand{\protocolValidateESP}{\ensuremath{\genericProtocol_
{\mathit{IC}}}\xspace}

\newcommand{\protocolEmbed}{\protocolEmbedESP}
\newcommand{\protocolCommit}{\protocolCommitESP}
\newcommand{\protocolValidate}{\protocolValidateESP}
\newcommand{\protocolOLE}{\ensuremath{\genericProtocol_{\mathsf{OLE}}}\xspace}

\NewDocumentCommand{\protocolOPRF}{o}
{
	\IfNoValueTF{#1}
	{\ensuremath{\genericProtocol_{\mathit{OPRF}}}\xspace}
	{\ensuremath{\genericProtocol_{\mathit{OPRF}#1}}\xspace}
}

\newcommand{\protocolAKE}{\ensuremath{\genericProtocol_{\mathit{AKE}}}\xspace}

\newcommand{\embedAndMap}{\ensuremath{\mathcal{F}_{\mathsf{EM}}}\xspace}
\newcommand{\testAndCommit}{\ensuremath{\mathcal{F}_{\mathsf{TC}}}\xspace}
\newcommand{\authentication}{\ensuremath{\mathcal{F}_{\mathsf{IC}}}\xspace}
\newcommand{\editDistDefn}{\ensuremath{\vectorNorm{\mathsf{Edit}}{\cdot}}\xspace}
\newcommand{\genericMetricSpaceDist}{\ensuremath{D}\xspace}

\NewDocumentCommand{\genericPRPKey}{g}{\ensuremath{\langle
\PRPKeyVec[1]\IfNoValueF{#1}{_{#1}},
\PRPKeyVec[2]\IfNoValueF{#1}{_{#1}}\rangle}\xspace}

\begin{document}

\title{Blocklisted Oblivious Pseudorandom Functions}
\author{
Xinyuan Zhang \\ 
Duke University
\and
Anrin Chakraborti \\ 
University of Illinois at Chicago
\and
Michael Reiter \\ 
Duke University}
\maketitle

\begin{abstract}
An oblivious pseudorandom function (OPRF) is a protocol by which a
client and server interact to evaluate a pseudorandom function on a
key provided by the server and an input provided by the client,
without divulging the key or input to the other party.  We extend this
notion by enabling the server to specify a blocklist, such that OPRF
evaluation succeeds only if the client's input is not on the
blocklist.  More specifically, our design gains performance by
embedding the client input into a metric space, where evaluation
continues only if this embedding does not cluster with blocklist
elements.  Our framework exploits this structure to separate the
embedding and blocklist check to enable efficient implementations of
each, but then must stitch these phases together through cryptographic
means.  Our framework also supports subsequent evaluation of the OPRF
on the same input more efficiently.  We demonstrate the use of our
design for password blocklisting in augmented password-authenticated
key exchange, and to MAC only executables that are not similar to ones
on a blocklist of known malware.
\end{abstract}


\section{Introduction}
\label{sec:intro}

An oblivious pseudorandom function (OPRF) enables a 
\textit{client} to provide an input \cInput to an interactive protocol 
with a \textit{server}, and receive $\PRF(\cInput)$ in return, 
for a pseudorandom function family $\PRF[1]$ with key known only to 
the server.  The server does not learn $\cInput$, and the client does 
not learn the server's key. OPRFs are the main building block for many 
cryptographic primitives, e.g., for private set intersection (PSI) 
(e.g., \cite{Pinkas2015:HashingPSI, Kolesnikov2016:Efficient}),
deduplicating file storage (e.g., \cite{Bellare2013:DupLESS,
Duan2014:Distributed}), and private searching
(e.g., \cite{Freedman2005:OPRF, Cash2013:SSE}).  
OPRFs have been extensively used to authenticate clients to remote servers. 
For instance, OPAQUE~\cite{Jarecki2018:OPAQUE} integrates an OPRF with
key exchange to establish a password-based login system that is robust
against pre-computation attacks. Privacy Pass~\cite{Davidson2018:PP}
makes use of a OPRF variation to issue credentials to honest clients
and to guarantee that only clients deemed honest gain access to
further web resources.

A limitation of OPRFs is that the server cannot check the client's
input against policies that it wants to enforce, since the client
input is kept hidden from the server. So, we initiate the study of a
{\it policy-enforced} OPRF and focus on policies where adherence
implies that the client's input is not in a set of blocklisted items
compiled by the server. For such a blocklist policy enforcement, we
define a new primitive called \primitiveName (\primitiveNameShort). A
\primitiveNameShort protocol evaluates and outputs an OPRF on the
client's input only if this input is not on a pre-determined
blocklist, while hiding the client input from the server and keeping
the server's policy list secret from the client.  We define the
\primitiveNameShort ideal functionality, provide a provably secure
construction, and demonstrate its practicality in two
applications. One is privacy-preserving password blocklisting that can
be directly integrated with any existing augmented
password-authenticated key exchange (aPAKE). The other is a
blocklisted MAC where an authentication token is produced only if the
input file is not on a malware blocklist.

A natural way to approach this problem would be for the OPRF sender
(i.e. the server) to preface the OPRF evaluation with a private
membership test~\cite{Tamrakar17:PMT, Ram17:PMT, Kulshrestha2021}, or
to ``program'' the OPRF value so that the functionality outputs a
known value when the input is in the blocklist, and a pseudorandom
value otherwise~\cite{Chandran2022:OPPRF}.  In some applications,
however, the blocklist might be quite large, rendering either of the
approaches expensive. For this reason, we explore a version of
\primitiveNameShort in which the user's input and the blocklist
elements are embedded into a metric space wherein similar elements
cluster, and the OPRF evaluation proceeds only if the user's secret
input, once embedded, does not land in any of the blocklist elements'
clusters in the embedding space.  Provided that the blocklist elements
embed into relatively few clusters in the metric space, this can
substantially improve the efficiency of the check, albeit while
introducing quantifiable risks of false detections and acceptances
that result from the embedding.

This general strategy raises a technical challenge, however: if the
policy check is performed on an \textit{embedding} of the client
input, then the server needs to ensure that the embedding it
checks---which it also does not get to see---actually corresponds to
the client's input.  The most direct approach to do so would be to
perform the embedding and check together in a single immutable
computation, for which a monolithic garbled circuit is a natural
candidate (e.g.,~\cite{Katz2016:Input2PC}).  As we will show in
\secref{sec:eval}, however, this approach does not scale well, either,
in the number of blocklist elements (or their clusters).  Our
\primitiveNameShort construction scales better than the approaches
outlined above, while maintaining strong security guarantees against
deviating parties.


Our exploration of \primitiveNameShort{}s reveals another
improvement, as well, in applications where a client will
repeatedly evaluate the \primitiveNameShort on the same input.  Once
the policy check has been passed on a client input, our framework can
store a token at the server that permits the client to re-evaluate the
\primitiveNameShort on the same input without repeating the policy
check.  We refer to this method of evaluating the \primitiveNameShort
as an \textit{\implicitCheckPhase}, as it only verifies that the
client provided an input that previously passed a policy check.  We
henceforth refer to the protocol in which the client first evaluates
the \primitiveNameShort as an \textit{\explicitCheckPhase}.

We now provide more intuition on the practical utility of
\primitiveNameShort{}s and applications we will
demonstrate in \secref{sec:app}.

\myparagraph{Augmented password-authenticated key exchange}
aPAKEs~\cite{Bellovin1993:aPAKE,Boyko2000:aPAKE,Gentry2006:aPAKE} are
an elegant mechanism to build password-based authentication
systems. An aPAKE server uses a secret derived from a user's password
$\pw$ to authenticate the user, without receiving the user's password
itself.  Because the server is unable to observe $\pw$, it cannot
check if $\pw$ follows any of the password strength policies that the
server would like to enforce. The OPAQUE (the state-of-the-art aPAKE) IETF
draft~\cite{IETF2025:OPAQUE} notes that OPAQUE's security goals
interfere with checking all but very limited policies (e.g., detecting
password reuse).  Assuming the client is honest and does not want to
degrade security, and the server's policy is public (e.g. a blocklist
of the top 10,000 passwords), a naive approach would be to prepend a
password strength meter to an OPRF evaluation. However, this approach
would trust the client to perform this check; a client who values
convenience more than security (i.e., most
clients~\cite{Florencio2007, Shay2010}) might then omit this check.
Even if the client can be trusted to perform this check, some servers
use customized rules targeting specific security
goals~\cite{Florencio2010:secPolicy} and so might not want to make the
policy public. Therefore, server-side mitigating practices are more
appropriate for this problem setting.

We show in \secref{sec:pake} that a \primitiveNameShort can be used in 
aPAKEs to enforce that the client's password satisfies the server's blocklist 
policy. Ensuring that a client's password is not similar to any of the most
commonly used ones is an effective way to protect against most online
attacks (e.g.,~\cite{Weinert2019:Password}).  With a common password
blocklist that embeds into 1,500 clusters, registration with our
protocol (an \explicitCheckPhase) completes in 1.42 seconds, and
authentication (an \implicitCheckPhase) takes 11 milliseconds.  We
improve the communication cost of both phases by a factor of
$1.53\times$, compared to the monolithic garbled-circuit design outlined
above, while achieving a reasonably low false acceptance rate and 
false rejection rate. 

\myparagraph{Message Authentication Code} \primitiveNameShort can be
used to build a blocklisted MAC functionality that provides private
checking on the client input against a predetermined blocklist and
produces a MAC tag showing that the input is compliant with this
blocklist policy. A blocklisted MAC can be useful in many
applications, such as spam filtering and malware defense. For example,
Secure Email Gateways (SEGs)~\cite{SEG} scan email attachments to
enforce corporate email policies. It is standard to integrate SEGs
with DomainKeys Identified Mail (DKIM) and add cryptographic signatures
to outbound email headers to indicate that the scanning has occurred
and the document is policy-compliant~\cite{SEGAuth}.  A blocklisted
MAC can enable the sending organization to have its outbound emails
scanned by a third-party provider that issues a MAC for the email if,
say, the attachments are not on a malware blocklist; this would
constitute an \textit{explicit check}.  The receiving organization,
then, would validate the MAC on the inbound email by interacting with
the third party via an \textit{implicit check}.  Critically, the third
party would never learn the email contents; the third party's
blocklist would remain secret; and the receiving organization, who
might not even be a customer of the third-party provider, would not
need to maintain a trusted signature verification key for the
provider.

We evaluate the performance of a \primitiveNameShort-based blocklisted
MAC for this application. When instantiated with a blocklist of
$1{,}000$ executable embeddings, the \explicitCheckPhase from the
sending client to the policy server completes in 41.34 seconds and the
\implicitCheckPhase between the receiving client and the policy server
takes 4.67 seconds. On average the latency of a monolithic garbled-circuit is 
$1.69\times$ the latency of this tailored solution, and the total network traffic 
of the garbled-circuit baseline is $2.33\times$ the traffic cost of the tailored 
protocol.

To summarize, our contributions are as follows.
\begin{itemize}[nosep,leftmargin=1em,labelwidth=*,align=left]
\item We formally define the \primitiveNameShort primitive, and provide 
a general framework to design \primitiveNameShort protocols. 

\item We provide 
efficient \primitiveNameShort implementations for blocklists that can be embedded 
into a metric space based on Hamming distance. 

\item We demonstrate two different applications of a \primitiveNameShort.  
  First, we augment the state-of-the-art aPAKE
  protocol~\cite{Jarecki2018:OPAQUE} and introduce 
  mechanisms to prevent client registration with weak passwords, 
  {\it without diminishing security and performance}.  
  Second, we use our framework to design an efficient
  three-party blocklisted MAC for similarity-based malware defense.
  
\end{itemize}

\section{Related Work}
\label{sec:related}

Below we discuss tools that are relevant to our task, which include
OPRF variants, PSI protocols, and others.

\myparagraph{OPRF variants}
OPRFs have been extensively researched since \citet{naor2004number} 
formalized them; see~\citet{casacuberta2022sok}. Some of the variations 
and corresponding deployments bear similarity to our \primitiveNameShort. 
Programmable OPRFs (OPPRF)~\cite{Chandran2022:OPPRF} and Oblivious 
Key-Value Store (OKVS)~\cite{Garimella2021:OKVS} both outputs an associated 
value that is known to server if $\cInput$ matches a server list item, while the output 
is pseudorandom elsewhere. Another more distantly related variant is partially oblivious 
PRFs (POPRF)~\cite{Tyagi2022:POPRF}, which extends the OPRF functionality by including 
a public tag \POPRFtag and outputs $\PRF(\POPRFtag, \cInput)$ to the client. 
OPPRF, OKVS, and POPRF can all support verifiable token generating protocols,
but none of them explicitly considers any policy enforcement over the client input, while
\primitiveNameShort checks an input based on a blocklist and outputs
a verifiable token for that exact valid input.

We note that OPPRFs and POPRFs bring non-trivial drawbacks when adapting to 
the \primitiveNameShort setting. First, OPPRFs work in a model where the receiver is 
semi-honest, while in \primitiveNameShort, \textit{we consider a threat model where 
the receiver can be malicious}. Second, if the blocklist is large, programming the OPRF 
is expensive as the costs scale linearly in the size of blocklist (see \tblref{tab:comp_protocols}). 
Although it is possible to first embed the blocklist in a metric space (as we plan to do 
in our protocol) and then program the OPRF, this would introduce additional mechanisms
to ensure that the client submits a correctly embedded input to the subsequent OPRF.

\begin{table}[t]
{\small
\begin{tabular}{@{}l@{}c@{\hspace{0.5em}}c@{\hspace{0.5em}}c@{\hspace{0.5em}}c@{}}
\toprule
\multicolumn{1}{c}{\multirow{2}{*}{Protocol}} &
  Computation & Comms.\ & Input & \multirow{2}{*}{Sec.} \\
& Cost & Cost & Assum.\ \\
\midrule
\citet{Wang2015:approx_editDist} & $\bigO{\setSize{\policyLan}}^\starNot$ & \bigO{1} & yes & \semiHonest \\
OKVS \cite{Garimella2021:OKVS} & \bigO{\setSize{\blockedElmts}} & \bigO{\setSize{\blockedElmts}} & no & \malicious \\
Programmable OPRF \cite{Chandran2022:OPPRF} & \bigO{\setSize{\blockedElmts}} & \bigO{\setSize{\blockedElmts}} & no & \semiHonest \\
Similarity-aware C3 \cite{Pal2022:C3} & $\bigO{\setSize{\policyLan}}^*$ & $\bigO{\setSize{\policyLan}}^*$ & no & \semiHonest \\
Structure-aware PSI \cite{Garimella2022:structurePSI, Garimella2024:structurePSI} & \bigO{\setSize{\policyLan}} & \bigO{\setSize{\policyLan}} & yes & \malicious \\
Distance-aware PSI \cite{Chakraborti2023:DAPSI,Uzun2021:FuzzyPSI} & \bigO{\setSize{\policyLan}} & \bigO{\setSize{\policyLan}} & no & \semiHonest \\
Approximate PSI \cite{Cho2024:approxPSI} & \bigO{\setSize{\policyLan}} & \bigO{\setSize{\policyLan}} & yes & \semiHonest \\
\primitiveNameShort (this paper) & \bigO{\setSize{\policyLan}} & \bigO{\setSize{\policyLan}} & no & \maliciousC \\
\bottomrule
\end{tabular}
}
\caption{Protocol comparison.
  \policyLan: embedded blocklist; \blockedElmts: original blocklist;
  \universe: input universe; \starNot: $\setSize{\policyLan} \ge
  \setSize{\blockedElmts}$; Input Assumptions: `yes' if the efficiency
  of the scheme relies on assumptions about the input distribution,
  `no' otherwise; Security: \semiHonest = semi-honest, \malicious =
  malicious, \maliciousC = malicious client or semi-honest server.}
\label{tab:comp_protocols}
\end{table}


\myparagraph{PSI-based mechanisms}
An alternative design would be to preface a plain OPRF with a private 
membership test (e.g., using a PSI like functionality). As discussed before, 
running a general PSI over the entire blocklist might be too costly due to 
the blocklist size. Therefore, we only focus on fuzzy PSI mechanisms. The most 
notable works are listed in \tblref{tab:comp_protocols}\footnote{All asymptotic 
costs listed are for a client with a single element set input. We have omitted 
factors showing reliance on the distance metric for the fuzzy PSI designs since 
the protocols work for different metric spaces under different data assumptions.}. 
Structure-aware PSIs~\cite{Garimella2022:structurePSI, Garimella2024:structurePSI} 
partition the points in the metric space into \textit{tiles} that can be then queried using 
OKVS. However, spatial hashing guarantees reasonable compute costs only under 
certain assumptions about the data in one (or both) of the input sets. 
Approximate PSI~\cite{Cho2024:approxPSI} is a similarly motivated concept that 
also works only under certain distribution assumptions. As it is unclear if these 
distribution assumptions hold across a variety of applications, we cannot 
generally use a structure-aware PSI or an approximate PSI as a building block 
for a \primitiveNameShort. 

\textit{Distance-aware} PSI~\cite{Chakraborti2023:DAPSI} performs fuzzy matching 
over two sets \textit{without any data distribution assumptions}. While this comes at 
a higher cost than both structure-aware PSI and approximate PSI, distance-aware PSI 
is more general and more applicable because they are agnostic to the type of 
application, data, etc. We will use \cite{Chakraborti2023:DAPSI} as a building block, as 
we will describe in \secref{sec:background}. 

Beyond general application-agnostic designs, \citet{Wang2015:approx_editDist} 
realizes an approximate edit-distance check on genome data. Similar to a 
\primitiveNameShort, they utilize an embedding phase and a test phase to 
compare entries at the client and the server. Similarly, the similarity-aware 
compromised credential checking (C3)~\cite{Pal2022:C3} uses a generative 
model to ``guess'' possible tweaks for a stuffed credential.
However, these solution assume a semi-honest client, while \primitiveNameShort 
does not trust the client to provide the correct embedding and detects when the 
client submits inconsistent inputs to different phases. In addition, 
\primitiveNameShort removes the assumptions on the structure of input data by
\citet{Wang2015:approx_editDist}, that are more specific to genome data 
and may not be applicable to other settings. 




Aside from general OPRF constructions and approximate matching techniques, 
\primitiveNameShort can compare to primitives
that consider similar design goals, as described below.

\myparagraph{Enforcing Policies on Protocol Inputs}
Privacy-preserving input validation has been considered in other
contexts. For example, Private Hash Matching \cite{Kulshrestha2021} is 
a service designed for end-to-end encrypted messaging, where a server 
matches a client's media content against a set of images (e.g., child sexual 
abuse material) while hiding the server's hash set or nonmatching 
content from the client. It does not
consider setting up a shared secret for verification later. 

Several protocols (e.g.,~\cite{Camenisch2009:CertSet,
  Blanton2015:genom, Zhang2017:Input}) address the general problem of
ensuring that inputs to a multiparty computation satisfy some
predicate through a trusted third party, but in our applications,
finding a trusted third party is problematic. Others
(e.g.,~\cite{Chang2021:stat, Prio, Zombie, Reef, Luo2023:rexMatching})
prove in zero knowledge certain properties of an input. However, the
proof complexity increases as a function of the property that needs to
be checked. Besides, zero-knowledge policy-enforcement designs work
only in settings where the matching pattern and policy circuit are
public, while \primitiveNameShort aims at keeping the server’s
blocklist private.  Secure computation tools, e.g., garbled circuits,
provide a viable alternative. \citet{Katz2016:Input2PC} proposed a
scheme where a garbled circuit is preceded by a predicate-checking
circuit to ensure that the input follows certain properties.  A
similar construction is proposed by
\citet{Baum2016:GC}. \citet{Agrawal2021:covert} proposed a MPC scheme
by which a party can commit to a value and use this value subsequently
in a secure computation. However, a garbled circuit may not be the
most efficient primitive for performing a certain task privately
\cite{Garg2018:pkGC}, especially when dealing with sets of items or
computing public-key commitments.

\myparagraph{Password Policy Enforcement}
Various methods can check a password string against certain policies
during registration without revealing the password to the server
(e.g., \cite{Kiefer2014:ZKPPC, Dong2015:SPC, Kiefer2016:BPR,
  Nguyen2018:lattice}). However, these techniques are
restricted to password length requirements and limited character-level
checks. As we will show, a \primitiveNameShort protocol is a more powerful 
tool since it allows us to model password strength in the form of 
approximate matches against a reference ``easy'' password list, 
which has been used in real-world password strength meters.
Unlike popular password strength meters~\cite{zxcvbn, Melicher16:Password},
which are based on a banned list, \primitiveNameShort further allows entities to 
customize the blocklist, which is considered as a better security practice in many 
organizations~\cite{Microsoft2024:Entra}.

\myparagraph{Malware Defense} 
We also apply \primitiveNameShort to implement blocklisted MACs, with
a driving application being generating MACs only for email attachments
that a third-party service provider confirms are not on its malware
blocklist.  Our framework is best suited to malware detection by fuzzy
hashing~\cite{Roussev10:sdhash, Li15:FHReview} and provides the
content privacy necessitated by outsourcing this scanning to a
third-party service.  \citet{Sun17:PriMal} proposed a
privacy-preserving cloud-based malware detection scheme that is
perhaps closest to our envisioned design.  However, it works by
exactly matching client input files with malware
signatures, which does not scale as well as our approach.

\section{Framework}
\label{sec:framework}

In this section, we introduce our threat model and the framework. 
We start by defining the necessary notations and cryptographic primitives. 

\subsection{Preliminaries}
\label{sec:prelim}

\myparagraph{Notation} 
Let $\genericFnFamily: \genericFnKeyspace \times \genericFnDomain
\rightarrow \genericFnRange$ denote a function family with
\textit{keyspace} \genericFnKeyspace, \textit{domain}
\genericFnDomain, and \textit{range} \genericFnRange.  For
$\genericFnKey \in \genericFnKeyspace$, we denote by
$\genericFnFamily[\genericFnKey]: \genericFnDomain \rightarrow
\genericFnRange$ the function with
$\genericFnFamily[\genericFnKey](\genericFnDomainElmt) =
\genericFnFamily(\genericFnKey, \genericFnDomainElmt)$ for each
$\genericFnDomainElmt \in \genericFnDomain$.  Let \nonnegativeReals
denote the non-negative real numbers. 
For the description of the cryptographic primitives, we 
use $\finiteField$ to denote a finite field, and $\polyNotation{p} \in
\polyField$ to denote a polynomial whose coefficients are drawn from
$\finiteField$. A random polynomial $\genericPoly \getsr \polyField$
is a polynomial with coefficients uniformly sampled from
$\finiteField$. Given two vectors $\genericVec[1], \genericVec[2] \in 
\finiteField^{\numberOfPointsSim}$, $\genericVec[1] \hadamardProd \genericVec[2] \in 
\finiteField^{\numberOfPointsSim}$, and ``\hadamardProd'' is the Hadamard product. 
We use $\secParam$ to denote the security parameter
for our schemes.

\myparagraph{OPRF}
OPRF \cite{Freedman2005:OPRF} is a cryptographic primitive involving a 
two-party protocol where one party, the client, holds an input \cInput and 
the other party, the server, holds a secret key \PRFKey. The goal is for the 
client to learn $\PRF[][\PRFKey](\cInput)$, the output of a PRF evaluated 
at \cInput with the key \PRFKey while ensuring:

\begin{itemize}[nosep,leftmargin=1em,labelwidth=*,align=left]
	
	\item Pseudorandomness: The function $\PRF[][\PRFKey](\cdot)$ behaves
	indistinguishably from a random function to any computationally
	limited adversary who does not know \PRFKey.
	
	\item Obliviousness: The server learns nothing about the input $\cInput$, 
	and the client learns nothing about the key \PRFKey.
	
\end{itemize}

\myparagraph{Oblivious Linear Evaluation ($\oleFunc$)} OLE is a two-party
cryptographic primitive between a server and a client.  
The client's input is $\OLEAliceInput \in
\finiteField$; the server's inputs are $\OLEBobInputOne,
\OLEBobInputTwo \in \finiteField$. The ideal function $\oleFunc$
returns to the client $\OLEBobInputOne \OLEAliceInput +
\OLEBobInputTwo$ \textit{without revealing \OLEBobInputOne and
	\OLEBobInputTwo}.  No information is revealed to the server. In our
protocols, we will use the maliciously secure OLE construction by
\citet{Ghosh2017:OLE}.


\myparagraph{Noisy Polynomial Addition ($\npaFunc$)}
\citet{Ghosh2019communication} introduced a {\it noisy polynomial addition}
functionality \npaFunc that takes as input pairs of vectors $\randomVec[1], \cInputVec \in 
\finiteField^{\numberOfPointsSim}$ from one 
party, say a client, and another pair of vectors $\randomVec[2], \cInputVecAlt \in 
\finiteField^{\numberOfPointsSim}$ from another party, say a server, and computes a 
linear combination of these vectors. 
Specifically, the functionality returns to either one or both of the parties the vector 
$\randomVec[2] \hadamardProd \cInputVec + \randomVec[1] \hadamardProd \cInputVecAlt$, 
where ``\hadamardProd'' is the Hadamard product. The functionality does not reveal 
$\randomVec[1], \cInputVec$ to the server and $\randomVec[2], \cInputVecAlt $ to the client (beyond 
what can be learned from the linear combination).

\myparagraph{One-time pairwise unpredictable function}
We will require a function that is {\it one-time pairwise unpredictable}. 
Intuitively, a keyed-function $\genericFnFamily: \genericFnKeyspace \times \genericFnDomain \rightarrow \genericFnRange$  
is one-time pairwise unpredictable if given the output at a specific point 
under a randomly chosen key, an adversary cannot guess the output at another distinct point. 
The function is one-time in the sense that each new invocation of the function 
requires a 
new random key selection to preserve unpredictable outputs.

We also extend this notion to a {\it one-time pairwise unpredictable permutation} 
over a metric space for our construction. We present the formal
definition in \appref{app:defn_prp}.

\subsection{Goals}
We are concerned with building secure mechanisms for applications where 
a client authenticates itself to a server using a token that is
derived off the client's input $\cInput$ {\it only if $\cInput$ is compliant with 
a blocklist policy that the server would like to enforce}. Therefore, we will introduce a new 
cryptographic primitive called a \primitiveName (\primitiveNameShort 
in short). Intuitively, \primitiveNameShort is a two-party primitive that 
takes an input $\cInput \in \universe$ from a client $\client$, where \universe is 
the input universe, and a list of items $\blockedElmts \subset \universe$
and a key for a PRF \PRFKey from a server \server. The primitive should
produce $\PRF[][\PRFKey](\cInput)$ \textit{only if} $\cInput
\not\in \blockedElmts$; if $\cInput \in \blockedElmts$, the protocol
should abort.

That said, there are many scenarios where the blocklist \blockedElmts is too large 
to be specified directly as an input to a \primitiveNameShort protocol (as we will show
in \secref{sec:eval}). Therefore, we focus on settings where the set \blockedElmts
can be embedded into a metric space \metricSpace with distance metric
$\distance: \metricSpace \times \metricSpace \rightarrow
\nonnegativeReals$, and the embedded set can be summarized by a much smaller set
$\policyLan \subset \metricSpace$.  That is, in lieu of specifying
\blockedElmts to the protocol directly, the server instead specifies a
blocklist $\policyLan \subset \metricSpace$, $\setSize{\policyLan}
\ll \setSize{\blockedElmts}$, and a threshold \threshold such that
elements of \blockedElmts, once embedded in $\metricSpace$, should fall
within distance \threshold of some elements of \policyLan. Since ``similarity'' is 
defined for distance metrics, we use metric embedding, which is standard for 
information retrieval and approximating text data~\cite{Broder97:Sim}. Mathematically, for an 
output \cInputEmbedding embedded from an element in \blockedElmts,
\begin{align*}
	\blocked{}{\policyLan}{\threshold}(\cInputEmbedding)
	& \defeq \left(\exists \lanElmt \in \policyLan: \distance\left(\cInputEmbedding, \lanElmt\right) \le \threshold\right)
\end{align*}
should be true.

Naturally, such an embedding will not be perfect, and will introduce non-zero false-accept and false-reject rates. Here we consider a randomized embedding function family of the form $\embed: \embedKeySpace \times \universe \rightarrow \metricSpace$.
For instance, the keyspace \embedKeySpace might include the random
vectors used for projections in locality-sensitive hashing algorithms
(e.g.,~\cite{datar2004locality}).  We then define true- and false-reject
rates as follows:
{\small
\begin{align*}
	\trueRejectRate{\blockedElmts}{\policyLan}{\threshold}
	& \defeq \min_{\cInput \in \blockedElmts} 
	\cprob{\Big}{\blocked{\policyLan}{\threshold}(\cInputEmbedding)}{\embedKey \getsr 
	\embedKeySpace, \cInputEmbedding \gets \embed[\embedKey](\cInput)} \\
	\falseRejectRate{\blockedElmts}{\policyLan}{\threshold}
	& \defeq \max_{\cInput \in \universe\setminus\blockedElmts} \cprob{\Big}{\blocked{}{\policyLan}{\threshold}(\cInputEmbedding)}{\embedKey \getsr \embedKeySpace, \cInputEmbedding \gets \embed[\embedKey](\cInput)}
\end{align*}
}

In words, \trueRejectRate{\blockedElmts}{\policyLan}{\threshold} is a
lower bound on the probability of blocking an input \cInput that
should be blocked, and
\falseRejectRate{\blockedElmts}{\policyLan}{\threshold} is an upper
bound on the probability of blocking an input \cInput that should not
be blocked.  As usual, false- and true-accept rates can be defined by
$\falseAcceptRate{\blockedElmts}{\policyLan}{\threshold} \defeq 1 -
\trueRejectRate{\blockedElmts}{\policyLan}{\threshold}$ and
$\trueAcceptRate{\blockedElmts}{\policyLan}{\threshold} \defeq 1 -
\falseRejectRate{\blockedElmts}{\policyLan}{\threshold}$,
respectively.  Naturally, we strive to maximize
\trueRejectRate{\blockedElmts}{\policyLan}{\threshold} and
\trueAcceptRate{\blockedElmts}{\policyLan}{\threshold}.
The definition implies that inputs are not uniformly distributed, while 
keys are. This corresponds to the applications, where inputs are user-chosen.

\begin{defn}[\PrimitiveName]
  \label{dfn:primitive}
  A \primitiveName for blocklist \blockedElmts is a two-party protocol
  between a client and a server, consisting of two phases:
  \explicitCheckPhase and \implicitCheckPhase.
  \begin{itemize}[nosep,leftmargin=1em,labelwidth=*,align=left]
  \item {\bf \ExplicitCheckPhaseHeading:} The client inputs $\cInput \in
    \universe$ and the server inputs $\blockedElmts \subset \universe$,
    $\policyLan \subset \metricSpace$, $\threshold \in
    \nonnegativeReals$ (with both \policyLan and \threshold computed
    from \blockedElmts), and $\PRFKey \in \PRFKeySpace$.  This phase
    \textit{succeeds} with probability at least
    \trueAcceptRate{\blockedElmts}{\policyLan}{\threshold} if $\cInput
    \not\in\blockedElmts$ and with probability at most
    \falseAcceptRate{\blockedElmts}{\policyLan}{\threshold} otherwise.
    If this phase succeeds, then the client outputs
    $\PRF[][\PRFKey](\cInput)$ and the server outputs state
    \serverState (containing \PRFKey).  Otherwise, the server aborts.
    The client outputs $\bot$.

   \item {\bf \ImplicitCheckPhaseHeading:} The client inputs $\cInput 
     \in \universe$ and the server inputs \serverState. This
     phase \textit{succeeds} if \serverState was the output in a
     successful \explicitCheckPhase using $\cInput$.
     If so, the client outputs
     $\PRF[][\PRFKey](\cInput)$ again.  Otherwise, both output $\bot$.
   \end{itemize}
\end{defn}
We will define the guarantees that we want to provide from
these steps depending on the following threat models.


\subsection{Threat model}
\label{sec:framework:threat-model}

The threat models we consider for our applications in \secref{sec:app} are:
\begin{enumerate}[label=\textbf{T\arabic*},nosep,align=left,labelwidth=*,leftmargin=1.7em]
\item The server is honest, but the client is honest-but-curious
  during both \explicitCheckPhase and \implicitCheckPhase.  In this
  case, our primary concern is leaking more information about
  \blockedElmts to the client than the decision ($\cInput \in
  \blockedElmts$ or not) implies. We note that the server needs 
  to limit the number of \explicitCheckPhase attempts. Otherwise, the
  client can online guess the blocklist entries.
  \label{model:client-hbc}
\item The server is honest, but the client is malicious during both
  \explicitCheckPhase and \implicitCheckPhase, where the client
  deviates from the protocol to circumvent the check.  
  \label{model:client-mal}
\item The client is honest, but the server is honest-but-curious
  during \explicitCheckPhase and malicious during \implicitCheckPhase.
  This is possible since \explicitCheckPhase is one-time but
  \implicitCheckPhase can be repeated an arbitrary number of times;
  the server may be compromised at any time after \explicitCheckPhase.
  In this case, our concern is leaking information about \cInput to
  the server: the server should not be able to learn $\cInput$ with
  any additional advantage over performing a dictionary attack on
  $\cInput$.
  \label{model:server-mal}
\end{enumerate}
We do not consider a threat model in which the server is always
malicious, simply because we do not consider that reasonable in our
applications.

\subsection{Building Blocks}
\label{sec:framework:building-blocks}

We are now ready to describe the ideal functionalities for implementing our general 
framework. Recall that our \primitiveNameShort primitive has two phases
that share states, an \explicitCheckPhase and an \implicitCheckPhase. Furthermore, to 
accommodate the threat models above, we will further split the \explicitCheckPhase functionality 
into two separate ideal functionalities. The rationale is that the \explicitCheckPhase can be viewed 
as a two-step process: i) embed the client $\client$'s input into the metric space \metricSpace, 
and ii) perform the predicate check $\blocked{}{\policyLan}{\threshold}(\cdot)$ on the embedded 
input against the embedded blocklist $\policyLan$ provided by the server \server. 
Below we present the details for the three ideal functionalities: i)  \textit{Embed-and-Map} 
$\embedAndMap$,
ii) \textit{Test-and-Commit} $\testAndCommit$,
and iii) \textit{\ImplicitCheckPhase} $\authentication$. 
generation on a previously checked input.

%
%
%


\noindent{\bf Embed-and-map:} Given an embedding function 
$\embed$, $\client$ input $\cInput$, and \server input $\embedKey$, 
the functionality first computes $\embed[\embedKey](\cInput)$ and 
$\hash(\cInput,\embed[\embedKey](\cInput), \embedKey)$ where 
$\hash$ is a random oracle; we will explain the need for $\hash(\cInput,
\embed[\embedKey](\cInput), \embedKey)$ later. These values are expected 
as subsequent inputs to the Test-and-Commit functionality. However, since a 
malicious $\client$ may not follow the protocol, we make Embed-and-Map output 
{\it one-time pairwise unpredictable} permutations of these values instead with
permutation keys provided by \server. The rationale is that if $\client$ provides 
incorrect values to Test-and-Commit, then their inverses, which will be used in further 
computations, will be unpredictable, preventing $\client$ from learning 
the correct OPRF output and/or succeeding in an \implicitCheckPhase later.

\noindent{\bf Test-and-Commit:}
During Test-and-Commit, the functionality \testAndCommit inverts the outputs 
of $\client$'s preceding invocation of \embedAndMap. Note that the inversions 
are unpredictable if the client deviates. Then, the functionality checks whether
the inverted value is blocklisted. If not, it produces $\authToken \assign 
\embed[\embedKey](\cInput) \linearComb \hash(\cInput,\embed[\embedKey]
(\cInput), \embedKey)$, which serves as the \textit{authentication token} that will 
be used later for the \implicitCheckPhase.  It outputs $\PRF[][\PRFKey](\cInput) = 
\PRF[1][\PRFKey[1]]\left(\authToken\right)$ to $\client$ and $\authToken$ to \server. 
Again, note that if the client deviates, $\authToken$ is unpredictable to the client, 
and therefore hard to generate later for the \implicitCheckPhase . \server locally 
stores \authToken, and the keys. Finally, note that \authToken ``blinds'' 
$\embed[\embedKey](\cInput)$ from \server using $\hash(\cInput,
\embed[\embedKey](\cInput), \embedKey)$, i.e.,  due to the random oracle $\hash$, 
\server learns $\embed[\embedKey](\cInput)$ from \authToken only if it is able to 
guess $\cInput$. This addresses the security requirements for a semi-honest 
\server (threat model \ref{model:server-mal}).

\noindent{\bf \ImplicitCheckPhase:}
The blocklist check introduces non-trivial overhead over the OPRF 
generation. In many applications, it is necessary for $\client$ to regenerate 
$\PRF[][\PRFKey](\cInput)$ for the same input \cInput repeatedly.  
More specifically, $\client$ should only be able to re-generate
$\PRF[][\PRFKey](\cInput)$ if and only if \explicitCheckPhase 
has previously succeeded with $\cInput$.  To accomplish this, the server will 
retain state $\serverState = \langle \embedKey, \PRFKey[1], \authToken \rangle$ 
after a successful \explicitCheckPhase, though the client need not; it only must 
remember $\cInput$. During \implicitCheckPhase, $\client$ can be provided the 
embedding key \embedKey in the plain: each \explicitCheckPhase requires a 
new embedding key, and once $\client$ has set up the authentication token during 
the \explicitCheckPhase, the embedding key does not provide any additional 
information/advantage to a malicious $\client$ in circumventing the policy check. 
If $\client$ correctly generates $\authToken \assign \embed[\embedKey](\cInput) 
\linearComb \hash(\cInput,\embed[\embedKey](\cInput), \embedKey)$, it learns 
$\PRF[][\PRFKey[]](\cInput)$, otherwise it learns nothing.  


\begin{figure}[h]
\begin{oframed}
\noindent\small

\smallskip\noindent \textbf{Parameters:~} An embedding function
$\embed: \embedKeySpace \times \universe \rightarrow \metricSpace$; a one-time 
pairwise unpredictable 
permutation family over the metric space $\PRP: \PRPKeySpace \times 
\metricSpace\rightarrow \metricSpace$; a random oracle $\hash: \{0,1\}^{\ast}
\rightarrow \metricSpace$; and a function family
$\PRF[1]: \PRFKeySpace[1] \times \metricSpace \rightarrow \rangePRF$.

\noindent\underline{$\embedAndMap$: Embed-and-Map functionality}

	\smallskip\noindent \textbf{Inputs:~} $\client$ inputs $\cInput
\in \universe$. \server inputs $\embedKey \in
\embedKeySpace$, and $\PRPKey[1], \PRPKey[2]
\in \PRPKeySpace$.

\smallskip\noindent \textbf{Output:~} $\client$ learns
$\prpOutputEmbedding \assign
\PRP[\PRPKey[1]]\left(\embed[\embedKey](\cInput)\right)$ and
$\prpOutputHash \assign \PRP[\PRPKey[2]]\left(\hash(\cInput,
\embed[\embedKey](\cInput), \embedKey)\right)$.  \server learns
nothing.

\noindent\underline{$\testAndCommit$: Test-and-Commit Functionality}

\smallskip\noindent \textbf{Inputs:~} $\client$ inputs
$\prpOutputEmbedding$, $\prpOutputHash \in \metricSpace$.
\server inputs $\policyLan \subset \metricSpace$, $\threshold
\in \nonnegativeReals$, keys $\PRPKey[1]$, $\PRPKey[2]
\in \PRPKeySpace$, and $\PRFKey[1] \in \PRFKeySpace[1]$.

\smallskip\noindent \textbf{Output:~} If
$\neg\blocked{\policyLan}{\threshold}(\PRP[\PRPKey[1]]^{-1}
(\prpOutputEmbedding))$, then $\client$ learns $\prfOutput \assign 
\PRF[1][\PRFKey[1]]\left(\authToken\right)$ for $\authToken \gets
\PRP[\PRPKey[1]]^{-1}(\prpOutputEmbedding) \linearComb
\PRP[\PRPKey[2]]^{-1}(\prpOutputHash)$, and \server learns
$\authToken$. Otherwise, \server aborts and learns $\{\lanElmt \in \blockList: \distance\left(\PRP[\PRPKey[1]]^{-1}(\prpOutputEmbedding), \lanElmt\right) \le \threshold\}$\footnotemark. $\client$ outputs $\bot$. 

\noindent\underline{$\authentication$: \ImplicitCheckPhaseHeading Functionality}

\smallskip\noindent \textbf{Inputs:~} $\client$ inputs
$\authTokenAlt \in \metricSpace$.  \server inputs key $\PRFKey[1]
\in \PRFKeySpace[1]$ and $\authToken \in \metricSpace$.

\smallskip\noindent \textbf{Output:~} If $\authToken = \authTokenAlt$
then $\client$ receives $\prfOutput \assign
\PRF[1][\PRFKey[1]](\authTokenAlt)$; otherwise, $\client$ learns
$\bot$. \server learns $\bot$.

\end{oframed}
\captionof{figure}{Ideal Functionalities used in Implementation}
\end{figure}

\footnotetext{Note that the server can implicitly 
learn this information even if the functionality did not return it, as an abort would imply 
\cInput is close to one of the server's inputs.}

We present the framework utilizing these functionalities in
\figref{fig:frameworkSH}, i.e., assuming that \embedAndMap,
\testAndCommit, and \authentication are secure in whichever threat
model in \secref{sec:framework:threat-model} is in force.  We present 
a proof of security of the framework in \appref{app:proof_framework}, 
relative to the properties of $\PRF[1]$, $\PRP$ and
$\hash$. Note that in \stepref{frameworkSH:commit}, $\client$ learns \prfOutput
even though $\client$ neither uses \prfOutput subsequently in
\explicitCheckPhase nor stores it.  (Recall that we require that
$\client$ be able to perform \implicitCheckPhase knowing \textit{only}
$\cInput$.) This is done so that $\client$ can use \prfOutput in other protocol 
steps.

\begin{figure}[h]
\begin{oframed}
{\small
\noindent\underline{\primitiveNameShort Framework}
		

\smallskip\noindent \textbf{Inputs:~} $\client$ inputs $\cInput \in
\universe$.  \server inputs $\policyLan \subset \metricSpace$
and $\threshold \in \nonnegativeReals$.

\smallskip\noindent
\textbf{\ExplicitCheckPhase:}
\begin{enumerate}[topsep=0.5em,leftmargin=1.75em,labelwidth=*,align=left]

\item \server selects $\embedKey \getsr \embedKeySpace$, $\PRFKey[1]
  \getsr \PRFKeySpace[1]$, and $\PRPKey[1], \PRPKey[2]
  \getsr \PRPKeySpace$.

\item $\client$ and \server call \embedAndMap. $\client$ inputs
  $\cInput$. \server inputs \embedKey, $\PRPKey[1]$, and
  $\PRPKey[2]$.  $\client$ learns $\prpOutputEmbedding \assign
  \PRP[\PRPKey[1]](\embed[\embedKey](\cInput))$ and $\prpOutputHash
  \assign \PRP[\PRPKey[2]](\hash(\cInput, \embed[\embedKey](\cInput),
  \embedKey))$. \server gets no output.
  \label{frameworkSH:embed}
			
\item $\client$ sets $\prpOutputEmbeddingAlt \gets \prpOutputEmbedding$
  and $\prpOutputHashAlt \assign \prpOutputHash$.
  \label{frameworkSH:glue}
  
\item $\client$ and \server call \testAndCommit. $\client$ inputs
  $\prpOutputEmbeddingAlt$ and $\prpOutputHashAlt$. \server inputs
  \policyLan, \threshold, $\PRPKey[1]$, $\PRPKey[2]$ and $\PRFKey[1]$.
  If
  $\neg\blocked{\policyLan}{\threshold}(\PRP[\PRPKey[1]]^{-1}(\prpOutputEmbeddingAlt))$
  then $\client$ learns $\prfOutput \assign
  \PRF[1][\PRFKey[1]]\left(\authToken\right)$ for $\authToken \gets
  \PRP[\PRPKey[1]]^{-1}(\prpOutputEmbeddingAlt) \linearComb
  \PRP[\PRPKey[2]]^{-1}(\prpOutputHashAlt)$, and \server learns
  $\authToken$. Otherwise, \server aborts and learns $\{\lanElmt \in \blockList: \distance\left(\PRP[\PRPKey[1]]^{-1}(\prpOutputEmbedding), \lanElmt\right) \le \threshold\}$. $\client$ outputs $\bot$. \label{frameworkSH:commit}

\end{enumerate}

\textbf{\ImplicitCheckPhase:}
\begin{enumerate}[topsep=0.5em,leftmargin=1.75em,labelwidth=*,align=left]
\setcounter{enumi}{4}

\item \server sets $\embedKeyAlt \assign \embedKey$ and sends
  \embedKeyAlt to $\client$.

\item $\client$ sets $\cInputAlt \assign \cInput$ and $\authTokenAlt
  \assign \embed[\embedKeyAlt](\cInputAlt) \linearComb \hash(\cInputAlt,
  \embed[\embedKeyAlt](\cInputAlt), \embedKeyAlt)$.
  \label{frameworkSH:client_input}

\item \server sets $\PRFKeyAlt[1] \gets \PRFKey[1]$ and
  $\authTokenDblAlt \gets \authToken$.
  
\item $\client$ and \server call \authentication. $\client$ inputs
  $\authTokenAlt$, and \server inputs $\PRFKeyAlt[1]$ and
  $\authTokenDblAlt$.  $\client$ learns $\prfOutputAlt \assign
  \PRF[1][\PRFKeyAlt[1]](\authTokenAlt)$ if $\authTokenAlt =
  \authTokenDblAlt$. Otherwise, $\client$ and \server output $\bot$.
  \label{frameworkSH:authenticate}
\end{enumerate}
}
\end{oframed}
\captionof{figure}{Framework for implementing \primitiveNameShort}
\label{fig:frameworkSH}	
\end{figure}

\section{Background}
\label{sec:background}

As we will describe later, \primitiveNameShort will embed 
Hamming distance\footnote{We choose Hamming distance as 
it is a popular metric for embedding and similarity searching on large sets of 
data~\cite{kushilevitz1998efficient,gionis1999similarity,andoni2008near}, 
and many such embedding functions exist~\cite{Charikar2002:SimHash, 
datar2004locality}.} into a metric space defined over the finite
field $\finiteField$ for the blocklist \blockedElmts. We describe in this
section the definitions and primitives related to this embedding.

\subsection{Symmetric Uncommon Roots}
\label{sec:bb:metric}
We define the distance metric {\it symmetric uncommon 
roots} (\symmUR) as follows. For polynomial 
$\genericPoly \in \polyField$, let $\uniqueRoots{\genericPoly}$ be 
its set of (unique) roots. Let $\genericSetVar[1] 
\symmDiff \genericSetVar[2]$ denote the symmetric difference 
between sets $\genericSetVar[1]$ and $\genericSetVar[2]$, i.e., 
$(\genericSetVar[1] \setminus \genericSetVar[2]) \cup
(\genericSetVar[2] \setminus \genericSetVar[1])$.

\begin{defn}[Symmetric Uncommon Roots]
Let $\setOfEvalPts \assign \{\evalPt[1], \dots, \evalPt[\numberOfPoints] \}$ be 
a set $\numberOfPoints$ unique, fixed elements in $\finiteField$.  
For vector $\genericVec = \langle \genericVecComp[1], \dots, 
\genericVecComp[\numberOfPoints] \rangle \in 
\finiteField^{\numberOfPoints}$, let $\polyFromVec{\genericVec} 
\in \polyField$ be the unique polynomial of degree $\le 2 
\degreeOfKey{1}$ that interpolates the set of points 
$\{ (\evalPt[\ptIdx], \genericVecComp[\ptIdx]) \}_{\ptIdx \in \numberOfPoints}$
Then, for $\genericVec[1], 
\genericVec[2] \in \finiteField^{\numberOfPoints}$,
\begin{align*}
    \symmURDist{\genericVec[1]}{\genericVec[2]}
    & ~\defeq~ \setSize{\uniqueRoots{\polyFromVec{\genericVec[1]}} 
    \symmDiff \uniqueRoots{\polyFromVec{\genericVec[2]}}} \\
    & ~=~ \setSize{\uniqueRoots{\polyFromVec{\genericVec[1]}}} + 
    \setSize{\uniqueRoots{\polyFromVec{\genericVec[2]}}} -
    2\cdot\setSize{\uniqueRoots{\genericPoly}}
\end{align*}    
where $\genericPoly = \gcd(\polyFromVec{\genericVec[1]}, 
\polyFromVec{\genericVec[2]})$ is the greatest common 
divisor of \polyFromVec{\genericVec[1]} and 
\polyFromVec{\genericVec[2]}.
\end{defn}

Since the cardinality of symmetric difference between two sets is a
metric, \symmUR is also a metric.

\subsection{Hamming-distance-aware private queries }
\label{sec:background:HammingPSI}
We will then use a two-party Hamming-distance-aware query 
protocol~\cite{Chakraborti2023:DAPSI} to privately check
if the Hamming distance of the input vectors, 
$\hammingDist{\cInputVec}{\cInputVecAlt}$, is within a certain 
threshold $\distInMaskedSets$.  
At a high level, the protocol has two steps:

\smallskip
{\it \underline{Mapping step}:}
The parties first locally map their respective inputs, $\cInputVec$ and
$\cInputVecAlt$, to the metric space $(\finiteField^{\numberOfPointsSim}, 
\symmURDefn)$ using the injective map $\vecMap:\{0,1\}^{\degreeOfKey{1}} 
\rightarrow \finiteField^{\numberOfPointsSim}$.  $\vecMap$ takes a 
set $\setOfEvalPts$ as a parameter, which both parties first agree 
on. $\genericVec \assign \vecMap(\cInputVec)$ and $\genericVecAlt 
\assign \vecMap(\cInputVecAlt)$ are the mapped embeddings. 

\smallskip
{\it \underline{Matching step}:}
To check whether $\hammingDist{\cInputVec}{\cInputVecAlt} \le
\distInMaskedSets$, the parties interactively and privately check
that $\symmURDist{\genericVec}{\genericVecAlt} \le \threshold = 
2\distInMaskedSets$. As pointed out in \citet{Chakraborti2023:DAPSI},
$\hammingDist{\cInputVec}{\cInputVecAlt} \le
\distInMaskedSets$ is equivalent to $\symmURDist{\genericVec}{\genericVecAlt} \le \threshold$
since $\vecMap$ is injective. We prove this equivalence and provide more protocol description 
in \appref{app:HammingPSI}; further 
details can be found in \citet{Chakraborti2023:DAPSI}.

\section{Blocklisted OPRF for Hamming Distance}
\label{sec:protocol}

In this section, we will describe a \primitiveName protocol for the 
Hamming distance using the framework described in \figref{fig:frameworkSH}. We
will use this construction as a building block to design the applications in \secref{sec:app}.

\myparagraph{Problem}
Given a universe of items $\universe$, we want to design a protocol between a
client ($\client$) and a server (\server)
such that when \server inputs a blocklist of items $\blockedElmts \subseteq
\universe$, and the key for a \primitiveNameShort,
$\PRFKey = \langle \embedKey,
\PRFKey[1] \rangle$, and $\client$ inputs an item
$\cInput\in\metricSpace$, 
$\client$ learns $\PRF[][\PRFKey](\cInput) =
\PRF[1][\PRFKey[1]](\embed[\embedKey](\cInput)
\linearComb \hash(\cInput, \embed[\embedKey](\cInput), \embedKey))$
iff $\blocked{\blockedElmts}{\policyLan}{\threshold}(\embedESP(\cInput)) =
\fals$.  Here $\embed$ is a an embedding function that maps items in
$\universe$ to a
metric space $(\metricSpace, \genericMetricSpaceDist)$ with distance metric
$\genericMetricSpaceDist$.  This means that if $\client$ learns
$\PRF[][\PRFKey](\cInput)$
from the protocol then with probability at least $1
- \falseAcceptRate{\blockedElmts}{\policyLan}{\threshold}$, there is no element
$\cInputAlt \in \blockedElmts$ such that
$\distanceNotation[\genericMetricSpaceDist][\cInput][\cInputAlt] \le
\distInMaskedSets$ where $\distInMaskedSets$ is a distance 
threshold.

\myparagraph{Protocol intuition}
Consider first the (flawed) approach where \embedAndMap embeds
inputs into Hamming space, i.e., $\embed:\embedKeySpace\times \universe 
\rightarrow \{0,1\}^{\degreeOfKey{1}}$. Then, during \testAndCommit, we 
implement the Hamming distance-aware query protocol described in 
\secref{sec:background:HammingPSI}. If $\client$'s input is not close to any 
of the embedded \server inputs, $\client$ learns $\PRF[][\PRFKey](\cInput)$. 
Unfortunately, this idea poses a problem in the context of the framework. 
Specifically, $\client$ learns $\PRP[\PRPKey](\embed[\embedKey](\cInput))$ 
from \embedAndMap, where \PRP is an unpredictable permutation. Due to this, 
the output of \embedAndMap is incompatible as an input to the Hamming 
distance-aware query protocol. 

To solve this problem we make the following crucial observation. Instead of 
embedding the input into the Hamming space $(\{0,1\}^{\degreeOfKey{1}}, 
\hammingDist{\cdot})$, we will directly embed the input into 
$(\metricSpaceESP, \symmURDefn)$. This is indeed possible due to the 
injective vector map \vecMap from the Hamming space to $\metricSpaceESP$ 
described before in \secref{sec:bb:metric}. More specifically, our embedding 
function $\embed:\embedKeySpace\times \universe \rightarrow 
\metricSpaceESP$ first embeds the inputs from \universe into bit-vectors, and
then injectively maps them using \vecMap to vectors in \metricSpaceESP. As a 
result, computing $\distanceNotation[\genericMetricSpaceDist][\cInput][\cInputAlt]$ 
reduces to computing $\symmURDist{\embed[\embedKey](\cInput)}
{\embed[\embedKey](\cInputAlt)}$. This is identical to the computation that the 
Hamming distance-aware query protocol performs using the \npaFunc functionality. 

After the embedding, we need a one-time pairwise unpredictable permutation \PRP 
over $\finiteField^{\numberOfPointsSim}$ such that given
$\cInputVec = \PRP(\embed[\embedKey](\cInput))$ and some arbitrary
$\lanElmtVec \in \metricSpaceESP$,
we can {\it efficiently} compute
$\symmURDist{\cInputVec}{\lanElmtVec}$ using \npaFunc similar to distance-aware 
query protocol. In that case, the output of \embedAndMap can be
used directly as an input to \testAndCommit.
We prove that a random linear combination $\PRP[\langle\PRPKeyVec[1], \PRPKeyVec[2]\rangle](\cInputVec)
=   \PRPKeyVec[1] \hadamardProd\cInputVec + \PRPKeyVec[2]$, where $\PRPKeyVec[1]$ and $\PRPKeyVec[2]$ are randomly selected, 
 is such a one-time pairwise unpredictable permutation in \appref{app:proof_prp}.
So, given $\PRP[\langle \PRPKeyVec[1], 
\PRPKeyVec[2] \rangle](\cInputVec)$ and
$\lanElmtVec$, it is possible to produce another linear combination 
$\randomVec[1] \hadamardProd \cInputVec + \randomVec[2] \hadamardProd
\lanElmtVec$ using a series of linear operations (calls to \oleFunc), where 
$\randomVec[1],\randomVec[2]$ are random vectors. Due to 
\eqref{eq:sur_random}, given $\randomVec[1] \hadamardProd
\cInputVec + \randomVec[2] \hadamardProd \lanElmtVec$, \server can 
compute $\symmURDist{\cInputVec}{\lanElmtVec}$ and perform the matching 
step of the Hamming
distance-aware query protocol. Then, our protocol for \testAndCommit will
compute this
linear combination using calls to \npaFunc without revealing
\cInputVec to \server, or
\lanElmtVec to $\client$.

\subsection{Embed-and-Map}

The protocol \protocolEmbed (\figref{fig:esp_em}) realizing \embedAndMap
performs two tasks: i) it embeds $\client$'s input into 
$\metricSpaceESP$ (\stepref{esp:em:embed}), and ii) it computes the 
one-time pairwise-independent permutation \PRP
on the mapped input (\stepref{esp:em:map}). 
In this process, \server selects keys for the embedding
and the permutation. \server acts as the garbler, uses the keys as inputs to the 
garbled circuit, and sends the circuit to $\client$ (evaluator). \protocolEmbed
utilizes a garbled circuit as the malicious client is not trusted to honestly 
embed its inputs. 

\begin{oframed}
{\small

	\smallskip\noindent\textbf{Parameters:} A keyed one-time pairwise-unpredictable permutation 
	scheme	$\PRP: \PRPKeySpace \times \metricSpaceESP \rightarrow
	\metricSpaceESP$; a random oracle $\hash:
	\{0,1\}^{\ast} \rightarrow \metricSpaceESP$; an embedding function 
	$\embed:\embedKeySpace\times \universe
	\rightarrow \metricSpaceESP$; and a set of elements $\setOfEvalPts \assign \{\evalPt[1], \dots,
	\evalPt[\numberOfPointsSim] \} \subset \finiteField$.
	
	\smallskip\noindent\textbf{Inputs:} $\client$'s input is
	$\cInput \in \universe$. \server's input is a $\PRPKey[1],
	\PRPKey[2] \in \PRPKeySpace$, $\embedKey \in \embedKeySpace$.
	
	\smallskip\noindent\textbf{Protocol:}
	\server generates a garbled circuit and $\client$ executes this circuit
	as the evaluator. The circuit does the following:
	
	\begin{enumerate}
		\item \label{esp:em:embed} Compute $\cInputVec \assign \embedESP(\cInput)$ 
		
		\item \label{esp:em:map} Output $\prpOutputEmbedding \assign
		\PRP[\PRPKey[1]] (\cInputVec)$ and
		$\prpOutputHash \assign
		\PRP[\PRPKey[2]](\hash(\cInput, \embedESP(\cInput),
		\embedKey))$ to $\client$.
		\label{em:output}
	\end{enumerate}
	
}
\end{oframed}
\captionof{figure}{\protocolEmbed: Embed-and-Map protocol  \label{fig:esp_em}}

\subsection{Test-and-Commit}
The protocol \protocolCommit (\figref{fig:esp_tc}) realizes \testAndCommit, where
\server first locally computes his input set \policyLan by mapping each element
in \blockedElmts to a vector in $\finiteField^{\numberOfPointsSim}$ with $\embed$. 
Then,  the protocol checks $\symmURDist{\cInputVec}{\lElmtVecComp{\setIdx}} \le
\threshold$ for all $\lElmtVecComp{\setIdx} \in \policyLan$ 
(\stepsref{esp:tc:lan_S}{esp:tc:interpolate}).
\server can learn the result from \npaFunc (\stepref{esp:tc:compute_comb_poly}) 
iff \server is able to reverse $\client$'s input $\prpOutputEmbeddingVec$ with 
$\PRPKey[1]$, which indicates that the output from \protocolEmbed and the 
input to \protocolCommit are consistent. 

During commit phase, $\client$ and \server interactively compute the summation of 
$\commitVec[\setIdx] \gets \serversRandomPolyEvalVec[\setIdx] \hadamardProd
\cInputVec + \clientsRandomPolyEvalVec[\setIdx] \hadamardProd
\lElmtVecComp{\setIdx}$ and $\hash(\cInput,
\embed[\embedKey](\cInput), \embedKey)$ using a series of calls to \oleFunc.
They start by eliminating $\clientsRandomPolyEvalVec[]$ ($\client$'s input, in 
\stepsref{esp:tc:remove_client_random}{esp:tc:client_comb}) and
$\serversRandomPolyEvalVec[],
\lanElmtVec$ (\server's inputs, in
\stepsref{esp:tc:server_unblind_1}{esp:tc:output}), without
revealing  either of $\cInputVec$ and 
$\hash(\cInput, \embed[\embedKey](\cInput), \embedKey)$ individually to
\server. $\client$ and \server proceed with computing the {\it authentication token}
$\authToken \assign \cInputVec
+ \hash(\cInput, \embed[\embedKey](\cInput), \embedKey)$ (\stepref{esp:tc:output_1}). 
\server locally stores $\authToken$, and provides to $\client$ 
$\PRF[1][\PRFKey[1]](\authToken)$ as the output from the \explicitCheckPhase if 
$\blocked{\blockedElmts}{\policyLan}{\threshold}(\embedESP(\cInput)) =
\fals$ (\stepref{esp:tc:output}).

\begin{oframed}
	{\small	
		
		\smallskip\noindent\textbf{Parameters:} 
		A PRF scheme $\PRF[1]: \PRFKeySpace[1] \times \finiteField^
		{\numberOfPointsSim} \rightarrow \rangePRF$.
		
		\smallskip\noindent\textbf{Inputs:} \client's input is 
		$\prpOutputEmbeddingVec \gets \PRP[\PRPKey[1]] (\cInputVec) = \langle
		\prpOutputEmbeddingComp{1}, \dots, \prpOutputEmbeddingComp{\numberOfPointsSim}
		\rangle$, and $\prpOutputHashVec \gets \PRP[\PRPKey[2]](\hash(\cInput,
		\embedESP(\cInput), \embedKey)) = \langle \prpOutputHashComp{1},
		\dots,\prpOutputHashComp{\numberOfPointsSim} \rangle$
		obtained from \protocolEmbed. 
		\server's input is a set $\policyLan \assign \{\embed[\embedKey](\lanElmt)~|~ \lanElmt 
		\in 
		\blockedElmts \} \subset \metricSpaceESP$  of size
		$\blockListSize$ derived from \blockedElmts;
		a distance threshold $\threshold$, keys $\PRPKey[1], \PRPKey[2] \in
		\PRPKeySpace$ where $\PRPKey[1] = \langle \PRPKeyEmbeddingVec[1],
		\PRPKeyEmbeddingVec[2] \rangle$ and $\PRPKey[2] = \langle \PRPKeyHashVec[1],
		\PRPKeyEmbeddingVec[2] \rangle$, a key
		$\PRFKey[1] \in \PRFKeySpace[1]$.

		{\small \smallskip\noindent\textbf{Test Phase}}
		\begin{enumerate}[nosep, leftmargin=1.7em,labelwidth=*,align=left]
			
			{\small \item \label{esp:tc:lan_S} For $\lElmtVecComp{\setIdx} =  \langle  \lElmtVecComp{\setIdx}{1},
				\dots, \lElmtVecComp{\setIdx}{\numberOfPointsSim}  \rangle  \in \policyLan$,
				\server computes  $\polyFromEmbedding[\lElmtVecComp{\setIdx}]$ by
				interpolating
				the
				set of points $\{ (\evalPt[1], \lElmtVecComp{\setIdx}{1}), \dots,
				(\evalPt[\numberOfPointsSim], \lElmtVecComp{\setIdx}{\numberOfPointsSim} ) \}$.
				
				\item \label{esp:tc:rand_poly_C} $\client$ selects $\blockListSize$ random
				polynomials $\clientsRandomPoly[1], \dots, \clientsRandomPoly[\blockListSize]$
				of degree $\numberOfMasks$, and computes
				$\clientsRandomPolyEvalVecDefn{\setIdx} = \langle
				\clientsRandomPolyEvalVec[\setIdx][1], \dots,
				\clientsRandomPolyEvalVec[\setIdx][\numberOfPointsSim] \rangle$ for $\setIdx
				\in \nats[\blockListSize]$.
				
				\item \label{esp:tc:rand_poly_S} \server selects $\blockListSize$ random
				polynomials $\serversRandomPoly[1], \dots, \serversRandomPoly[\blockListSize]$
				of degree $\numberOfMasks$, and computes
				$\serversRandomPolyEvalVecDefn{\setIdx} = \langle
				\serversRandomPolyEvalVec[\setIdx][1], \dots,
				\serversRandomPolyEvalVec[\setIdx][\numberOfPointsSim] \rangle$ for $\setIdx
				\in \nats[\blockListSize]$.

				\item \label{esp:tc:compute_comb_poly} For $\setIdx \in [1, \blockListSize]$,
				\server and $\client$ call \npaFunc:
				
				\begin{enumerate}[nosep, leftmargin=1em,labelwidth=*,align=left]
					\item $\client$'s inputs are $\clientsRandomPolyEvalVec[\setIdx]$ and
					$\PRP[\PRPKey[1]](\embedESP(\cInput))$.

					\item \server's inputs are $\serversRandomPolyEvalVec[\setIdx]$ and
					$\PRPKeyEmbeddingVec[1]\hadamardProd \lElmtVecComp{\setIdx}$.
					
					\item \server learns 
					\vspace{-0.1cm}
					\begin{align*}
						& \left(\serversRandomPolyEvalVec[\setIdx] \hadamardProd 
						\PRP[\PRPKey[1]](\embedESP(\cInput))\right) + \left(
						\clientsRandomPolyEvalVec[\setIdx] \hadamardProd
						(\PRPKeyEmbeddingVec[1]\hadamardProd \lElmtVecComp{\setIdx})\right)\\
						& = \left(\serversRandomPolyEvalVec[\setIdx] \hadamardProd 
						(\PRPKeyEmbeddingVec[1] \hadamardProd \embedESP(\cInput) +
						\PRPKeyEmbeddingVec[2])\right) + \left(\clientsRandomPolyEvalVec[\setIdx]
						\hadamardProd (\PRPKeyEmbeddingVec[1]\hadamardProd
						\lElmtVecComp{\setIdx})\right)
					\end{align*}
				\end{enumerate}
				
				\item \label{esp:tc:compute_vec} Using $\PRPKeyEmbeddingVec[1]$,
				$\serversRandomPolyEvalVec[\setIdx]$ and $ \PRPKeyEmbeddingVec[2]$,  \server
				computes
				\vspace{-0.1cm}
				\begin{align*}
					\commitVec[\setIdx] & \gets \serversRandomPolyEvalVec[\setIdx] \hadamardProd
					\embedESP(\cInput) + \clientsRandomPolyEvalVec[\setIdx] \hadamardProd
					\lElmtVecComp{\setIdx} \\
					& = \left\langle \serversRandomPoly[\setIdx][\evalPt] \cdot
					\polyFromEmbedding[\cInputVec][\evalPt]  +
					\clientsRandomPoly[\setIdx][\evalPt]
					\cdot \polyFromEmbedding[\lElmtVecComp{\setIdx}][\evalPt] \right\rangle_{\evalPt \in
						\setOfEvalPts}.
				\end{align*}

				\item \label{esp:tc:interpolate} For $\setIdx \in [1, \blockListSize]$, \server tries to compute
				$\gcd(\polyFromEmbedding[\cInputVec],
				\polyFromEmbedding[\lElmtVecComp{\setIdx}])$
				from $\commitVec[\setIdx]$ by interpolating
				the points \cite{Chakraborti2023:DAPSI}. If the interpolation is successful,
				\server sets
				$\blocked{\policyLan}{\threshold}(\cInputVec) = \tru$. Otherwise, it sets
				$\blocked{\policyLan}{\threshold}(\cInputVec) = \fals$.}

%
		\end{enumerate}
		
		\smallskip\noindent\textbf{Commit Phase}
		
		\begin{enumerate}[nosep, leftmargin=1.7em,labelwidth=*,align=left,start=7]
			
			{\small \item \server computes $\randomPolyComb \assign \sum\limits_{\setIdx
					\in \nats[\blockListSize]} \serversRandomPoly[\setIdx]$, and
				$\randomPolyCombEvalVec \gets \langle \randomPolyComb[\evalPt]
				\rangle_{\evalPt
					\in \setOfEvalPts}$.
				

				\item \label{esp:tc:remove_client_random} \server samples
				$\tmpVarAltVec[\setIdx] = \langle \tmpVarAltVec[\setIdx][1], \dots,
				\tmpVarAltVec[\setIdx][\numberOfPointsSim] \rangle \getsr \metricSpaceESP$.
				For $\setIdx \in \nats[\blockListSize], \ptIdx \in \nats[\numberOfPointsSim]$,
				$\client$ and \server call \oleFunc where client's input is $-
				\clientsRandomPolyEvalVec[\setIdx][\ptIdx]$ and \server's inputs are
				$\lElmtVecComp{\setIdx}{\ptIdx}$
				and $\tmpVarAltVec[\setIdx][\ptIdx]$. After all the calls,  \client learns
				$\tmpVarAltVec[\setIdx] - \clientsRandomPolyEvalVec[\setIdx] \hadamardProd
				\lElmtVecComp{\setIdx}$ for $\setIdx \in \nats[\blockListSize]$.

				\item \label{esp:tc:prp_inv_hash}  \server samples $\tmpVarVec[\setIdx] =
				\langle \tmpVarVec[\setIdx][1], \dots, \tmpVarVec[\setIdx][\numberOfPointsSim]
				\rangle \getsr \metricSpaceESP$. For
				$\setIdx \in \nats[\blockListSize], \ptIdx \in \nats[\numberOfPointsSim]$,
				\server and $\client$
				call \oleFunc, where $\client$'s input is $\prpOutputHashComp{\ptIdx}$, and
				\server's inputs are
				$\frac{\randomPolyComb[\evalPt[\ptIdx]]}{\PRPKeyEmbeddingVec[1][\ptIdx]'}$.
				After all calls, $\client$ learns 
				$\PRPKeyHashVec[1]^{-1} \hadamardProd \randomPolyCombEvalVec \hadamardProd
				\prpOutputHashVec + \tmpVarVec[\setIdx]$ for $\setIdx \in
				\nats[\blockListSize]$.
				
				\item \label{esp:tc:client_comb} From the step above, $\client$ computes and
				sends to \server
				
				\[
				\genericVec' \gets \sum\limits_{\setIdx \in \nats[\blockListSize]}
				\tmpVarAltVec[\setIdx] + \tmpVarVec[\setIdx]-
				\clientsRandomPolyEvalVec[\setIdx] \hadamardProd \lElmtVecComp{\setIdx} +
				\PRPKeyHashVec[1]^{-1} \hadamardProd \randomPolyCombEvalVec \hadamardProd
				\prpOutputHashVec.
				\]}

			\item \label{esp:tc:server_unblind_1} \server computes $\genericVec
			\gets (\sum\limits_{\setIdx \in \nats[\blockListSize]} \tmpVarAltVec[\setIdx]
			+ \tmpVarVec[\setIdx]) + \blockListSize \cdot (\PRPKeyHashVec[1]^{-1}
			\hadamardProd \randomPolyCombEvalVec \hadamardProd \PRPKeyHashVec[2])$.

			\item \label{esp:tc:output_1} \server computes 
			
			\vspace{-0.5cm}
			\begin{align*}
				& \authToken \assign \frac{1}{\blockListSize} \cdot
				\randomPolyCombEvalVec^{-1}
				\hadamardProd \left(\sum\limits_{\setIdx = 1}^{\blockListSize}
				\combVector_\setIdx
				+ \genericVec' - \genericVec \right) \\
				& =  \frac{1}{\blockListSize} \cdot \randomPolyCombEvalVec^{-1}
				\hadamardProd
				\left(\blockListSize \cdot \randomPolyCombEvalVec \hadamardProd
				(\embedESP(\cInput) + \PRPKeyHashVec[1]^{-1} \hadamardProd \prpOutputHashVec -
				\PRPKeyHashVec[1]^{-1} \hadamardProd \PRPKeyHashVec[2]\right) \\
				& = \embedESP(\cInput) + \hash(\cInput, \embedESP(\cInput), \embedKey).
			\end{align*}
			
			\item \label{esp:tc:output} If $\blocked{\policyLan}{\threshold}(\cInputVec) = \fals$, \server stores
			$\langle \PRFKey[1], \authToken, \embedKey \rangle$ and outputs $\prfOutput
			\assign \PRF[\PRFKey](\authToken)$ to $\client$. Otherwise, \server aborts. 
			
		\end{enumerate}
	}
\end{oframed}
\captionof{figure}{\protocolCommitESP: Test-and-Commit protocol}\label{fig:esp_tc}

\subsection{\ImplicitCheckPhaseHeading}
The protocol \protocolValidate (\figref{fig:esp_tc}) realizes \authentication. 
$\client$ first recreates the output of \protocolCommit with his input to the 
\implicitCheckPhase and \embedKey obtained from \server 
(\stepsref{esp:auth:key}{esp:auth:hash}). 
\server may reveal $\embedKey$ to $\client$ because $\embedKey$ is randomly 
chosen for each $\client$ and an authentication token $\authToken$
for that $\client$ will only be stored at \server if \explicitCheckPhase goes through.
Then, \server secret shares $\prfOutput = \PRF[\PRFKey](\authToken)$ 
(\stepref{esp:auth:serverSS}).  $\client$ and \server evaluate \oleFunc with $\client$
inputting shares of $\authTokenAlt$ and \server inputting shares of $\prfOutput$
and $\authToken$ (\stepref{esp:auth:ole}). $\client$ will be able to reconstruct 
$\prfOutput$ from the shares iff $\authTokenAlt$ matches $\authToken$ (\stepref{esp:auth:prf}).

\begin{oframed}
{\small

		\smallskip\noindent\textbf{Parameters:} 
		A $\numberOfPointsSim$-out-of-$\numberOfPointsSim$ secret
		sharing scheme.
		
		\smallskip\noindent\textbf{Inputs:} $\client$'s input is $\cInputAlt$.
		\server's input is embedding key
		$\embedKey \in \embedKeySpace$, $\prfOutput \in \rangePRF$, and $\authToken \in
		\metricSpaceESP$.
		
		\smallskip\noindent\textbf{Protocol}
		\begin{enumerate}[nosep, leftmargin=1em,labelwidth=*,align=left]
			\item \label{esp:auth:key} \server sends \embedKey to $\client$.
			
			\item \label{esp:auth:hash} $\client$ computes $\embedESP(\cInputAlt)$ and
			$\hash(\cInputAlt, \embedESP(\cInputAlt), \embedKey)$.

			\item \label{esp:auth:startSS} $\client$ computes 
			\vspace{-0.1cm}
			\[
			\langle \authTokenAltComp{1}, \dots, \authTokenAltComp{\numberOfPointsSim}
			\rangle \assign \embedESP(\cInputAlt) + \hash(\cInputAlt,
			\embedESP(\cInputAlt), \embedKey).
			\]
			
			\item \label{esp:auth:serverSS} \server secret-shares $\prfOutput$ into $\numberOfPointsSim$ shares
			$\secretShare{1}, \dots, \secretShare{\numberOfPointsSim}$.

			\item \server selects \numberOfPointsSim random elements $\genericRand[1],
			\dots, \genericRand[\numberOfPointsSim] \getsr \finiteField$.

			\item \label{esp:auth:ole} Let $ \authToken = \langle \authTokenComp{1}, \dots,
			\authTokenComp{\numberOfPointsSim} \rangle$. For $\ptIdx \in [1,
			\numberOfPointsSim]$, $\client$ and \server call \oleFunc.
			
			\begin{enumerate}
				\item $\client$'s input is \authTokenAltComp{\ptIdx}.
				\item \server's input is \genericRand[\ptIdx], and $(\secretShare{\ptIdx} -
				\genericRand[\ptIdx] \cdot \authTokenComp{\ptIdx})$.
				\item $\client$ learns $\secretShare{\ptIdx}' \assign \genericRand[\ptIdx] \cdot
				(\authTokenAltComp{\ptIdx} - \authTokenComp{\ptIdx}) + \secretShare{\ptIdx}$.
			\end{enumerate}
			\item \label{esp:auth:prf} $\client$ combines the shares $\secretShare{1}', \dots,
			\secretShare{\ptIdx}'$.
			and obtains $\prfOutput$ if $\authTokenAlt \equals \authToken$. 
		\end{enumerate}

}
\end{oframed}
	\captionof{figure}{\protocolValidateESP: \ImplicitCheckPhase protocol \label{fig:esp_authenticate}}

\section{Applications}
\label{sec:app}

In \secref{sec:intro}, we identified two applications of the
\primitiveNameShort protocol described in \secref{sec:protocol}. Here
we present the policy-enforced aPAKEs and the blocklisted MACs in details. 


\subsection{Policy-enforced aPAKEs}
\label{sec:pake}

\myparagraph{Problem}
Augmented Password-Authenticated Key Exchange (aPAKE) protocols
\cite{Bellovin1993:aPAKE,Boyko2000:aPAKE,Gentry2006:aPAKE} allow a
client ($\client$) and a server (\server) who share a low-entropy
password to establish a strong cryptographic key, without disclosing
values that a network observer could use to confirm an offline guess
at the password. Additionally, aPAKE security ensures that an attacker
who compromises \server still must succeed in an offline dictionary
attack to recover the password.  In general, aPAKE protocols have two
phases. In the {\it registration} phase, $\client$ registers a new
account at \server (e.g., a website) by providing a one-way mapping of
the password to \server. \server stores the mapping as
$\client$'s authentication token. During {\it login} phase, $\client$ and
\server will then be able to derive a common session key \textit{if and only
  if $\client$ uses the same password that was used during the
  registration}. Otherwise, $\client$ and \server cannot set up a
session.

We augment existing password authenticated key exchange protocols
(aPAKEs) to facilitate policy enforcement before session
establishment.  More specifically, our protocol enables \server to
enforce policies on password inputs during $\client$ registration,
\textit{without being provided the password in the plain}. The
protocol is secure against a malicious $\client$ who attempts to
circumvent the checks, and can be integrated with the state-of-the-art
aPAKE protocol (OPAQUE~\cite{Jarecki2018:OPAQUE}) \textit{without
  altering any guarantees of the original protocol}.  Informally, our
protocol ensures that: i) a malicious $\client$ cannot register with a
blocklisted password, ii) an honest-but-curious \server does not learn
anything more than what it learns from the underlying aPAKE protocol,
and iii) if \server is breached after registration, then the malicious
\server cannot do better than a dictionary attack in order to learn a
registered password.  Note that standard aPAKE protocols
(without password policy checks) assume $\client$ is honest during
password registration, and thus are secure against threat model
\ref{model:server-mal} described in
\secref{sec:framework:threat-model}. Since, $\client$ may deviate from
the protocol to circumvent the policy checks, our protocol {\it
  additionally} provides security against threat model
\ref{model:client-mal}.

\myparagraph{High-level idea}
The policy enforced aPAKE protocol uses \primitiveNameShort
to facilitate a ``blocklist'' policy. The use of blocklists is
commonplace among service providers, since as long as a user's
password is not one of the most commonly used passwords, defenses
against online password-guessing attacks are much more
effective~\cite{Weinert2019:Password}. In light of this observation,
we consider the following setup: in an \textit{\explicitCheckPhase}, 
\server has a list of passwords
that it considers ``easy'', denoted by $\blockedElmts \subset
\universe$.  \server wants to check if $\client$ is registering with
a password $\cInput \in \blockedElmts$. If so, \server aborts the
registration. Otherwise, similar to an aPAKE scheme, $\client$ and
\server set up a shared pseudorandom authentication token, which 
concludes the \explicitCheckPhase. During an \textit{\implicitCheckPhase},
$\client$ authenticates itself to \server with the pseudorandom token 
from \explicitCheckPhase. 

\myparagraph{Protocol Description}
The client $\client$ and the server \server executes \primitiveNameShort
by interactively evaluating in sequence \embedAndMap, 
\testAndCommit, and \authentication listed in \secref{sec:protocol}. 

We start by describing the embedding function.
Since the blocklist \blockedElmts can be large, we embed the inputs 
into $(\Sigma^{\ast}, \editDistDefn)$.
That is, we are going to cluster passwords based 
on the edit-distance between them, and create an embedded blocklist \policyList.
Then, for any given input $\cInput$, we will compare it against the items in 
$\policyList$ using the edit-distance between them as the metric. We are 
interested in edit distance since it provides good estimates of password 
similarity~\cite{Pal2019:wordEmbed}. To create \policyList, \server first embeds each 
item in \blockedElmts to $(\Sigma^{\ast}, \editDistDefn)$. Then, \server selects a 
subset of this embedded list that provides {\it maximum coverage} of the items 
in \blockedElmts. Specifically,  all the embedded elements are ranked by the number 
of items of the original blocklist \blockedElmts that lie within distance-$\distInUniverse$ 
edit distance balls centered around that element, and then the top $\setSize{\policyLan}$ 
are included in \policyList. The value \setSize{\policyLan} is fixed to optimize 
$\falseAcceptRate{\blockedElmts}{\policyLan}{\threshold}$ and 
$\falseRejectRate{\blockedElmts}{\policyLan}{\threshold}$ (see \secref{sec:eval}). 

Given such an embedded list, we then {\it approximate} edit distance with Hamming 
distance~\cite{Ostrovsky07:edit, Yen08:edit} as described in \secref{sec:protocol}.
For this, \server embeds the 
items in \policyList into $(\metricSpaceESP, \symmURDefn)$ to create the embedded list \policyLan. 
Thus, the embedding function $\embed: \embedKeySpace \times 
\universe \rightarrow \metricSpaceESP$ implemented by \embedAndMap
is identical to the one described in \secref{sec:protocol}. 
Also, the rest of the steps are identical to the protocol described 
before. Finally, the output $\authToken = \PRF[][\PRFKey[]](\cInput)$
of \protocolCommit is used as 
the pseudorandom token for setting up a session key with
\server, similar to other aPAKE designs~\cite{Jarecki2018:OPAQUE}. Note that this 
design is modular, i.e., beyond setting up the authentication token using the \primitiveNameShort 
protocol from the client's password, no further changes are required to the actual authenticated key exchange mechanism. Therefore, our protocol is not specific to any particular aPAKE design.

\myparagraph{Addressing pre-computation attacks}
A pre-computation attack adversary pre-computes a table of values
on some candidate passwords offline and can almost instantaneously learn 
client credentials upon server compromise. 
Strong aPAKE protocols, e.g., OPAQUE~\cite{Jarecki2018:OPAQUE}, 
necessitates that aPAKEs protect against such pre-computation attacks. 
A \primitiveNameShort-enhanced aPAKE follows the standard aPAKE 
security and is thus vulnerable to pre-computation attacks as the authentication 
token \authToken can be precomputed. In \appref{app:opaque} \figref{fig:protocol_pePAKE}, we show how to adapt our \primitiveNameShort protocol for a strong blocklist policy-enforced aPAKE. 
The key idea is that 
we will use an additional $\aPAKEOPRF$ keyed by the server to compute 
$\authToken \assign  \embedESP(\cInput) \linearComb 
\hash\left(\aPAKEOPRF[\aPAKEOPRFKey]\left(\cInput, \embedESP(\cInput), 
\embedKey \right)\right)$; \protocolEmbed is modified accordingly. 
In effect, an adversary cannot precompute $\authToken$ without the 
key $\aPAKEOPRFKey$. The additional steps due to the
\primitiveNameShort protocol can be directly applied to the OPAQUE 
workflow without significant changes (see \appref{app:opaque}).

\subsection{Blocklisted MACs}
\label{sec:mac}

\myparagraph{Problem}
Consider the scenario where a sending client ($\client[1]$) wants to
store/execute software at a receiving client ($\client[2]$), but
$\client[2]$ wants to enforce that the file is not among a list of
malware through outsourcing the check to a policy server
($\policyServer$).  Since the benign file can contain sensitive
content (company IP, patent, etc.), $\policyServer$ should complete
the check without obtaining the file in plain. Once $\policyServer$
confirms that the file does not match a malware, it provides a MAC
for it that permits \client[2] to confirm this check was performed.

We demonstrate and evaluate a solution using \primitiveNameShort,
where $\policyServer$ provides a blocklist $\blockedElmts$ of malware
to compare $\client[1]$'s file against. With \primitiveNameShort, we
not only perform the checks in a privacy-preserving way, but also
adapt a more flexible definition for $\blockedElmts$ that includes not
only the known malware but also files that can be considered similar
to a known malware family.  For this, we will map all the files in
$\blockedElmts$ to a metric space using an embedding function where
similar executables cluster and can be compared using the metric
distance. If $\client[1]$'s input file is similar to any of the files
in $\blockedElmts$, then $\policyServer$ will abort. Otherwise, the
pseudorandom output of the \primitiveNameShort protocol functions like
an unforgeable MAC for the pushed software, which $\policyServer$
stores locally. This allows $\client[2]$ to later verify at
$\policyServer$ that a received executed has been scanned for malware
similarity.

\myparagraph{High-level idea}
For the malware similarity clustering, we choose among the fuzzy
hashing schemes commonly used in malware similarity
tests. Sdhash~\cite{Roussev10:sdhash} uses feature extraction to embed
executables as a Bloom filter, which matches our Hamming-distance
based design. The more similar the filter, the lower the Hamming
distance and the more features two executables share.  Our protocol
will check if the vector derived from $\client[1]$'s input is close to
the vectors derived from the malware in a blocklist stored at
$\policyServer$.  To perform this check privately, we use the protocol
described in \secref{sec:protocol}.

Upon being provided an input $\cInput \in \maxFileLen$, and a key 
$\embedKey \in \embedKeySpace$, the function computes
$\embed[\embedKey](\cInput) = \vecMap(\sdHash[\embedKey](\cInput))$
where $\sdHash: \embedKeySpace \times \{0,1\}^{\ast} \rightarrow \{0,1\}^{\degreeOfKey{1}}$. 
\policyServer wants to check that
$\hammingDist{\sdHash[\embedKey](\cInput)}{\sdHash[\embedKey](\listElmt)}
> \distInUniverse$ for all $\listElmt \in \blockedElmts$ and for
some Hamming distance threshold $\distInUniverse$. If
$\blocked{\blockedElmts}{\policyLan}{\threshold}(\embedSim(\cInput)) =
\tru$, which implies that there is some $\lanElmtVec \in \policyLan:
\symmURDist{\embedSim(\cInput)}{\lanElmtVec} \le \threshold$ due to 
\eqnref{eq:hamming_distance_sur}, then we consider that $\client[1]$'s
input is similar to one (or more) of the files in
$\blockedElmts$, and should be blocked by \policyServer. 

In this three-party setting, we can simplify the \embedAndMap step due
to the threat model. Specifically, since $\client[2]$ receives
$\client[1]$'s input and the output tag, we can trust that
$\client[1]$ will correctly compute the embedding of its input given
the key \embedKey, {\it and} will correctly provide this embedding to
the protocol implementing \testAndCommit.  Otherwise, $\client[2]$ can
compute the embedding by itself and will notice that the value
provided by $\client[1]$ does not match.  Therefore, the unpredictable
permutation is not necessary, or in other words, we can set the
permutation to the identity, which $\client[1]$ computes
implicitly. In essence, $\embedAndMap$ reduces to $\policyServer$
providing \embedKey to $\client[1]$, and $\client[1]$ computing
$\embed[\embedKey](\cInput)$ to use as an input to the protocol for
\testAndCommit.

\testAndCommit is almost identical to the one described in
\figref{fig:esp_tc}. The only caveat is that the permutation keys will
be set to produce the identity permutation. The protocol outputs
$\prfOutput \assign \PRF[1][\PRFKey[1]](\authToken)$ to $\client[1]$
and the authentication token $\authToken$ to $\policyServer$ where
$\authToken \assign \embedSim(\cInput) + \hash(\cInput,
\embedSim(\cInput), \embedKey)$ if
$\blocked{\blockedElmts}{\policyLan}{\threshold}(\cInput) = \fals$.
$\client[1]$ forwards file $\cInput$ and $\prfOutput$ to $\client[2]$,
and $\policyServer$ stores $\langle \embedKey, \PRFKey[1], \authToken
\rangle$.

The \implicitCheckPhase is between $\policyServer$ and $\client[2]$.
$\client[2]$ has input the file $\cInputAlt$ that was previously sent
by $\client[1]$.  $\client[2]$ obtains $\embedKey$ from
$\policyServer$, and locally computes $\authTokenAlt \assign
\embedSim(\cInputAlt) + \hash(\cInputAlt, \embedSim(\cInputAlt),
\embedKey))$.  Subsequently, $\client[2]$ and $\policyServer$ check
that $\authTokenAlt = \authToken$ with $\protocolValidateESP$.  If so,
$\client[2]$ learns $\PRF[1][\PRFKey[1]](\authToken)$ and
authenticates the token \prfOutput stored along with the file. If the
implicit check is not successful, $\client[2]$ discards the software.
If $\authTokenAlt \neq \authToken$, then $\client[2]$ learns a random
value which does not authenticate $\prfOutput$ with high probability.

\section{Evaluation}
\label{sec:eval}

We implemented the policy-enforced aPAKE and blocklisted MAC in 
C++11. We 
benchmarked these implementations on a
Ubuntu 22.04 machine with 12 cores and 32GB of RAM. We used the
emp-toolkit\footnote{\url{https://github.com/emp-toolkit}} for OT
and garbled circuit operations. We used the
NTL\footnote{\url{https://libntl.org/}} library for
finite field arithmetic. The operations were performed in a
128-bit prime order field.  We implemented ESP~\cite{Cormode07:ESP},
Ghosh et al.'s OLE~\cite{Ghosh2017:OLE}, and Ghosh et al.'s
PSI~\cite{Ghosh2019:PSI} as building blocks for
\primitiveNameShort. Our construction works for any metrics that can
be embedded into Hamming distance.  

\myparagraph{Baseline Algorithm} To our knowledge, there is no
existing primitive tailored for privacy-preserving input policy
enforcement that, after an \explicitCheckPhase, leaves state at the
server to enable subsequent \implicitCheckPhase{}s, with the security
properties we seek. We consider the nearest solution to be that by
\citet{Katz2016:Input2PC}, where a predicate check is executed before
a secure 2PC computation, utilizing garbled circuits.  We implemented
this solution, denoted \baselineEmbed.  In addition, we compare 
\primitiveNameShort to a PSI \cite{Rindal2021:volepsi} on $\PSIList$, 
which contains a fraction of elements from $\blockedElmts$. This 
corresponds to a trivial solution (denoted \baselineNoEmbed) where 
\server keeps a subset of $\blockedElmts$ as a hashtable and is trusted 
to create the \implicitCheckPhase state.

\setlength{\textfloatsep}{5pt}
\subsection{Policy-enforced aPAKE}

For the policy-enforced aPAKE, we picked edit distance and Simhash,
as they are common metrics for measuring text similarity~\cite{Broder97:Sim}.
We generated $\blockedElmts$ by including everything within edit 
distance 2 of the 100 most common passwords~\cite{toppw}. Such a 
$\blockedElmts$ is of size 342.7MB. $\policyLan$ consists of the
embeddings of everything within edit distance 1 of the 100 most 
common passwords, which is of size 937.5KB. As was described in 
\secref{sec:pake},we rank in decreasing order each $\listElmt$ in 
$\policyList$ by the number of items from $\blockedElmts$ within 
the edit-distance balls of \listElmt and form $\policyLan$ according 
to this rank. Through reducing the size of $\policyLan$ under careful 
selection, we improve our efficiency at the loss of some accuracy. 

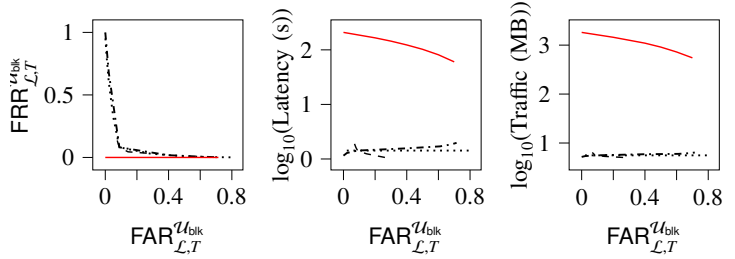
\begin{figure*}[t]
\centering
\begin{tikzpicture}[scale=0.9]

\definecolor{darkgray176}{RGB}{176,176,176}
\definecolor{lightgray204}{RGB}{204,204,204}
\definecolor{mygray1}{RGB}{96,96,96}
\definecolor{mygray2}{RGB}{128,128,128}
\definecolor{mygray3}{RGB}{169,169,169}

\begin{groupplot}[
group style={group name=my plots, group size=4 by 1, horizontal sep=1.2cm}, 
height=4cm, 
width=\columnwidth,
legend style={at={(-0.55, 0.5)},
anchor=east,
draw=none,
legend cell align={left}},
]
\nextgroupplot[
scaled x ticks=manual:{}{\pgfmathparse{#1}},
tick align=outside,
tick pos=left,
x grid style={darkgray176},
xtick style={color=black},
xlabel={\falseAcceptRate{\blockedElmts}{\policyLan}{\threshold}},
xticklabels={$0$,$0$, ,$0.4$, ,$0.8$},
xticklabel style={/pgf/number format/fixed},
y grid style={darkgray176},
ylabel={\falseRejectRate{\blockedElmts}{\policyLan}{\threshold} },
ylabel style={yshift=-1ex},
ytick style={color=black},
height=3.8cm,
width=3.8cm,
]
\addplot [thick, black, dotted]
table {%
0.791 0.001
0.659 0.003
0.493 0.01
0.287 0.03
0.089 0.0814
0.05 0.325
0.022 0.604
0.007 0.829
0.001 0.95
0 0.991
};
\addlegendentry{{\small\parbox{3.55em}{\primitiveNameShort}: \parbox{5.4em}{\centering $0 \le \threshold \le 10$}; \parbox{8em}{\centering $\setSize{\policyLan} = 1.5\text{e}3$}; \parbox{6em}{\centering $\numberOfPointsSim = 65$}}}
\addplot [semithick, black, dashed]
table {%
0.262 0.034
0.144 0.045
0.089 0.081
0.079 0.12
0.073 0.149
0.067 0.181
};
\addlegendentry{{\small\parbox{3.55em}{\primitiveNameShort}: \parbox{5.4em}{\centering $\threshold = 1$}; \parbox{8em}{\centering $5\text{e}2 \le \setSize{\policyLan} \le 3\text{e}3$}; \parbox{6em}{\centering $\numberOfPointsSim = 65$}}}
\addplot [thick, black, dashdotted]
table {%
0 1
0.022 0.68
0.089 0.081
0.398 0.018
0.61 0.006
0.68 0.003
0.715 0.002
};
\addlegendentry{{\small\parbox{3.55em}{\primitiveNameShort}: \parbox{5.4em}{\centering $\threshold = 1$}; \parbox{8em}{\centering $\setSize{\policyLan} = 1.5\text{e}3$}; \parbox{5.8em}{\centering $33 \le \numberOfPointsSim \le 129$}}}
\addplot [semithick, red]
table {%
0 0
0.715 0
};
\addlegendentry{{\small\parbox{3.55em}{\baselineNoEmbed}: $\setSize{\PSIList} = \setSize{\blockedElmts} * (1 - \falseAcceptRate{\blockedElmts}{\policyLan}{\threshold})$}}
\nextgroupplot[
scaled x ticks=manual:{}{\pgfmathparse{#1}},
tick align=outside,
tick pos=left,
x grid style={darkgray176},
xtick style={color=black},
xticklabel style={/pgf/number format/fixed},
xlabel={\falseAcceptRate{\blockedElmts}{\policyLan}{\threshold}},
xticklabels={$0$,$0$, ,$0.4$, ,$0.8$},
y grid style={darkgray176},
ylabel={$\log_{10}(\text{Latency (s)})$},
ylabel style={yshift=-1ex},
ytick style={color=black},
height=3.8cm,
width=3.8cm,
xshift=0.1cm
]
\addplot [thick, black, dotted]
table {%
0.791 0.154
0.659 0.154
0.493 0.154
0.287 0.154
0.089 0.154
0.05 0.154
0.022 0.154
0.007 0.154
0.001 0.154
0 0.154
};
\addplot [semithick, black, dashed]
table {%
0.262 0.028
0.144 0.096
0.089 0.154
0.079 0.208
0.073 0.253
0.067 0.299
};
\addplot [thick, black, dashdotted]
table {%
0 0.061
0.022 0.11
0.089 0.154
0.398 0.195
0.61 0.231
0.68 0.266
0.715 0.296
};
\addplot [semithick, red]
table {%
0.00 2.32
0.1 2.27
0.2 2.22
0.3 2.16
0.4 2.09
0.5 2.01
0.6 1.91
0.7 1.78
};
\nextgroupplot[
scaled x ticks=manual:{}{\pgfmathparse{#1}},
tick align=outside,
tick pos=left,
x grid style={darkgray176},
xtick style={color=black},
xticklabel style={/pgf/number format/fixed},
xlabel={\falseAcceptRate{\blockedElmts}{\policyLan}{\threshold}},
xticklabels={$0$,$0$, ,$0.4$, ,$0.8$},
y grid style={darkgray176},
ylabel={{$\log_{10}(\text{Traffic (MB)})$}},
ylabel style={yshift=-1ex},
ytick style={color=black},
height=3.8cm,
width=3.8cm,
xshift=0.1cm
]
\addplot [thick, black, dotted]
table {%
0.791 0.748
0.659 0.748
0.493 0.748
0.287 0.748
0.089 0.748
0.05 0.748
0.022 0.748
0.007 0.748
0.001 0.748
0 0.748
};
\addplot [semithick, black, dashed]
table {%
0.262 0.705
0.144 0.727
0.089 0.748
0.079 0.768
0.073 0.788
0.067 0.806
};
\addplot [thick, black, dashdotted]
table {%
0 0.716
0.022 0.733
0.089 0.748
0.398 0.771
0.61 0.778
0.68 0.792
0.715 0.806
};
\addplot [semithick, red]
table {%
0 3.26
0.1 3.21
0.2 3.16
0.3 3.1
0.4 3.04
0.5 2.96
0.6 2.86
0.7 2.74
};

\end{groupplot}


\end{tikzpicture}
    \caption{False-reject rate (left), latency (middle), and
      communication volume (right) for policy-enforced aPAKE
      \explicitCheckPhase using our \primitiveNameShort framework
      versus the \baselineNoEmbed baseline, as a function of
      false-accept rate.  \primitiveNameShort curves created by
      varying \threshold ($\cdot \cdot$), \setSize{\policyLan} (- -),
      and \numberOfPointsSim ($\cdot$ -).}
    \label{fig:accuracy}
\end{figure*}

The dataset for accuracy measurement is based on real-world leaked
passwords\footnote{\url{https://github.com/philipperemy/tensorflow-1.4-billion-password-analysis}}.  
In \figref{fig:accuracy}, we compare the
$\falseRejectRate{\blockedElmts}{\policyLan}{\threshold}$, client
latency, and network traffic.  The solid line, which represents
\baselineNoEmbed, had a
$\falseRejectRate{\blockedElmts}{\policyLan}{\threshold} = 0$ since it
performs a direct comparison to $\blockedElmts$. The other three lines
are \primitiveNameShort while varying threshold $\threshold$, the size
of blocklist $\setSize{\policyLan}$, and the size of the metric space
$\numberOfPointsSim$. The cross-point is the optimal trade-off with
$\threshold = 4$, $\setSize{\policyLan} = 1{,}500$, and
$\numberOfPointsSim = 65$. With these parameters, \primitiveNameShort
showed $\falseAcceptRate{\blockedElmts}{\policyLan}{\threshold} =
8.85\%$, $\falseRejectRate{\blockedElmts}{\policyLan}{\threshold} =
8.14\%$ with a client latency of 1.43 seconds and a total network
traffic of 5.6MB. At the same 
$\falseAcceptRate{\blockedElmts}{\policyLan}{\threshold}$, 
\baselineNoEmbed requires 162 seconds and a communication 
cost of 1.51GB.

\begin{figure*}[t]
\centering
\begin{tikzpicture}[scale=0.8]

\definecolor{darkgray176}{RGB}{176,176,176}
\definecolor{lightgray204}{RGB}{204,204,204}
\definecolor{mygray1}{RGB}{96,96,96}
\definecolor{mygray2}{RGB}{128,128,128}
\definecolor{mygray3}{RGB}{169,169,169}

\begin{groupplot}[group style={group size=4 by 1, horizontal sep=1.5cm}, height=5cm, width=\textwidth]
\nextgroupplot[
title={A},
legend cell align={left},
legend style={
	fill opacity=0.8,
	draw opacity=1,
	text opacity=1,
	at={(0.03,0.97)},
	anchor=north west,
	draw=lightgray204,
	font=\small,
},
scaled x ticks=manual:{}{\pgfmathparse{#1}},
tick align=outside,
tick pos=left,
x grid style={darkgray176},
xtick style={color=black},
xlabel={\setSize{\policyLan}},
xticklabels={$0$,$0$,$5e3$, $1e4$}, 
ymin=-0.5,
ymax=20,
y grid style={darkgray176},
ylabel={Client Latency (s)},
ylabel style={yshift=-1ex},
ytick style={color=black},
height=4.5cm,
width=4cm,
xshift=0.1cm,
]
\addplot [semithick, black]
table {%
1000 1.25
2000 1.62
3000 1.99
4000 2.37
5000 2.74
6000 3.11
7000 3.49
8000 3.86
9000 4.24
10000 4.54
};
\addlegendentry{\primitiveNameShort}
\addplot [semithick, black, dashed]
table {%
1000 1.48
2000 2.6
3000 3.66
4000 4.94
5000  6.42
6000 7.13
7000 8.88
8000 9.39
9000 10.41
10000 11.23
};
\addlegendentry{\baselineEmbed}
\addplot [semithick, black, dotted]
table {%
1000 0.416
2000 0.416
3000 0.416
4000 0.416
5000 0.416
6000 0.416
7000 0.416
8000 0.416
9000 0.416
10000 0.416
};
\addlegendentry{\protocolOPRF}
\nextgroupplot[
title={B},
legend cell align={left},
legend style={
	fill opacity=0.8,
	draw opacity=1,
	text opacity=1,
	at={(0.03,0.97)},
	anchor=north west,
	draw=lightgray204,
	font=\small,
},
scaled x ticks=manual:{}{\pgfmathparse{#1}},
tick align=outside,
tick pos=left,
x grid style={darkgray176},
xtick style={color=black},
xlabel={\setSize{\policyLan}},
xticklabels={$0$,$0$,$5e3$, $1e4$}, 
ymin=-2,
ymax=45,
y grid style={darkgray176},
ylabel={Total Comm. (MB)},
ylabel style={yshift=-1ex},
ytick style={color=black},
height=4.5cm,
width=4cm,
xshift=0.1cm,
]
\addplot [semithick, black]
table {%
1000 5.338
2000 5.869
3000 6.399
4000 6.93
5000 7.461
6000 7.992
7000 8.523
8000 9.053
9000 9.584
10000 10.115
};
\addlegendentry{\primitiveNameShort}
\addplot [semithick, black, dashed]
table {%
1000 5.475
2000 8.783
3000 12.091
4000 15.399
5000 18.708
6000 22.016
7000 25.483
8000 28.952
9000 32.42
10000 35.889
};
\addlegendentry{\baselineEmbed}
\addplot [semithick, black, dotted]
table {%
1000 0.000288
2000 0.000288
3000 0.000288
4000 0.000288
5000 0.000288
6000 0.000288
7000 0.000288
8000 0.000288
9000 0.000288
10000 0.000288
};
\addlegendentry{\protocolOPRF}
\nextgroupplot[
title={C},
ybar stacked,
ymin=0,
ymax=45,
bar width=0.3cm,
symbolic x coords={\primitiveNameShort \\ \setSize{\policyLan}=$1\text{e}3$,\primitiveNameShort \\ \setSize{\policyLan}=$5\text{e}3$,\primitiveNameShort \\ \setSize{\policyLan}=$1\text{e}4$,spacer, \baselineEmbed\\ \setSize{\policyLan}=$1\text{e}3$,\baselineEmbed\\ \setSize{\policyLan}=$5\text{e}3$,\baselineEmbed\\ \setSize{\policyLan}=$1\text{e}4$},
xtick=data,
xticklabels={~\\$1\text{e}3$, \primitiveNameShort w/\\ $5\text{e}3$,  ~\\$1\text{e}4$, , ~\\$1\text{e}3$, \baselineEmbed w/ \\ $5\text{e}3$,  ~\\$1\text{e}4$},
x tick label style={align=center,font=\small, text height=1.5ex, yshift=-10pt},
ylabel style={yshift=-1ex},
y grid style={darkgray176},
ylabel={Total Comm. (MB)},
ytick style={color=black},
legend style={at={(0.03,0.97)},
anchor=north west,
legend columns=1,,
draw=none,
legend cell align={left}},
height=4.5cm, 
width=6.5cm,
xshift=-0.1cm,
after end axis/.code={
        \node[anchor=north west, font=\small] at (rel axis cs:-0.15, -0.13) {$\setSize{\policyLan}=$};
 	\draw[dashed,] 
            (axis cs:spacer,0) -- 
            (axis cs:spacer,45);
},
]
\addplot[ybar,fill=mygray1,draw=none] plot coordinates {(\primitiveNameShort \\ \setSize{\policyLan}=$1\text{e}3$,2.6) (\primitiveNameShort \\ \setSize{\policyLan}=$5\text{e}3$,2.6) (\primitiveNameShort \\ \setSize{\policyLan}=$1\text{e}4$,2.6) (spacer, 0) (\baselineEmbed\\ \setSize{\policyLan}=$1\text{e}3$,2.99) (\baselineEmbed\\ \setSize{\policyLan}=$5\text{e}3$,13.49) (\baselineEmbed\\ \setSize{\policyLan}=$1\text{e}4$,27.25)};
\addlegendentry{Circuit Size}
\addplot[ybar,fill=mygray2,draw=none] plot coordinates {(\primitiveNameShort \\ \setSize{\policyLan}=$1\text{e}3$,2.21) (\primitiveNameShort \\ \setSize{\policyLan}=$5\text{e}3$,2.21) (\primitiveNameShort \\ \setSize{\policyLan}=$1\text{e}4$,2.21) (spacer, 0) (\baselineEmbed\\ \setSize{\policyLan}=$1\text{e}3$,2.48) (\baselineEmbed\\ \setSize{\policyLan}=$5\text{e}3$,5.22) (\baselineEmbed\\ \setSize{\policyLan}=$1\text{e}4$,8.64)};
\addlegendentry{OT Cost}
\addplot[ybar,fill=mygray3,draw=none] plot coordinates {(\primitiveNameShort \\ \setSize{\policyLan}=$1\text{e}3$,0.53) (\primitiveNameShort \\ \setSize{\policyLan}=$5\text{e}3$,2.66) (\primitiveNameShort \\ \setSize{\policyLan}=$1\text{e}4$,5.31) (spacer, 0) (\baselineEmbed\\ \setSize{\policyLan}=$1\text{e}3$,0) (\baselineEmbed\\ \setSize{\policyLan}=$5\text{e}3$,0) (\baselineEmbed\\ \setSize{\policyLan}=$1\text{e}4$,0)};
\addlegendentry{OLE Cost}

\end{groupplot}

\node (tab) [anchor=west] at ($(group c3r1.east)+(0.5cm,0)$) {
{\small
\begin{tabular}{c@{\hskip 6pt}c@{\hskip 6pt}c}
\toprule
 & \textbf{Comm.} &  {\textbf{$\client$ Latency}} \\
\midrule
\protocolValidateESP & 21.63KB & 11ms \\
GC & 2.72MB & 457ms \\
\protocolValidateESP + \protocolAKE & 22.01KB & 484ms \\
GC w/ \protocolAKE & 15.8MB & 2.14s \\
\bottomrule
\end{tabular}
}
};

\node [anchor=south, align=center] at (tab.north) {{\small D}};

\end{tikzpicture}
    \caption{Cost comparison for policy-enforced aPAKE; explicit check (A--C) and implicit check (D)}
    \label{fig:pake}
\end{figure*}
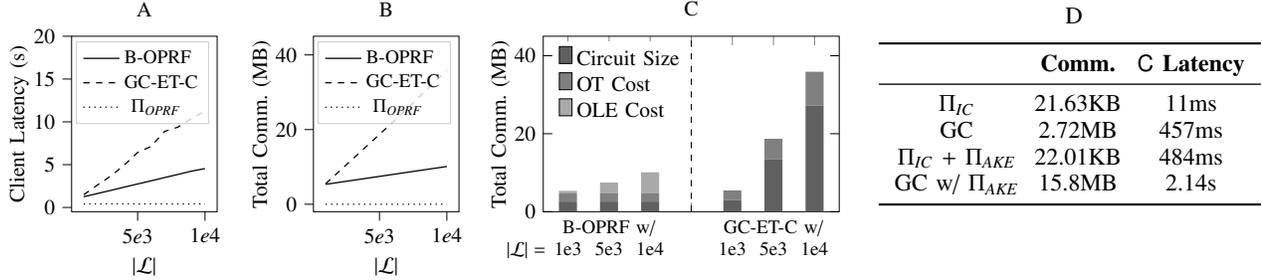

\baselineEmbed has the same accuracy as \primitiveNameShort. In
\figref{fig:pake} (A) we report the latency required for client to
complete the \explicitCheckPhase phase while increasing
$\setSize{\policyLan}$. Client latency was measured as the time
between \client starting and ending the \explicitCheckPhase, and so
includes all server computation, as well.  The
running time in \baselineEmbed grew rapidly, since the blocklist is
encoded into the garbled circuit.  In contrast, in
\primitiveNameShort, the latency and communication cost (B) scaled
much better, without the need to encode the list inside the garbled
circuit, and the \protocolOLE took only milliseconds. The latency of
\baselineEmbed was $1.19 \times$ the latency of \primitiveNameShort
when $\setSize{\policyLan} = 10^3$ and $2.48 \times$ when
$\setSize{\policyLan} = 10^4$.  In addition, we mark the latency and
network traffic for $\protocolOPRF$ and visualize the overhead
introduced by the policy enforcement.  
Without the blocklist checking, a plain OPRF from the OPAQUE IETF
draft\footnote{https://github.com/facebook/opaque-ke/} has a 416~ms
latency for client and a total communication size of 288 bytes.

In \figref{fig:pake} (C), we further visualize the communication cost
details for \primitiveNameShort and \baselineEmbed during
\explicitCheckPhase. Compared to \baselineEmbed, \primitiveNameShort
had OLE cost outside of the garbled circuit. When
$\setSize{\policyLan} = 10^3$, \primitiveNameShort had 136.71KB less
communication. The total communication size of \baselineEmbed was
$2.51\times$ the size of \primitiveNameShort when
$\setSize{\policyLan} = 5 \times 10^3$ and $3.55\times$ when
$\setSize{\policyLan} = 10^4$.

In \figref{fig:pake} (D) we measure all metrics for
\implicitCheckPhase and the same steps with
garbled circuits (GC).  In \implicitCheckPhase, the check is an
equality test and is independent of the blocklist size.  So both
solutions have fixed latency and communication cost, but those
of \protocolValidateESP are much smaller. We also mark the costs
of \implicitCheckPhase when combined with a follow-up 
authenticated key exchange \protocolAKE and a GC that contains 
\protocolAKE.

\subsection{Blocklisted MAC}

For the malware detection case, we use sdHash to embed
executables. \citet{Li15:FHReview} has demonstrated that the precision
and recall for malware clustering with sdHash is around 87\% with
fingerprint size of 2048 bits per block. We borrow the parameters for
our performance test from \citet{Li15:FHReview} for the
\primitiveNameShort-based blocklisted MAC.

\begin{figure}[h]
\centering
\begin{tikzpicture}[scale=0.8]

\definecolor{darkgray176}{RGB}{176,176,176}
\definecolor{lightgray204}{RGB}{204,204,204}

\begin{groupplot}[group style={group size=2 by 1, horizontal sep=1.8cm}, height=6cm, width=0.63\columnwidth]
\nextgroupplot[
title={A},
legend cell align={left},
legend style={
	fill opacity=0.8,
	draw opacity=1,
	text opacity=1,
	at={(0.03,0.97)},
	anchor=north west,
	draw=lightgray204
},
scaled x ticks=manual:{}{\pgfmathparse{#1}},
tick align=outside,
tick pos=left,
x grid style={darkgray176},
xlabel={\setSize{\policyLan}},
xtick style={color=black},
xticklabels={$0$,$0$,$5e2$, $1e3$}, 
y grid style={darkgray176},
ylabel={Client Latency (s)},
ylabel style={at={(axis description cs:-0.4,0.5)}, anchor=south},
ymin=-2, ymax=80,
ytick style={color=black},
height=3.8cm,
width=3cm,
]
\addplot [semithick, black]
table {%
100 4.347
200 8.481
300 12.618
400 16.856
500 20.894
600 25.03
700 29.174
800 33.308
900 37.442
1000 41.581
};
\addplot [semithick, black, dashed]
table {%
100 6.934
200 14.089
300 20.61
400 27.77
500 34.773
600 45.189
700 49.701
800 55.056
900 62.636
1000 68.974
};
\nextgroupplot[
title={B},
legend cell align={left},
legend style={
	fill opacity=0.8,
	draw opacity=1,
	text opacity=1,
	at={(0.03,0.97)},
	anchor=north west,
	draw=lightgray204
},
scaled x ticks=manual:{}{\pgfmathparse{#1}},
tick align=outside,
tick pos=left,
x grid style={darkgray176},
xtick style={color=black},
xlabel={\setSize{\policyLan}},
xticklabels={$0$,$0$,$5e2$, $1e3$}, 
y grid style={darkgray176},
ylabel={Total Comm. (MB)},
ylabel style={at={(axis description cs:-0.5,0.5)}, anchor=south},
ymin=-1.998, ymax=250,
ytick style={color=black},
height=3.8cm,
width=3cm,
xshift=-0.3cm,
]
\addplot [semithick, black]
table {%
100 10.191
200 20.383
300 30.574
400 40.766
500 50.957
600 61.149
700 71.34
800 81.531
900 91.722
1000 101.914
};
\addplot [semithick, black, dashed]
table {%
100 21.157
200 44.049
300 66.913
400 89.514
500 112.379
600 134.98
700 157.844
800 180.446
900 203.309
1000 227.191
};
\end{groupplot}

\node (tab) [anchor=west] at ($(group c2r1.east)+(0.3cm,0)$) {
{\small
\begin{tabular}{c@{\hskip 4pt}c@{\hskip 4pt}c}
\toprule
 & \textbf{Comm.} &  {\textbf{$\client[2]$ Latency}} \\
\midrule
\protocolValidateESP & 49KB & 73ms \\
GC & 10.3MB & 1052ms \\
\bottomrule
\end{tabular}
}
};

\node [anchor=south, align=center] at (tab.north) {{\small C}};

\end{tikzpicture}
    \vspace{-10pt}
    \caption{\client[1] latency (A), total communication (B) for 
      \explicitCheckPhase, and \implicitCheckPhase cost summary
      (C) between \primitiveNameShort (---) and \baselineEmbed (- -)
      in blocklisted MAC.}
    \label{fig:file}
\end{figure}

In \figref{fig:file} (A), we show the latency for the sending client
$\client[1]$ (left) to complete the \explicitCheckPhase (which
includes \server computation time, as well), with respect to the
$\setSize{\policyLan}$ growth.  The solid line represents the
\primitiveNameShort based blocklisted MAC and the dashed line
represents \baselineEmbed. \primitiveNameShort strictly outperformed
\baselineEmbed and the gap increased when the blocklist size
increased. The total communication volume is shown in
\figref{fig:file} (B).  \baselineEmbed requires a garbled circuit and
the circuit size dominated the communication cost. In the
\primitiveNameShort solution, both metrics grew linearly with the
expanding blocklist since the cost is dominated by the OLE operations.
With the blocklist size ranging from $100$ to $1{,}000$, the
communication cost of \baselineEmbed is $2.21\times$ the communication
size of \primitiveNameShort.

In \figref{fig:file} (C) we measure metrics for an
\implicitCheckPhase.  Similar to the policy-enforced aPAKE, both
solutions have fixed latency and communication cost. The network
traffic of GC is $209.14\times$ the network traffic of
\primitiveNameShort based blocklisted MAC.
The latency of GC is $14.41\times$ the latency of \primitiveNameShort
for $\client[2]$ during \implicitCheckPhase.

\section{Conclusion}

This paper presents a \primitiveName (\primitiveNameShort) which
produces an output from an OPRF only if its input is not in a blocklist. 
We built \primitiveNameShort protocols and 
demonstrated applications that augment aPAKEs and MACs
with lower overheads than a hashtable or a general garbled-circuit
baseline. From an intellectual perspective, the paper initiates the
study of general policy-enforced OPRF protocols, and addresses a
special case. This primitive opens avenues for additional application 
scenarios and broader generalization.

\bibliographystyle{IEEEtranSN}
\bibliography{bib/bib.bib}

\appendices

\section{One-Time Pairwise Unpredictable-Permutation}
\label{app:prp}
%

\subsection{Definition}
\label{app:defn_prp}

\begin{defn} 
	
	A keyed function $\genericFnFamily: \genericFnKeyspace \times \genericFnDomain 
	\rightarrow 
	\genericFnRange$ is one-time pairwise unpredictable if under a random choice of a key 
	$\genericFnKey \in \genericFnKeyspace$, and for any two distinct pairs of inputs $\cInput, 
	\cInputAlt 
	\in 
	\genericFnDomain$, and outputs $\genericFieldElmt, \genericFieldElmtAlt \in 
	\genericFnRange$
	
	\begin{eqnarray}
	 & \prob{\genericFnFamily[\genericFnKey](\cInput)} = \frac{1}{\setSize{\genericFnRange}}\\
		& \cprob{}{\genericFnFamily[\genericFnKey](\cInput) = \genericFieldElmt} 
		{\genericFnFamily[\genericFnKey](\cInputAlt) = \genericFieldElmtAlt} =
	\frac{\setSize{\genericFnRange}}{\setSize{\genericFnKeyspace}}  
	\end{eqnarray}

\noindent
where the probability is over the choice of the key. 
\end{defn}

The first condition in the definition implies that under a random choice of the key the function acts like a perfectly-secret cipher. 
Thus, it must be that $\setSize{\genericFnKeyspace} \ge \setSize{\genericFnRange}$. The second condition implies that given the evaluation at a specific point, the value at a different point can be predicted with probability at most $\frac{\setSize{\genericFnRange}}{\setSize{\genericFnKeyspace}}$. If $\setSize{\genericFnKeyspace} >> \setSize{\genericFnRange}$, then this probability is small and implies that the 
value is hard to predict.

We now extend this notion to a {\it one-time pairwise unpredictable permutation} 
over a metric space. We define the properties required from such a permutation below. 

\begin{defn}
	Let $\metricSpace$ be a metric space where 
	$\numberOfPointsSim \ge 1$. A keyed-function $\genericFnFamily:
	\genericFnKeyspace^{\numberOfPointsSim} \times \metricSpace \rightarrow \metricSpace$ is
	one-time pairwise unpredictable over the metric space \metricSpace if
	for any pair of input $\cInputVec,\cInputVecAlt
	\in \metricSpace$ for which $| \cInputVec \Delta \cInputVecAlt | \ge \diffInComps$ , and outputs 
	$\vecNotation{\genericFieldElmt},
	\vecNotation{\genericFieldElmtAlt} \in \metricSpace$, 
	
	\begin{eqnarray}
	& \prob{\genericFnFamily[\genericFnKey](\cInputVec) = \vecNotation{\genericFieldElmt}} \le \frac{1}{\setSize{\finiteField}^{\numberOfPointsSim}} \\
	&\cprob{}{\genericFnFamily[\genericFnKey](\cInputVec) = \vecNotation{\genericFieldElmt}}{
		\genericFnFamily[\genericFnKey](\cInputVecAlt) = \vecNotation{\genericFieldElmtAlt}}
	\le
	\left(\frac{\setSize{\finiteField}}{\setSize{\genericFnKeyspace}}\right)^{\diffInComps}
	\end{eqnarray}

\noindent
where the probability is over the choice of the key $\genericFnKey
\in \genericFnKeyspace^{\numberOfPointsSim}$. Here $| \cInputVec
\Delta \cInputVecAlt | = \diffInComps$ means that \cInputVec and
\cInputVecAlt differ in at least $\diffInComps$ dimensions.  

	
\end{defn}

Intuitively, this definition implies that the if two vectors disagree on \diffInComps components, then the evaluations at these components remain unpredictable. If there is a one-time pairwise unpredictable function 
$\genericFnFamily: \genericFnKeyspace \times \finiteField \rightarrow \finiteField$, 
then 
we can construct a one-time pairwise unpredictable permutation over the metric space 
$\metricSpace$ by using \numberOfPointsSim 
instances of 
$\genericFnFamily$, one for each dimension in $\metricSpace$.

\subsection{Security Proof}
\label{app:proof_prp}

\begin{theorem*}
	For a random selection of a key $\langle \PRPKeyVec[1], \PRPKeyVec[2]
	\rangle \in (\finiteField\setminus\{0\})^{\numberOfPointsSim} \times
	(\finiteField)^{\numberOfPointsSim}$, the function $\PRP[\langle
	\PRPKeyVec[1], \PRPKeyVec[2]\rangle](\cInputVec) = \PRPKeyVec[1] \hadamardProd
	\cInputVec + \PRPKeyVec[2]$ is a one-time pairwise-unpredictable permutation
	over \metricSpaceESP.
\end{theorem*}

\begin{proof}

	First note that $\PRP$ is a permutation. For a fixed key 
$\PRPKey \assign \langle \PRPKeyVec[1], \PRPKeyVec[2] \rangle$, 
if there are two vectors $\genericVec[1] \assign \langle 
\genericVecComp[1][1], \dots, \genericVecComp[1][\numberOfPointsSim] 
\rangle$ and $\genericVec[2] \assign \langle \genericVecComp[2][1], \dots, 
\genericVecComp[2][\numberOfPointsSim] \rangle$ such that $\genericVec[1] 
\neq \genericVec[2]$ and $\PRP[\genericPRPKey](\genericVec[1]) = 
\PRP[\genericPRPKey](\genericVec[2])$, then there is at least one $\ptIdx 
\in [1, \metricSpaceDimGen]$ where  $\genericVecComp[1][\ptIdx] \neq 
\genericVecComp[2][\ptIdx]$ but $\PRPKeyVec[1][\ptIdx] \cdot 
\genericVecComp[1][\ptIdx] + \PRPKeyVec[2][\ptIdx] = \PRPKeyVec[1][\ptIdx] 
\cdot \genericVecComp[2][\ptIdx] + \PRPKeyVec[2][\ptIdx]$. This leads to a 
contradiction that \finiteField is a field. Also, $\PRP$ is surjective because 
for any vector $\genericVec$, $\PRP[\genericPRPKey]^{-1}(\genericVec) = 
\langle \frac{\genericVecComp[1] - \PRPKeyVec[2][1]}{\PRPKeyVec[1][1]}, 
\dots, \frac{\genericVecComp[\metricSpaceDimGen] - 
\PRPKeyVec[2][\metricSpaceDimGen]}{\PRPKeyVec[1][\metricSpaceDimGen]} 
\rangle \in \metricSpaceGen$. Note, $\PRPKeyVec[1] \neq \langle 0, \dots, 0 
\rangle$.

Now, it is clear that $\prob{\PRP[\langle \PRPKeyVec[1], \PRPKeyVec[2]\rangle](\cInputVec) = \PRPKeyVec[1] \hadamardProd
\cInputVec + \PRPKeyVec[2] = \genericVec} = \frac{1}{\setSize{\finiteField}^{\numberOfPointsSim}}$ due to the random choice 
of $\PRPKeyVec[1]$ and $\PRPKeyVec[2]$ from $\metricSpace$. 
Consider another arbitrary input $\cInputVecAlt \in \metricSpaceESP$ such 
that $\setSize{\cInputVec \Delta  \cInputVecAlt} = \diffInComps$ 
where $\diffInComps \in [1, \numberOfPointsSim]$, i.e., 
 $\setSize{\cInputVec \Delta  \cInputVecAlt}$ is the 
 number of dimensions where \cInputVec and \cInputVecAlt 
 differ. Consider any one such dimension $\rootIdx \in [1, \numberOfPointsSim]$, and 
 two arbitrary vectors $\genericVec, \genericVecAlt \in \finiteField$. Then,

	
	\begin{align*}
	\prob{\PRPKeyVec[1][\rootIdx] \cdot \cInputVec[\rootIdx] + \PRPKeyVec[2][\rootIdx] = 
	\genericVecComp[\rootIdx]
	 \land \PRPKeyVec[1][\rootIdx] \cdot \cInputVecAlt[\rootIdx] + \PRPKeyVec[2][\rootIdx] = 
	 \genericVecAltComp[\rootIdx]}\\
	= \prob{\PRPKeyVec[2][\rootIdx] = 
	\genericVecComp[\rootIdx] - \PRPKeyVec[1][\rootIdx] \cdot \cInputVec[\rootIdx]
	\land \PRPKeyVec[1][\rootIdx] = \frac{\genericVecAltComp[\rootIdx] - 
	\genericVecComp[\rootIdx]}{\cInputVecAlt[\rootIdx] - \cInputVec[\rootIdx]}}
	= \frac{1}{\setSize{\finiteField \setminus 0} \cdot 
\setSize{\finiteField}}
	\end{align*}
	
The probability 
reflects the fact that it is always possible to produce one pair $\PRPKeyVec[1][\rootIdx], 
\PRPKeyVec[2][\rootIdx] \in \{\finiteField \setminus 0\} \times 
\finiteField$ to satisfy both equations, and since $\PRPKeyVec[1]$ and $\PRPKeyVec[2]$
are selected randomly, any such pair is equally likely. 
Also, for any other dimension $\setIdx \in [1, \numberOfPointsSim], \setIdx \neq \rootIdx$, 
where \cInputVec and \cInputVecAlt differ, $\prob{\PRPKeyVec[1][\rootIdx] \cdot 
\cInputVec[\rootIdx] + \PRPKeyVec[2][\rootIdx] = 
	\genericVecComp[\rootIdx]
	\land \PRPKeyVec[1][\rootIdx] \cdot \cInputVecAlt[\rootIdx] + \PRPKeyVec[2][\rootIdx] = 
	\genericVecAltComp[\rootIdx]}$ is independent of $\prob{\PRPKeyVec[1][\setIdx] \cdot 
	\cInputVec[\setIdx] + \PRPKeyVec[2][\setIdx] = 
	\genericVecComp[\setIdx]
	\land \PRPKeyVec[1][\setIdx] \cdot \cInputVecAlt[\setIdx] + \PRPKeyVec[2][\setIdx] = 
	\genericVecAltComp[\setIdx]} $ because each component of \PRPKeyVec[1]
	and \PRPKeyVec[2] are selected independently. Thus, we have 
	
\begin{align*}
	\prob{\PRP[\genericPRPKey](\cInputVec) = \genericVec \land 
	\PRP[\genericPRPKey](\cInputVecAlt) = \genericVecAlt} \\
	= \frac{1}{(\setSize{\finiteField \setminus 
	0} \cdot 
		\setSize{\finiteField})^{\setSize{\cInputVec \Delta \cInputVecAlt}}} \cdot \left(\frac{1}{\setSize{\finiteField}}\right)^{\numberOfPointsSim - \diffInComps} = \frac{1}{\setSize{\finiteField 
		\setminus 
			0}^{\diffInComps} \cdot 
		\setSize{\finiteField}^{\numberOfPointsSim}} \end{align*}

By applying Bayes' theorem, we get the result

\begin{align*}
\cprob{}{\PRP[\genericPRPKey](\cInputVec) = \genericVec}{\PRP[\genericPRPKey](\cInputVecAlt) = \genericVecAlt} = \frac{1}{\setSize{\finiteField 
		\setminus 
			0}^{\diffInComps}}
\end{align*}


\end{proof}

\section{Hamming-distance-aware private queries}
\label{app:HammingPSI}
Following \citet{Chakraborti2023:DAPSI}, we define a metric space over 
the finite field $\finiteField$ to represent the embeddings of the blocklist
\blockedElmts. The rationale is that computation over finite-field elements will result
in efficient cryptographic protocols. The metric space is defined over vectors
$\finiteField^{\numberOfPoints}$, where $\degreeOfKey{1}$ is
application dependent. Here we describe the steps for computing hamming
distance from the embeddings.

\myparagraph{Computing Hamming distance from \symmURDefn}
Consider that there are two bit vectors $\cInputVec \in \{0,1\}^{\degreeOfKey{1}}$ 
and $\cInputVecAlt \in \{0,1\}^{\degreeOfKey{1}}$, and we want to compute the Hamming 
distance between them $\hammingDist{\cInputVec}{\cInputVecAlt}$. 
\citet{Chakraborti2023:DAPSI} defines an injective mapping function $\vecMap$ (which takes the 
set $\setOfEvalPts$ as a parameter) that allows us to compute 
$\hammingDist{\cInputVec}{\cInputVecAlt}$ from \symmURDefn. 
The mapping function $\vecMap:
\{0,1\}^{\degreeOfKey{1}} \rightarrow \finiteField^{\numberOfPointsSim}$ is parameterized by the set of arbitrary points  $\setOfEvalPts \assign \{\evalPt[1], \dots, \evalPt[\numberOfPointsSim]\} \subset \finiteField$ for $\numberOfPointsSim \ge \numberOfPoints$, and works as follows: 
\begin{enumerate}[nosep, leftmargin=1.6em,labelwidth=*,align=left]
	\item \label{step:hammingPSI:map} {\bf Map the input vector to a set:} Given
	$\cInputVec \in \{0,1\}^{\degreeOfKey{1}}$, use an injective function $\mapFn:
	\{0,1\} \times [1,\degreeOfKey{1}] \rightarrow \finiteField$ to create a
	set $\genericSetVar[\cInputVec] = \{\mapFn(\cInputVec[\rootIdx], \rootIdx)\}_{\rootIdx \in 
	\degreeOfKey{1 }}$, where
	$\cInputVec[\rootIdx]$ is the \rootIdx-th component of \cInputVec. 
	
	\item \label{step:hammingPSI:poly_create} {\bf Create a polynomial from the set
		\genericSetVar[\cInputVec]:}
	Compute a polynomial $\polyFromEmbedding[\cInputVec] \assign
	\prod\limits_{\genericSetElmt
		\in \genericSetVar[\cInputVec]} (\genericFieldElmt - \genericSetElmt)$, a polynomial with roots in $\genericSetVar[\cInputVec]$. 
	
	\item \label{step:hammingPSI:eval} {\bf Evaluate the polynomial:}
	Evaluate $\polyFromEmbedding[\cInputVec]$ at the points in \setOfEvalPts and output 
	a vector $\genericVec \assign \langle
	\polyFromEmbedding[\cInputVec][\evalPt[1]], \dots,
	\polyFromEmbedding[\cInputVec][\evalPt[\numberOfPointsSim]]\rangle$.

\end{enumerate}

Importantly, for any two vectors $\genericVec[1], \genericVec[2]$ derived as
such from the above process 

\begin{align}
	\hammingDist{\cInputVec}{\cInputVecAlt} \le \distInMaskedSets & \iff
	\setSize{\genericSetVar[\cInputVec] \cap \genericSetVar[\cInputVecAlt]} \ge
	(\degreeOfKey{1} - \distInMaskedSets) \label{eq:hamming_distance_diff} \\
	& \iff \frac{1}{2} \left(2 \numberOfMasks -
	\setSize{\genericSetVar[\cInputVec] \symmDiff
		\genericSetVar[\cInputVecAlt]}\right) \ge 	(\degreeOfKey{1} - \distInMaskedSets)  \label{eq:diff}\\
	& \iff \symmURDist{\genericVec}{\genericVecAlt} \le \threshold \label{eq:diff_sur}
\end{align}

where $\threshold =  2\distInMaskedSets$. The function 
\vecMap is injective for any value $\numberOfPointsSim \ge \degreeOfKey{1} + 1$ because the 
polynomial $\polyFromEmbedding[\cInputVec]$ has $\degreeOfKey{1}$ roots, and such a 
polynomial can be uniquely defined by $\degreeOfKey{1} + 1$ points. Thus, the equivalences in \label{eq:hamming_distance_sur}
follow by definition of Hamming distance \eqnref{eq:hamming_distance_diff}, symmetric difference \eqnref{eq:diff}, and the definition of 
\symmUR \eqnref{eq:diff_sur}. 

\myparagraph{\symmURDefn computation from noisy 
polynomial addition}
One advantage of using $\symmURDefn$ as the metric is that we can
compute the distance between two vectors contributed by mutually
untrusting parties {\it without explicitly revealing the vectors to
  each other}. The process is as follows. Let \genericVec[1]
and \genericVec[2] be two vectors contributed by $\client$ and \server,
respectively. They sample random polynomials $\randomPoly[1] \getsr
\polyField$ and $\randomPoly[2] \getsr \polyField$ respectively, of
degree $\degreeOfKey{1}$. 
Then, they locally compute $\randomVec[1]
\gets \langle \randomPolyEval[\evalPt][1] \rangle_{\evalPt \in
  \setOfEvalPts}$ and $\randomVec[2] \gets \langle
\randomPolyEval[\evalPt][2] \rangle_{\evalPt \in \setOfEvalPts}$.
Finally, using calls to \npaFunc (see \secref{sec:prelim}),
they compute $\randomVec[2] \hadamardProd \genericVec[1]
+ \randomVec[1] \hadamardProd \genericVec[2]$. Then,

\begin{align}
  \label{eq:sur_random}
& \symmURDist{\genericVec[1]}{(\randomVec[2] \hadamardProd \genericVec[1] +
\randomVec[1] \hadamardProd \genericVec[2])} \nonumber\\
& = \symmURDist{\genericVec[2]}{(\randomVec[2] \hadamardProd \genericVec[1] +
\randomVec[1] \hadamardProd \genericVec[2])} = \symmURDist{\genericVec[1]}{\genericVec[2]} 
& \mbox{(w.h.p.)} 
\end{align}

We prove \eqnref{eq:sur_random} below, and argues that since $\randomPoly[1]$ and
$\randomPoly[2]$ are random polynomials, with overwhelming
probability, they do not share common roots with the polynomials
\polyFromVec{\genericVec[1]} and \polyFromVec{\genericVec[2]} that interpolate
\genericVec[1] and \genericVec[2], respectively. Thus,
$\gcd(\polyFromVec{\genericVec[1]},\combPoly) =
\gcd(\polyFromVec{\genericVec[2]},\combPoly)
=\gcd(\polyFromVec{\genericVec[1]},\polyFromVec{\genericVec[2]})$ with high
probability, where $\combPoly \gets
\randomPoly[2] \cdot \polyFromVec{\genericVec[1]} + \polyFromVec{\genericVec[1]}
\cdot \randomPoly[1]$.

\begin{prop}
	\label{prop:symmURFromRand}
	Let $\genericVec[1], \genericVec[2] \in \finiteField^{\numberOfPoints}$.
	Let $\randomPoly[1], \randomPoly[2] \in \polyField$ be random
	polynomials of fixed degrees $\degreeOfPoly(\randomPoly[1]) 
	\ge \degreeOfPoly(\polyFromVec{\genericVec[1]})$ and
	$\degreeOfPoly(\randomPoly[2]) \ge \degreeOfPoly(\polyFromVec
	{\genericVec[2]})$. Then,
	{\small
	\begin{align*}
		\prob{\symmURDist{\genericVec[1]}{\genericVec[2]} \neq
			\setSize{\uniqueRoots{\polyFromVec{\genericVec[1]}}} + 
			\setSize{\uniqueRoots{\polyFromVec{\genericVec[2]}}} -
			2\cdot\setSize{\uniqueRoots{\genericPoly}}}
		& \le \frac{1}{\setSize{\finiteField}}\\
		\mathllap{\mbox{for $\genericPoly = \gcd(\polyFromVec
				{\genericVec[1]}, \randomPoly[1] \polyFromVec{\genericVec[2]})$}} \\
		\prob{\symmURDist{\genericVec[1]}{\genericVec[2]} \neq
			\setSize{\uniqueRoots{\polyFromVec{\genericVec[1]}}} + 
			\setSize{\uniqueRoots{\polyFromVec{\genericVec[2]}}} -
			2\cdot\setSize{\uniqueRoots{\genericPoly}}}
		& \le \frac{1}{\setSize{\finiteField}}\\
		\mathllap{\mbox{for $\genericPoly = \gcd(\polyFromVec
				{\genericVec[1]} \randomPoly[2], \polyFromVec{\genericVec[2]})$}} \\
		\prob{\symmURDist{\genericVec[1]}{\genericVec[2]} \neq
			\setSize{\uniqueRoots{\polyFromVec{\genericVec[1]}}} + 
			\setSize{\uniqueRoots{\polyFromVec{\genericVec[2]}}} -
			2\cdot\setSize{\uniqueRoots{\genericPoly}}}
		& \le \frac{1}{\setSize{\finiteField}} \\
		\mathllap{\mbox{for $\genericPoly = \gcd(\polyFromVec{\genericVec[1]} 
				\randomPoly[2], \randomPoly[1]\polyFromVec{\genericVec[2]})$}} \\
	\end{align*}
	}
\end{prop}

The proposition above follows from the result that given some fixed
polynomial \genericPoly and a random polynomial
$\randomPoly$, we have $\prob{\gcd(\genericPoly, \randomPoly) \neq 1} \le
1/\setSize{\finiteField}$~\cite{Ghosh2019communication}.

Additionally, we will use the following result by \citet{Kissner2005:PSI}.

\begin{lemma}[\cite{Kissner2005:PSI}]
	\label{lemma:kissener_uniformly_random}
	Given two polynomials $\genericPoly[1]$ and $\genericPoly[2]$, and two
	uniformly random polynomials $\randomPoly[1]$ and $\randomPoly[2]$ of
	fixed degrees $\degreeOfPoly(\randomPoly[1]) \ge \degreeOfPoly
	(\genericPoly[2])$ and $\degreeOfPoly(\randomPoly[2]) \ge 
	\degreeOfPoly(\genericPoly[1])$, the polynomial \randomPoly[3] satisfying the following 
	is a random polynomial. 
	\[
	\randomPoly[1] \cdot \genericPoly[1] + \randomPoly[2] \cdot \genericPoly[2] 
	= \gcd(\genericPoly[1], \genericPoly[2]) \cdot \randomPoly[3]
	\]
\end{lemma}

Thus, we can compute $\symmURDist{\genericVec[1]}{\genericVec[2]}$ from $\randomPoly[1] \cdot \polyFromVec{\genericVec[1]}
+ \randomPoly[2] \cdot \polyFromVec{\genericVec[2]}$, knowing either
$\polyFromVec{\genericVec[1]}$ or $\polyFromVec{\genericVec[2]}$, 
and the degrees of $\polyFromVec{\genericVec[1]}$,
$\polyFromVec{\genericVec[2]}$. This is because
$\gcd(\polyFromVec{\genericVec[1]}, \randomPoly[1] \cdot
\polyFromVec{\genericVec[1]} + \randomPoly[2] \cdot
\polyFromVec{\genericVec[2]}) = \gcd(\polyFromVec{\genericVec[1]}, 
\polyFromVec{\genericVec[2]})$ with overwhelming probability.

\myparagraph{Hamming-distance-aware private queries}
We will use the Hamming-distance-aware query protocol by 
\citet{Chakraborti2023:DAPSI}. It does not make any assumptions
about the input-vector distributions; in constrast to other options. 
Given two bit vectors $\cInputVec \in
\{0,1\}^{\degreeOfKey{1}}$ and $\cInputVecAlt \in
\{0,1\}^{\degreeOfKey{1}}$, the protocol checks whether
$\hammingDist{\cInputVec}{\cInputVecAlt} \le \distInMaskedSets$. If
$\hammingDist{\cInputVec}{\cInputVecAlt} > \distInMaskedSets$, then
neither of the parties learn any information about $\cInputVec$ and
$\cInputVecAlt$. Here, we will discuss the high level steps, which is
enough to understand its use in a \primitiveNameShort
construction. The protocol has two steps:

\smallskip
{\it \underline{Mapping step}:}
The parties first locally map their respective inputs to the metric space 
$(\finiteField^{\numberOfPointsSim}, \symmURDefn)$ using the injective map 
$\vecMap$ described above. For this, the parties first agree on the set of elements $\setOfEvalPts 
\subset \finiteField$. Let $\genericVec \assign \vecMap(\cInputVec)$ and $\genericVecAlt \assign 
\vecMap(\cInputVecAlt)$.

\smallskip
{\it \underline{Matching step}:}
To check whether $\hammingDist{\cInputVec}{\cInputVecAlt} \le
\distInMaskedSets$, the parties interactively and privately check
that $\symmURDist{\genericVec}{\genericVecAlt} \le \threshold$ using 
\eqnsref{eq:hamming_distance_diff}{eq:diff_sur}. 
For this, the protocol leverages the fact that \symmURDefn can be computed using 
\npaFunc due to \eqnref{eq:sur_random}. 

Crucially, the protocol sets $\numberOfPointsSim \in (\degreeOfKey{1}, 2\degreeOfKey{1} + 1)$ as a
function of $\degreeOfKey{1}$ and \threshold such that if
$\symmURDist{\genericVec}{\genericVecAlt} \le \threshold$, then the server
learns \genericVec  (and $\cInputVec$), otherwise {\it no} information is
revealed about $\cInputVec$ to the server. Intuitively, the idea is that with
less than $\numberOfPointsSim \le 2\degreeOfKey{1} + 1$ points, it is not
possible
to uniquely define and interpolate the polynomial $\randomPoly[2]
\polyFromEmbedding[\cInputVec] + \randomPoly[1]
\polyFromEmbedding[\cInputVecAlt]$
and thus $\randomVec[1], \randomVec[2]$,  unless
$\polyFromEmbedding[\cInputVec]$ and $\polyFromEmbedding[\cInputVecAlt]$ share
$2(\degreeOfKey{1} - \threshold)$ roots.

\section{Policy Enforced Strong aPAKE}
\label{app:opaque}

In \figref{fig:frameworkSH}, observe that $\client$ learns $\embedKey$ 
during \implicitCheckPhase. Thus, an adversary can also obtain \embedKey 
by pretending to be an honest $\client$, and pre-compute the values of 
$\embedESP(\pwAlt) \linearComb \hash(\pwAlt, \embedESP(\pwAlt), \embedKey)$ 
for some candidate passwords $\pwAlt \in \universe$ offline. Then, once it breaches \server, it 
can check in constant time whether there is some 
$\authTokenAlt = \embedESP(\pwAlt) \linearComb \hash(\pwAlt, \embedESP(\pwAlt), \embedKey)$ 
stored on \server, corresponding to one of the precomputed values of \pwAlt.
In fact, the problem remains even when \embedKey is kept secret
because the key space for the embedding function is not necessarily 
large, e.g., with 32-bit Hamming vectors, the key space may be limited to the space of all 32-bit 
strings, which does give enough security against guessing attacks. 


Fortunately, this problem can be easily solved with an 
additional pseudorandom function $\aPAKEOPRF$. More precisely, 
the modifications required are i) during \explicitCheckPhase the 
augmented functionalities $\embedAndMap'$ and $\testAndCommit'$ 
return $\PRF[1][\PRFKey[1]]\left( \embedESP(\cInput) \linearComb 
\hash\left(\aPAKEOPRF[\aPAKEOPRFKey]\left(\cInput, 
\embedESP(\cInput), \embedKey \right)\right)\right)$ 
to \client and  $\authToken \assign \embedESP(\cInput) \linearComb \hash\left(\aPAKEOPRF[\aPAKEOPRFKey]\left(\cInput, \embedESP(\cInput), \embedKey \right)\right)$ to \server, where 
$\PRF[1]$ and $\aPAKEOPRF$ are pseudorandom functions, 
ii) during \implicitCheckPhase, \client first learns 
$\aPAKEOPRF[\aPAKEOPRFKey](\cInput, \embedESP(\cInput), \embedKey)$ 
interactively using $\oprfFunc$, where \client's input 
is $(\cInput, \embedESP(\cInput), \embedKey)$ and \server's input is 
$\aPAKEOPRFKey$, and then reconstructs $\authToken$. 

Here we borrow the security definition of ``strong'' aPAKE from 
 \citet{Jarecki2018:OPAQUE} and 
present a ``policy-enforced'' strong aPAKE protocol with:

\begin{itemize}[topsep=0.5em,leftmargin=1em,labelwidth=*,align=left, nosep]
  \item \textit{Protection against precomputation attacks:} The 
  server is not provided the client's password or any deterministic 
  mapping of the password during registration, or subsequently 
  during login, that can be used for precomputation attacks.
	
  \item \textit{Protection against cheating clients:} If a client
  deviates from the protocol during the aPAKE registration, the
  protocol commits to a pseudorandom value and the client's
  advantage in obtaining this value in order to subsequently login is
  negligible. In other words, if the client cheats during registration, 
  it gets locked out of its own account.
	
  \item \textit{Confidentiality of policy/blocklist:} Clients do not
  learn about specific policies or contents of the blocklist that a
  server is enforcing. This property may be useful if the passwords 
  in the blocklist need to be kept confidential. 

\end{itemize}

\figref{fig:protocol_pePAKE} describes a strong aPAKE protocol (derived 
from OPAQUE~\cite{Jarecki2018:OPAQUE}) which performs a blocklist 
check to prevent the client from registering with a password in 
$\blockedElmts$ without learning the client's input. We mark the steps
that are unique to \primitiveNameShort in italic.

\begin{figure}
\begin{oframed}
{\small
		
		\smallskip\noindent \textbf{Public Parameters:} Universe $\universe$ 
		from which passwords are drawn; 
		$\ell = \setSize{\universe} + \setSize{\embedKeySpace} + \numberOfPointsSim \cdot \setSize{\finiteField}$; 
		An embedding function 
		$\embed:\embedKeySpace\times \universe
		\rightarrow \metricSpaceESP$;
		A hash function $\hash: \domainHash \rightarrow \finiteField^{\numberOfPointsSim}$; 
		Two pseudo-random functions $\PRF[1]: \PRFKeySpace[1] \times 
		\finiteField^{\numberOfPointsSim} \rightarrow \rangePRF$ and $\aPAKEOPRF: \aPAKEOPRFKeySpace \times \{0,1\}^{\ell} \rightarrow {\rangePRF}$;
		A keyed permutation $\PRP: \PRPKeySpace \times \finiteField^{\numberOfPointsSim} \rightarrow 
		\finiteField^{\numberOfPointsSim}$; An authenticated key exchange scheme \protocolAKE; An oblivious PRF functionality \oprfFunc for $\aPAKEOPRF$.
		
		\smallskip\noindent\textbf{Inputs:} The client \client's input is $\cInput 
		\in \universe$. The server \server's input is a blocklist $\policyLan \subseteq \finiteField^{\numberOfPointsSim}$ and a threshold $\threshold$ such that 
		if $\cInput \in \blockedElmts$ then $\blocked{\blockedElmts}{\policyLan}{\threshold}(\cInput) = \tru$ with probability at least 
		$\trueRejectRate{\blockedElmts}{\policyLan}{\threshold}$. 	
		\server also 
		has randomly sampled keys $\PRFKey[1] \in \PRFKeySpace[1], \aPAKEOPRFKey \in \aPAKEOPRFKeySpace$ and $\PRPKey[1], \PRPKey[2] \in \PRPKeySpace$, $\embedKey \in \embedKeySpace$.
		
		\smallskip\noindent\textbf{Registration (\explicitCheckPhase):}
		
		\begin{enumerate}[nosep, leftmargin=1em,labelwidth=*,align=left]
			\item \textit{\client and \server call 
				$\embedAndMap'$ and $\testAndCommit'$. \client's input is \cInput. 
				\server's input is $\LanPolicy, \PRFKey[1], \aPAKEOPRFKey, \PRPKey[1], \PRPKey[2], \embedKey$. }
			
			\textit{\server obtains $\authToken \assign  \embedESP(\cInput) \linearComb \hash\left(\aPAKEOPRF[\aPAKEOPRFKey]\left(\cInput, \embedESP(\cInput), \embedKey \right)\right)$}
			
			\textit{if $\blocked{\blockedElmts}{\policyLan}{\threshold}(\cInput) = \fals$. Otherwise, \server aborts.}

			\item  \server sets $\credKey \assign \PRF[1][\PRFKey[1]](\authToken)$ and generates two public-private key pairs 
			$\pakeClientPubKey,\pakeClientPrivKey$ and $\pakeServerPubKey, 
			\pakeServerPrivKey$ respectively, and computes $\Enc[\credKey]
			(\pakeClientPrivKey, \pakeClientPubKey, \pakeServerPubKey)$ 
			and stores $(\pakeServerPrivKey, \pakeServerPubKey, 
			\pakeClientPubKey, \clientID)$. 
		\end{enumerate}
		
		\smallskip\noindent\textbf{Authentication (\implicitCheckPhase):}
		
		\begin{enumerate}[nosep, leftmargin=1em,labelwidth=*,align=left,resume]
			
			\item \textit{\client learns \embedKey from \server}.
			
			\item \textit{\client and \server invoke $\oprfFunc$ with inputs $(\cInputAlt, \embedESP(\cInputAlt), \embedKey)$ and $\aPAKEOPRF$ respectively. \client learns $\aPAKEOPRF[\aPAKEOPRFKey]\left(\cInputAlt, \embedESP(\cInputAlt), \embedKey\right)$ }.
			
			\item \textit{\client computes $\authTokenAlt \assign \embedESP(\cInputAlt) \linearComb \hash\left(\aPAKEOPRF[\aPAKEOPRFKey]\left(\cInputAlt, \embedESP(\cInputAlt), \embedKey\right)\right)$}.
			
			\item   \textit{\client and \server call \authentication 
				with input $\authTokenAlt$, and $\authToken, \PRFKey[1]$ respectively. \client 
				obtains $\credKey \assign \PRF[1][\PRFKey[1]](\authToken)$ iff \client has registered with \server with the input $\cInput$}.
			
			\item \server sends  $\Enc[\credKey](\pakeClientPrivKey, 
			\pakeClientPubKey, \pakeServerPubKey)$ to \client who can retrieve 
			the public-private keys.
			
			\item \server and \client call \protocolAKE to set up a shared session 
			key. \protocolAKE takes as inputs $(\pakeClientPrivKey, 
			\pakeClientPubKey, \pakeServerPubKey)$ from \client, and 
			$(\pakeServerPrivKey, \pakeServerPubKey, \pakeClientPubKey)$ from 
			\server, and outputs a session ID $\sessionID$ to \client and \server. 
		\end{enumerate}

}
\end{oframed}		
	\caption{Policy-enforced strong aPAKE with OPRF and authenticated 
		key exchange. \label{fig:protocol_pePAKE}}
\end{figure}

\section{Proof of Framework Security}
\label{app:proof_framework}

In this appendix, we prove security for the framework of 
\figref{fig:frameworkSH}, under the assumption that the constituent 
functionalities $\embedAndMap, \testAndCommit, \authentication$ 
are secure.  (The proofs of security for these functionalities can be found in
\secref{app:proofs_pake}.)  We reduce the security of our framework to typical
properties of types of function families, which we recount below.  

\begin{defn}[Weak preimage resistance]
  \label{dfn:wpr}
  Let $\genericFnFamily: \genericFnKeyspace \times \genericFnDomain
  \rightarrow \genericFnRange$ be a family of functions, and let
  \Adv{} be an algorithm that takes an oracle $\genericFnDomain
  \rightarrow \genericFnRange$ and returns an element of
  \genericFnDomain.  Let
	
  \smallskip
  \indent\begin{minipage}[t]{0.5\textwidth}
  \begin{tabbing}
    ****\=****\=\kill
    Experiment $\ExptWPR{\genericFnFamily}(\Adv{})$ \\
    \> $\genericFnKey \getsr \genericFnKeyspace$ \\
    \> $\cInput \getsr \genericFnDomain$ \\
    \> $\prfOutput \gets \genericFnFamily[\genericFnKey](\cInput)$ \\
    \> $\cInputAlt \leftarrow \Adv{}^{\genericFnFamily
      [\genericFnKey](\cdot)}(\prfOutput)$ \\
    \> if $\genericFnFamily[\genericFnKey](\cInputAlt) = \prfOutput$ \\
    \> \> then return $1$ \\
    \> \> else return $0$
  \end{tabbing}
  \end{minipage}
	
  \smallskip
  \noindent Then,
  \begin{align*}
    \Advantage{\WPRsecdef}{\genericFnFamily}{\Adv{}} 
   & = \prob{\ExptPRFIND(\Adv{}) = 1} \\
    \Advantage{\WPRsecdef}{\genericFnFamily}{\timeBound, 
    \forwardOracleQueries} 
    & = \max_{\Adv{}} \Advantage
    {\WPRsecdef}{\genericFnFamily}{\Adv{}}
  \end{align*}
  where the maximum is taken over all algorithms \Adv{} that run in
  time at most \timeBound and make at most \forwardOracleQueries
  oracle queries.
\end{defn}

\begin{defn}[One-time pairwise unpredictability]
  \label{dfn:prp}
  Let $\genericFnFamily: \genericFnKeyspace \times \genericFnDomain
  \rightarrow \genericFnDomain$ be a family of permutations, and
  let \Adv{} be an algorithm that returns a pair of elements from
  $\genericFnDomain$.  Let
  
  \smallskip
  \indent\begin{minipage}[t]{0.5\textwidth}
  \begin{tabbing}
    ****\=****\=\kill
    Experiment $\ExptPRP{\genericFnFamily}(\Adv{})$ \\
    \> $\genericFnKey \getsr \genericFnKeyspace$ \\
    \> $\cInput \leftarrow \Adv{}()$\\
    \> $\genericFieldElmt \gets \genericFnFamily[\genericFnKey](\cInput)$ \\
    \> $(\cInputAlt, \genericFieldElmtAlt) \leftarrow \Adv{}(\genericFieldElmt)$ \\
    \> if $\genericFnFamily[\genericFnKey](\cInputAlt) = \genericFieldElmtAlt$ \\
    \> \> then return $1$ \\
    \> \> else return $0$
  \end{tabbing}
  \end{minipage}
  
   \smallskip
  \noindent Then,
  \begin{align*}
    \Advantage{\PRPsecdef}{\genericFnFamily}{\Adv{}} 
   & = \prob{\ExptPRP(\Adv{}) = 1} \\
    \Advantage{\PRPsecdef}{\genericFnFamily}{\timeBound} 
    & = \max_{\Adv{}} \Advantage
    {\PRPsecdef}{\genericFnFamily}{\Adv{}}
  \end{align*}
  where the maximum is taken over all algorithms \Adv{} that run in
  time at most \timeBound.
\end{defn}

\subsection{Security against \client}
\label{sec:proof_framework:client}

\subsubsection{In threat model \ref{model:client-hbc}}

Assuming \embedAndMap, \testAndCommit, and \authentication are secure, 
then no transcripts are leaked to \client and the only information \client 
learns are the outputs. Since \policyLan is not involved in the computation of 
any outputs, it is trivial that no information about \blockedElmts will be leaked 
to the client than the decision ($\cInput \in \blockedElmts$ or not)
implies in one execution of the \explicitCheckPhase phase.

\subsubsection{In threat model \ref{model:client-mal}}

In this threat model, \client is malicious and aims to have a
commitment stored at \server on some $\cInput \in \blockedElmts$. As
such, the goal of such a \client is to cause the following experiment
to return $1$, where \client is denoted as a triple of algorithms
$(\Adv{1}, \Adv{2}, \Adv{3})$.  For each call to the functionalities,
each input (output) pairs will be denoted in parenthesis, with the
\client input (output) as the first item and the \server input
(output) as the second item.

\smallskip

\begin{minipage}[t]{\columnwidth}
	\begin{tabbing}
		****\=****\=\kill
		Experiment $\ExptCC[\PRF[1],\PRP, \hash]
		(\Adv{1}, \Adv{2}, \Adv{3})$ \\
		\> $\langle \Advstate{1}, \cInput \rangle \gets \Adv{1}^{\hash(\cdot)}()$ \\
		\> $\PRPKey[1] \getsr \PRPKeySpace$,
		$\PRPKey[2] \getsr \PRPKeySpace$, \\
		\> $\PRFKey[1] \getsr \PRFKeySpace[1]$,
		$\embedKey \getsr \embedKeySpace$ \\
		\> $((\prpOutputEmbedding, \prpOutputHash), \cdot) \gets \embedAndMap^{\hash(\cdot)}(\cInput, (\PRPKey[1], \PRPKey[2], \embedKey))$ \\
		\> $\langle \Advstate{2}, \prpOutputEmbeddingAlt, \prpOutputHashAlt \rangle  \gets \Adv{2}^{\hash(\cdot)}(\Advstate{1}, \prpOutputEmbedding, \prpOutputHash)$ \\
		\> $(\prfOutput, \authToken) \gets \testAndCommit^{\hash(\cdot)}((\prpOutputEmbeddingAlt, \prpOutputHashAlt), (\PRPKey[1], \PRPKey[2], \PRFKey[1]))$ \\
		\> if $\prfOutput = \bot$ \\
		\> \> then return 0 \\	
		\> $\authTokenAlt \gets 
		\Adv{3}^{\hash(\cdot),\authentication^{\hash(\cdot)}(\cdot, (\authToken, \PRFKey[1]))}
		(\Advstate{2}, \prfOutput, \embedKey)$ \\ 
		\> if \= $\cInput \in \blockedElmts \wedge \authTokenAlt = \authToken$ \\
		****\=****\=\kill
		\> \> then return 1 \\ 
		\> \> else return 0
	\end{tabbing}
\end{minipage}

\noindent Then,
\begin{align*}
	\Advantage{\DCOPRFsecdef}{\PRF[1],\PRP, \hash}{\Adv{1}, \Adv{2}, \Adv{3}}
	& = \prob{\ExptCC[\PRF[1],\PRP, \hash]
		(\Adv{1}, \Adv{2}, \Adv{3}) = 1} \\
	\Advantage{\DCOPRFsecdef}{\PRF[1],\PRP, \hash}{\timeBound,\forwardOracleQueries}
	& = \max_{\Adv{1}, \Adv{2}, \Adv{3}} \Advantage{\DCOPRFsecdef}
	{\PRF[1],\PRP, \hash}{\Adv{1}, \Adv{2}, \Adv{3}}
\end{align*}
where the maximum is taken over all adversaries $(\Adv{1},
\Adv{2}, \Adv{3})$ running in total time \timeBound and making
at most \forwardOracleQueries oracle queries to
$\authentication(\cdot, (\authToken, \PRFKey[1]))$ from \Adv{3}.
\bigskip

The key distinctions between the prescribed algorithm in
\figref{fig:frameworkSH} and \ExptCC[\PRF[1],\PRP, \hash] are that (i)
the (malicious) client is permitted to \textit{compute}
$\authTokenAlt$ (in \Adv{3}) and $\prpOutputEmbeddingAlt,
\prpOutputHashAlt$ (in \Adv{2}), whereas a correct client simply sets
$\authTokenAlt \gets \embed[\embedKey](\cInput) \linearComb
\hash(\cInput, \embed[\embedKey](\cInput), \embedKey)$,
$\prpOutputEmbeddingAlt \gets \prpOutputEmbedding$, and
$\prpOutputHashAlt \gets \prpOutputHash$; and (ii) the (malicious)
client is given oracle access to $\authentication(\cdot, (\prfOutput,
\PRFKey[1]))$, to model repeated attempts at \implicitCheckPhase.

\begin{prop}
	\label{prop:framework}
	If $\embedAndMap, \testAndCommit, \authentication$ are secure 
	and output correctly, then
	\begin{align*}
		& \Advantage{\DCOPRFsecdef}{\PRF[1],\PRP, \hash}{\timeBound, \forwardOracleQueries} \\
		& \le \Advantage{\PRFINDsecdef}{\PRF[1]}{\timeBoundAlt, 1}
		+ \falseAcceptRate{\blockedElmts}{\policyLan}{\threshold}
		+ \frac{2}{\setSize{\PRPKeySpace}}
	\end{align*}
	for $\timeBoundAlt = \timeBound + \bigO{1}$.
\end{prop}

\begin{proof}
	
	Let $(\Adv{1}, \Adv{2}, \Adv{3})$ be a
	\primitiveNameShort adversary that runs in time \timeBound and makes 
	at most \forwardOracleQueries queries to $\authentication$ oracle.
	First note that
	{\small
	\begin{align*}
		\mathrlap{\Advantage{\DCOPRFsecdef}{\PRF[1],\PRP, \hash}{\Adv{1}, \Adv{2}, \Adv{3}}} \\
		& = \prob{\ExptCC[\PRF[1],\PRP, \hash](\Adv{1}, \Adv{2}, \Adv{3}) = 1} \\
		& = \cprob{\Bigg}{\ExptCC[\PRF[1],\PRP, \hash](\Adv{1}, \Adv{2}, \Adv{3}) = 1}{\begin{array}{@{}r@{}} \neg\blocked{\policyLan}{\threshold}(\embed[\embedKey](\cInput))\\ \wedge~\cInput\in\blockedElmts\end{array}} \\
		& \hspace{2em} \times \cprob{\big}{\neg\blocked{\policyLan}{\threshold}(\embed[\embedKey](\cInput))}{\cInput\in\blockedElmts} \times \prob{\cInput\in\blockedElmts}\\
		& \hspace{1em} + \cprob{\Bigg}{\ExptCC[\PRF[1],\PRP, \hash](\Adv{1}, \Adv{2}, \Adv{3}) = 1}{\begin{array}{@{}r@{}} \blocked{\policyLan}{\threshold}(\embed[\embedKey](\cInput))\\ \wedge~\cInput\in\blockedElmts\end{array}} \\
		& \hspace{2em} \times \cprob{\big}{\blocked{\policyLan}{\threshold}(\embed[\embedKey](\cInput))}{\cInput\in\blockedElmts} \times \prob{\cInput\in\blockedElmts}\\
		& \le \cprob{\big}{\neg\blocked{\policyLan}{\threshold}(\embed[\embedKey](\cInput))}{\cInput\in\blockedElmts} \\
		& \hspace{1em} + \cprob{\Bigg}{\ExptCC[\PRF[1],\PRP, \hash](\Adv{1}, \Adv{2}, \Adv{3}) = 1}{\begin{array}{@{}r@{}} \blocked{\policyLan}{\threshold}(\embed[\embedKey](\cInput))\\ \wedge~\cInput\in\blockedElmts\end{array}}
	\end{align*}
	}

The first term is just
$\falseAcceptRate{\blockedElmts}{\policyLan}{\threshold}$, and so for
the rest of the proof, we focus on bounding the second term.  Recall
that $\ExptCC[\PRF[1],\PRP, \hash] (\Adv{1}, \Adv{2}, \Adv{3}) = 1$
implies $\neg\blocked{\policyLan}
{\threshold}(\PRP[\PRPKey[1]]^{-1}(\prpOutputEmbeddingAlt))$ and
$\authToken = \authTokenAlt$. This implies
$\PRP[\PRPKey[1]]^{-1}(\prpOutputEmbeddingAlt) \neq
\embed[\embedKey](\cInput)$ because
$\blocked{\policyLan}{\threshold}(\embedESP(\cInput))$.
Since $\authToken = \PRP[\PRPKey[1]]^{-1}(\prpOutputEmbeddingAlt)
\linearComb \PRP[\PRPKey[2]]^{-1}(\prpOutputHashAlt)$ and 
$\authToken \in \finiteField^{\numberOfPointsSim}$. The mapping
from $\PRP[\PRPKey[1]]^{-1}(\prpOutputEmbeddingAlt)$ to $\authToken$
is injective. $\Adv{3}$ needs to
produce $\PRP[\PRPKey[1]]^{-1}(\prpOutputEmbeddingAlt)$ in order to
produce $\authTokenAlt = \authToken$. Note that $\authentication$
produces an output only if invoked with $\authToken$, otherwise it
returns $\bot$. Thus, having oracle access to $\authentication$
provides no advantage to $\Adv{3}$ since it receives an output only if
it already knows $\authToken$.
	
 Therefore, \Adv{3} can produce $\authToken$ only in two ways.
	
\smallskip\noindent Case (i): \Adv{3} produces
$\PRP[\PRPKey[1]]^{-1}(\prpOutputEmbeddingAlt)$ from
$\prpOutputEmbeddingAlt$ using $\prpOutputEmbedding,
\prpOutputHash$. First, note that $\Adv{1}$ gets one invocation to
$\PRP[\PRPKey[1]](\cdot)$ and one invocation to
$\PRP[\PRPKey[2]](\cdot)$ (both via \embedAndMap) to produce
$\prpOutputEmbedding$ and $\prpOutputHash$, respectively.  Though
$\Adv{2}$ generates $\prpOutputEmbeddingAlt$ and $\prpOutputHashAlt$
that are passed to $\PRP[\PRPKey[1]]^{-1}(\cdot)$ and
$\PRP[\PRPKey[2]]^{-1}(\cdot)$, $\Adv{3}$ never receives the results
(except as \prfOutput, which is covered in the next case).  Thus,
\Adv{3}'s advantage in producing
$\PRP[\PRPKey[1]]^{-1}(\prpOutputEmbeddingAlt)$ from
$\prpOutputEmbedding, \prpOutputHash$ is bounded by
$\Advantage{\PRPsecdef}{\PRP}{\timeBoundAlt} = \frac{2}{\setSize{\PRPKeySpace}}$ since 
$\PRP$ is an unpredictable function. Note that 
the keys \PRPKey[1] and \PRPKey[2] are chosen independently, and if $\PRPKey[2] \neq 
\PRPKey[1]$, \prpOutputHash does not provide any information about 
$\PRP[\PRPKey[1]]^{-1}(\prpOutputEmbeddingAlt)$. Due to the random key selection, the 
probability that $\PRPKey[2] = \PRPKey[1]$ is $\frac{1}{\setSize{\PRPKeySpace}}$. 

\smallskip\noindent Case (ii): \Adv{3} obtains $\authTokenAlt =
\PRP[\PRPKey[1]]^{-1}(\prpOutputEmbeddingAlt)
\linearComb~\PRP[\PRPKey[2]]^{-1}(\prpOutputHashAlt)$ from $\prfOutput$. If
\Adv{3} does not know $\PRP[\PRPKey[1]]^{-1}(\prpOutputEmbeddingAlt)$
from case (i) then $\PRP[\PRPKey[1]]^{-1}(\prpOutputEmbeddingAlt)
\oplus~\PRP[\PRPKey[2]]^{-1}(\prpOutputHashAlt)$ is unknown. Thus,
\Adv{2} has not made an oracle query to
$\PRF[1](\cdot)$ for $\PRP[\PRPKey[1]]^{-1}(\prpOutputEmbeddingAlt)
\linearComb~\PRP[\PRPKey[2]]^{-1}(\prpOutputHashAlt)$. In this case, the
advantage of \Adv{3} in obtaining $\authTokenAlt$ is bounded by
$\Advantage{\PRFINDsecdef}{\PRF[1]}{\timeBoundAlt, 1}$, the one oracle
query being that by \testAndCommit to make $\prfOutput$.
	 
From the above, we have 
\begin{align}
& \cprob{\Bigg}{\begin{array}{@{}r@{}}
  \cInputPRPOutAlt = \PRP[\PRPKey[1]]^{-1}(\prpOutputEmbeddingAlt) \linearComb~\PRP[\PRPKey[2]]^{-1}(\prpOutputHashAlt) \\
  \wedge~\ExptCC[\PRF[1],\PRP, \hash](\Adv{1}, \Adv{2}, \Adv{3}) = 1
  \end{array}}{\begin{array}{@{}r@{}}
  \blocked{\policyLan}{\threshold}(\embed[\embedKey](\cInput)) \\
  \wedge~\cInput \in \blockedElmts\end{array}} \nonumber\\
& \le \frac{2}{\setSize{\PRPKeySpace}} + \Advantage{\PRFINDsecdef}{\PRF[1]}{\timeBoundAlt, 1}
\end{align}
	
Summing up the two cases above, gives the result. 

\end{proof}

\subsection{Security against \server}
\label{sec:proof_framework:server}

We prove security for threat model \ref{model:server-mal} in this
section.
Assuming \embedAndMap, \testAndCommit, and \authentication are secure,
the method available to \server to attack the framework in
\figref{fig:frameworkSH} is to guess the input \cInput of \client.  As
such, the goal of such a \server is to cause the following experiment
to return $1$, where \server is denoted as algorithm $(\AdvS{1},
\AdvS{2})$.  In this experiment, we model $\hash$ as a random
oracle~\cite{Bellare1993:Random-oracles}.  Moreover, we stress that
in the experiment below, the statement ``$\cInput \getsr \universe$''
selects \cInput from \universe according to an application-specific
distribution that need not be uniform.

\smallskip

\begin{minipage}[t]{\columnwidth}
	\begin{tabbing}
	****\=\kill
	Experiment $\ExptS[\hash, \PRF[1]](\AdvS{1}, \AdvS{2})$ \\
	\> $\cInput \getsr \universe$ \\
	\> $\PRPKey[1] \getsr \PRPKeySpace$,
	   $\PRPKey[2] \getsr \PRPKeySpace$ \\
	\> $\PRFKey[1] \getsr \PRFKeySpace[1]$,
	   $\embedKey \getsr \embedKeySpace$ \\
	\> $((\prpOutputEmbedding, \prpOutputHash), \cdot) \gets \embedAndMap^{\hash(\cdot)}(\cInput, (\PRPKey[1], \PRPKey[2], \embedKey))$ \\
	\> $(\prfOutput, \authToken) \gets \testAndCommit^{\hash(\cdot)}((\prpOutputEmbedding, \prpOutputHash), (\PRPKey[1], \PRPKey[2], \PRFKey[1]))$ \\
	\> $\langle \Advstate{1}, \embedKeyAlt \rangle \gets \AdvS{1}^{\hash(\cdot)}(\authToken, \PRPKey[1], \PRPKey[2], \PRFKey[1], \embedKey)$\\
	\> $\authTokenAlt \gets \embed[\embedKeyAlt](\cInput) \linearComb \hash(\cInput, \embed[\embedKeyAlt](\cInput), \embedKeyAlt)$ \\
	\> $\cInputguess \gets \AdvS{2}^{\hash(\cdot), \authentication^{\hash(\cdot)}(\authTokenAlt, \cdot)}(\embedKeyAlt, \Advstate{1})$ \\
	\> if \= $\cInputguess = \cInput$ \\
	\> \> then return 1 \\
        \> \> else return 0
	\end{tabbing}
\end{minipage}

\smallskip

\noindent Then,
\begin{align*}
	\Advantage{\Ssecdef}{\hash, \PRF[1]}{\AdvS{1}, \AdvS{2}}
	& = \prob{\ExptS[\hash, \PRF[1]](\AdvS{1}, \AdvS{2}) = 1} \\
	\Advantage{\Ssecdef}{\hash, \PRF[1]}{\timeBound, 
	\forwardOracleQueries, \hashOracleQueries}
	& = \max_{\AdvS{1}, \AdvS{2}} \Advantage{\Ssecdef}
	{\hash, \PRF[1]}{\AdvS{1}, \AdvS{2}}
\end{align*}
where the maximum is taken over all adversaries $(\AdvS{1}, \AdvS{2})$
running in total time \timeBound, with at most \hashOracleQueries
queries to $\hash$, and at most \forwardOracleQueries oracle queries
to $\authentication^{\hash(\cdot)}(\cInputPRPOutAlt, \cdot)$ from
\AdvS{2}.

\bigskip

\newcommand{\probGuess}[1]{\ensuremath{\pi({#1})}\xspace}

To prove a result against such a server adversary, consider the
posterior distribution on $\cInput$, after observing $\neg
\blocked{\policyLan}{\threshold}(\embed[\embedKey](\cInput))$ for a
known \embedKey.  Then, let \probGuess{\hashOracleQueries} denote the
probability of guessing \cInput within \hashOracleQueries guesses,
when guessing values in order of nonincreasing probability according
to that posterior distribution.  No matter the properties of $\hash$,
the $(\AdvS{1}, \AdvS{2})$ adversary can win the above experiment with
probability at least \probGuess{\hashOracleQueries}, simply by
guessing elements $\cInputAlt$ of \universe in that order, and
checking if $\authToken = \embed[\embedKey](\cInputAlt) \linearComb
\hash(\cInputAlt, \embed[\embedKey](\cInputAlt), \embedKey)$.
\propref{prop:serverAdv} says that if $\hash$ is a random oracle, then
this is the best that $(\AdvS{1}, \AdvS{2})$ can do.

\begin{prop}
\label{prop:serverAdv}
If $\hash$ is a random oracle and $\embedAndMap, \testAndCommit,
\authentication$ are secure and output correctly, then
\begin{align*}
\Advantage{\Ssecdef}{\hash, \PRF[1]}{\timeBound, \forwardOracleQueries, 
\hashOracleQueries}
& \le \probGuess{\hashOracleQueries}
\end{align*}
for $\timeBoundAlt = \timeBound + \bigO{1}$.
\end{prop}

\begin{proof}
To win the experiment $\ExptS[\hash, \PRF[1]]$, $\AdvS{2}$ needs to
successfully guess $\cInput$.  First recall that $(\AdvS{1},
\AdvS{2})$ receives no output from \embedAndMap and so does not learn
$\prpOutputEmbedding$ or $\prpOutputHash$ individually.  It does learn
$\authToken = \PRP[\PRPKey[1]]^{-1}(\prpOutputEmbedding) \linearComb
\PRP[\PRPKey[2]]^{-1}(\prpOutputHash) = \embed[\embedKey](\cInput) 
\linearComb \hash(\cInput, \embed[\embedKey](\cInput), \embedKey)$. 
$\hash$ is a random oracle that maps $\{0,1\}^{\ast}$ into $\metricSpaceESP$,
where $\finiteField$ is a finite field. $\authToken \in \finiteField^{\numberOfPointsSim}$ 
is a vector of field elements, so the addition is a linear combination
in the finite field, which is a secure one-time-pad with perfect secrecy.
$\authToken$ carries no information about $\cInput$ or $\embed[\embedKey](\cInput)$.
\AdvS{1} or \AdvS{2} therefore has to invoke $\hash(\cInput, \embed[\embedKey](\cInput),
\embedKey)$. As such, the probability the adversary 
succeeds is as stated in the proposition.
\end{proof}

\section{Full Proofs of \secref{sec:protocol}}
\label{app:proofs_pake}

\begin{theorem}[\cite{Chakraborti2023:DAPSI}]
	\label{thm:dapsi}
	Let \finiteField be a finite field of order \fieldOrder. Fix polynomials $\genericPoly[1]$ and $\genericPoly[2]$ of degree $\degreeOfKey{1}$, and an arbitrary set of points $\setOfEvalPts \gets \{\evalPt[1], \dots, \evalPt[\numberOfPointsSim]\}$. Let $\simRandomPoly$ and $\simRandomPolyAlt$ be random polynomials of degree $\geq \degreeOfKey{1}$. Also, let $\genericPolyAlt \gets \prod\limits_{x_i \getsr \finiteField} (x - x_i)$. Then, when $\degreeOfPoly(\gcd(\genericPoly[1], \genericPoly[2])) < \degreeOfKey{1} - \frac{\threshold}{2}$ and $\numberOfPointsSim = \numberOfPointsSimDefn$,

	\begin{multline}
		\sum\limits_{Y \gets \finiteField^{\numberOfPointsSim}} | \prob{  \langle \simRandomPolyEval[\evalPt[\ptIdx]] \cdot \genericPolyEval[\evalPt[\ptIdx]][1] + \simRandomPolyAltEval[\evalPt[\ptIdx]] \cdot \genericPolyEval[\evalPt[\ptIdx]][2] \rangle_{\ptIdx = 1}^{\numberOfPointsSim} = Y} \\ 
		-   \prob{  \langle \simRandomPolyEval[\evalPt[\ptIdx]] \cdot \genericPolyEval[\evalPt[\ptIdx]][1] + \simRandomPolyAltEval[\evalPt[\ptIdx]] \cdot \genericPolyAltEval[\evalPt[\ptIdx]] \rangle_{\ptIdx = 1}^{\numberOfPointsSim} = Y} | \leq \frac{1}{\setSize{\finiteField}}
	\end{multline}
	
\end{theorem}

\begin{theorem*}
	Assuming that there is a protocol that securely realizes a garbled
	circuit, \protocolEmbed securely realizes \embedAndMap against a
	semi-honest server, and a malicious client (threat model
	\ref{model:client-mal} and \ref{model:server-mal} in
	\secref{sec:framework:threat-model}).
\end{theorem*}

\begin{proof}
	The entire functionality is implemented in a garbled circuit 
construction that is generated by the semi-honest \server. The 
party $\client$ playing the role of the evaluator is malicious, however 
it does not get outputs from the circuit unless it correctly evaluates 
it. Assuming that there is a construction realizing \otFunc with a
semi-honest garbler and a malicious evaluator, \protocolEmbedESP 
can be proved to be secure in the \otFunc-hybrid model. 
\end{proof}

\begin{theorem*}
	Assuming that there are protocols to securely realize \oleFunc and
	\npaFunc, \protocolCommit securely realized \testAndCommit against
	a semi-honest server, and a malicious client (threat models
	\ref{model:client-mal} and \ref{model:server-mal} described in
	\secref{sec:framework:threat-model}).
\end{theorem*}

\begin{proof}
	We will show that there is a polynomial time simulator that simulates $\client$'s and \server's views.
	
	\smallskip\noindent   
	{\bf When \server is corrupt:} The simulator obtains $\PRPKey[1] \assign \langle \PRPKeyVec[1], \PRPKeyVec[2] \rangle$  and $\PRPKey[2] \assign 
	\langle \PRPKeyHashVec[1], \PRPKeyHashVec[2] \rangle$,  and $\serversRandomPoly[1], \dots, \serversRandomPoly[\blockListSize]$  from \server's  random tape.
	Let $\PRPKeyVec[1] \assign \langle \PRPKeyVec[1][1], \dots, \PRPKeyVec[1][\numberOfPointsSim] \rangle$ and 
	$\PRPKeyVec[2] \assign \langle \PRPKeyVec[2][1], \dots, \PRPKeyVec[2][\numberOfPointsSim] \rangle$. Also, let 
	$\PRPKeyHashVec[1] \assign \langle \PRPKeyHashVec[1][1], \dots, \PRPKeyHashVec[1][\numberOfPointsSim] \rangle$ and 
	$\PRPKeyHashVec[2] \assign \langle \PRPKeyHashVec[2][1], \dots, \PRPKeyHashVec[2][\numberOfPointsSim] \rangle$.
	
	 For $\setIdx \in [1, \blockListSize]$, the simulator obtains $\lanElmt_\setIdx \in \policyLan$ from $\server$'s input tape; Recall \server is assumed to be semi-honest. 
	 Let $\lanElmt_\setIdx = \langle \lElmtVecComp{\setIdx}{1}, \dots, \lElmtVecComp{\setIdx}{\numberOfPointsSim} \rangle$ and let $\polyFromEmbedding[\lanElmt_\setIdx]$ be the polynomial obtained by interpolating the points $\{ (\evalPt[1], \lElmtVecComp{\setIdx}{1}), \dots, (\evalPt[\numberOfPointsSim], \lElmtVecComp{\setIdx}{\numberOfPointsSim}) \}$.

	First consider the case where \server's output tape has $\gcd(\polyFromEmbedding[\cInput], \polyFromEmbedding[\lanElmt_\setIdx])$ for $\setIdx \in [1, \blockListSize]$. There are two cases here 
	
	\begin{enumerate}
		\item \label{case:gcd_not_found} \textit{For $\rootIdx \neq \setIdx$:} When \server's output tape does not have $\gcd(\polyFromEmbedding[\cInput],  \polyFromEmbedding[\lanElmt_\rootIdx])$, then the degree of $\gcd(\polyFromEmbedding[\cInput],  \polyFromEmbedding[\lanElmt_\rootIdx]) < \degreeOfKey{1} - \threshold/2$. In this case, 
		the simulator sets $\simOp{\genericSetVar[\cInput]} \assign \{ \genericRand[1], \dots, \genericRand[\degreeOfKey{1}] \}$, where $\genericRand[1], \dots, \genericRand[\degreeOfKey{1}] \getsr \finiteField$, and the polynomial $\simOp{\polyFromEmbedding[\cInput]} \assign \prod\limits_{\genericSetElmt \in \simOp{\genericSetVar[\cInput]}} (x - \genericSetElmt)$. The simulator simulates \npaFunc returning evaluations of 
		$\langle \PRPKeyVec[1][\ptIdx] \cdot \left(   \serversRandomPoly[\rootIdx][\evalPt[\ptIdx]] \cdot \simOp{\polyFromEmbedding[\cInput][\evalPt[\ptIdx]]}  + \simOp{\clientsRandomPoly[\rootIdx][\evalPt[\ptIdx]]}  \cdot \polyFromEmbedding[\lanElmt_\rootIdx][\evalPt[\ptIdx]]  \right) \\
+ \serversRandomPoly[\rootIdx][\evalPt[\ptIdx]] \cdot \PRPKeyVec[2][\ptIdx] \rangle_{\ptIdx = 1}^{\numberOfPointsSim}$ 
		to \server, where $\simOp{\clientsRandomPoly[\rootIdx]}$ is a random polynomial of degree $\degreeOfKey{1}$. The simulation is indistinguishable from the 
		real protocol execution: $\simOp{\clientsRandomPoly[\rootIdx]}$ is sampled from the same distribution as $\clientsRandomPoly[\rootIdx]$, and 
		from  \thmref{thm:dapsi}, $\left\langle \serversRandomPoly[\rootIdx][\evalPt[\ptIdx]] \cdot \simOp{\polyFromEmbedding[\cInput][\evalPt[\ptIdx]]}  + \simOp{\clientsRandomPoly[\rootIdx][\evalPt[\ptIdx]]}  \cdot \polyFromEmbedding[\lanElmt_\rootIdx][\evalPt[\ptIdx]]  \right\rangle_{\ptIdx = 1}^{\numberOfPointsSim}$ is indistinguishable from $\left\langle \serversRandomPoly[\rootIdx][\evalPt[\ptIdx]] \cdot \polyFromEmbedding[\cInput] [\evalPt[\ptIdx]] + \clientsRandomPoly[\rootIdx][\evalPt[\ptIdx]] \cdot \polyFromEmbedding[\lanElmt_\rootIdx][\evalPt[\ptIdx]]  \right\rangle_{\ptIdx = 1}^{\numberOfPointsSim} $.

		\item \label{case:gcd_found} \textit{For $\rootIdx = \setIdx$:} When \server's output tape has $\gcd(\polyFromEmbedding[\cInput], \polyFromEmbedding[\lanElmt_\rootIdx])$, let $\genericSetVar[\cInput \cap \lanElmt_\rootIdx]$ be the set of the roots of $\gcd(\polyFromEmbedding[\cInput], \polyFromEmbedding[\lanElmt_\rootIdx])$ and $\setSize{\genericSetVar[\cInput \cap \lanElmt_\rootIdx]} = d$. Then, 
		the simulator computes a set $\simOp{\genericSetVar[\cInput]} \assign \{ \genericRand[1], \dots, \genericRand[\degreeOfKey{1} - d] \} \cup \genericSetVar[\cInput \cap \lanElmt_\rootIdx]$ where $\genericRand[1], \dots, \genericRand[\degreeOfKey{1} - d] \notin \genericSetVar[\cInput \cap \lanElmt_\rootIdx]$. The simulator sets $\simOp{\polyFromEmbedding[\cInput]} \assign \prod\limits_{\genericSetElmt \in \simOp{\genericSetVar[\cInput]}} (x - \genericSetElmt)$. The simulator simulates \npaFunc returning evaluations of 
		$\langle \PRPKeyVec[1][\ptIdx] \cdot \left(   \serversRandomPoly[\rootIdx][\evalPt[\ptIdx]] \cdot \simOp{\polyFromEmbedding[\cInput][\evalPt[\ptIdx]]}  + \simOp{\clientsRandomPoly[\rootIdx][\evalPt[\ptIdx]]}  \cdot \polyFromEmbedding[\lanElmt_\rootIdx][\evalPt[\ptIdx]]  \right) \\
+ \serversRandomPoly[\rootIdx][\evalPt[\ptIdx]] \cdot \PRPKeyVec[2][\ptIdx] \rangle_{\ptIdx = 1}^{\numberOfPointsSim}$ 
		to \server, where $\simOp{\clientsRandomPoly[\rootIdx]}$ is a random polynomial of degree $\degreeOfKey{1}$. Let $\serversRandomPoly[\rootIdx] \cdot \polyFromEmbedding[\cInput]  + \clientsRandomPoly[\rootIdx] \cdot \polyFromEmbedding[\lanElmt_\rootIdx]  = \gcd(\polyFromEmbedding[\cInput], \polyFromEmbedding[\lanElmt_\rootIdx]) \cdot R(x)$ and 
		$\serversRandomPoly[\rootIdx] \cdot \simOp{\polyFromEmbedding[\cInput]}  + \simOp{\clientsRandomPoly[\rootIdx]}  \cdot \polyFromEmbedding[\lanElmt_\rootIdx] = \gcd(\polyFromEmbedding[\cInput], \polyFromEmbedding[\lanElmt_\rootIdx]) \cdot \simOp{R(x)}$ 
		Since $\serversRandomPoly[\rootIdx]$, $\clientsRandomPoly[\rootIdx]$ and $\simOp{\clientsRandomPoly[\rootIdx]}$ are all random polynomials, 
		due to \lemmaref{lemma:kissener_uniformly_random}, $R(x)$ and  $\simOp{R(x)}$ are both random polynomials of the same degree, and are therefore indistinguishable. Thus,  $\serversRandomPoly[\rootIdx] \cdot \polyFromEmbedding[\cInput]  + \clientsRandomPoly[\rootIdx] \cdot \polyFromEmbedding[\lanElmt_\rootIdx]$ is indistinguishable from 	$\serversRandomPoly[\rootIdx] \cdot \simOp{\polyFromEmbedding[\cInput]}  + \simOp{\clientsRandomPoly[\rootIdx]}  \cdot \polyFromEmbedding[\lanElmt_\rootIdx]$, and consequently the simulation is indistinguishable 
		from the real execution. 
		
	\end{enumerate}
	
	In the case where there is no $\lanElmt_\setIdx$ where \server's output tape has $\gcd({\polyFromEmbedding[\cInput], \polyFromEmbedding[\lanElmt_\setIdx]})$, the simulator follows case \eqnref{case:gcd_not_found} for all $\setIdx \in [1, \blockListSize]$.

	In the commit phase, the simulator implements the random oracle, and sets $\langle \simOp{\hashComp{1}}, \dots, \simOp{\hashComp{\numberOfPointsSim}} \rangle \getsr 
	\metricSpaceESP$. The simulator computes $\langle  \simOp{\prpOutputHashComp{1}}, \dots, \simOp{\prpOutputHashComp{\numberOfPointsSim}}   \rangle \assign \PRP[\PRPKey[2]](\langle \simOp{\hashComp{1}}, \dots, \simOp{\hashComp{\numberOfPointsSim}} \rangle)$, where the simulator gets $\PRPKey[2]$ from the random tape of \server. The simulator simulates the calls to \oleFunc in \stepsref{esp:tc:remove_client_random}{esp:tc:prp_inv_hash} with inputs 
	$\simOp{\clientsRandomPoly[\setIdx][\evalPt[\ptIdx]]}$ and $\simOp{\prpOutputHashComp{\ptIdx}}$ respectively. The simulator does not need to simulate an output to \server. 
	
	Then, for \stepref{esp:tc:client_comb} the simulator returns 
	$\simOp{\genericVec'} \assign 
	\langle \sum\limits_{\setIdx = 1}^{\blockListSize} \tmpVar_{\setIdx, \ptIdx} + \tmpVar'_{\setIdx, \ptIdx} -  
	\simOp{\clientsRandomPoly[\setIdx][\evalPt[\ptIdx]]} \cdot \polyFromEmbedding[\lanElmt_\setIdx][\evalPt[\ptIdx]]+\frac{\randomPolyComb[\evalPt[\ptIdx]]}{\PRPKeyHashVec[1][\ptIdx]} \cdot \simOp{\prpOutputHashComp{\ptIdx}}  \rangle_{\ptIdx = 1}^{\numberOfPointsSim}$ to \server. $\simOp{\genericVec'}$ is indistinguishable from $\genericVec'$ as $\simOp{\clientsRandomPoly[\setIdx]}$ is identically distributed 
	to $\clientsRandomPoly[\setIdx]$, and $\langle \simOp{\hashComp{1}}, \dots, \simOp{\hashComp{\numberOfPointsSim}} \rangle$ is indistinguishable from $\langle \hashComp{1}, \dots, \hashComp{\numberOfPointsSim} \rangle$ assuming that $\hash(\cdot)$ is a random oracle.

	\smallskip\noindent
	{\bf When $\client$ is corrupt:} 
	Note that $\client$ is malicious, and so we will prove the protocol secure in the (\oleFunc, \npaFunc)-hybrid model. The simulator does the following
	
	\begin{enumerate}
		\item During \stepref{esp:tc:compute_comb_poly} of \protocolCommitESP, the simulator obtains \genericAdv's inputs
		$\langle \clientsRandomPoly[\setIdx][\evalPt[\ptIdx]] \rangle_{\ptIdx = 1}^{\numberOfPointsSim}$ and $\prpOutputEmbeddingAlt$. The simulator simulates 
		\npaFunc with $\client$'s inputs. There is no output to \genericAdv from this step.  
		
		\item In \stepref{esp:tc:remove_client_random}, the simulator gets 
		$\clientsRandomPoly[\setIdx][\evalPt[\ptIdx]]^{\ast}$ from \genericAdv's input to \oleFunc in  for $\setIdx \in [1, \blockListSize], \ptIdx \in [1, \numberOfPointsSim]$. If $\clientsRandomPoly[\setIdx][\evalPt[\ptIdx]]^{\ast} = \clientsRandomPoly[\setIdx][\evalPt[\ptIdx]]$ (obtained earlier), then the simulator sends $\clientsRandomPoly[\setIdx][\evalPt[\ptIdx]]$ to \oleFunc, obtains $out_{\setIdx, \ptIdx}$ from \oleFunc, and simulates \oleFunc returning $out_{\setIdx, \ptIdx}$ to $\client$. 
		Otherwise, the simulator sets $\simOp{out_{\setIdx, \ptIdx}} \getsr \finiteField$ and simulates \oleFunc returning $\simOp{out_{\setIdx, \ptIdx}}$ to $\client$. The simulator outputs whatever \genericAdv outputs. 
		
		\item In \stepref{esp:tc:prp_inv_hash}, the simulator gets $ \prpOutputHashComp{1}', \dots, \prpOutputHashComp{\numberOfPointsSim}'$ as inputs to \oleFunc.  The simulator simulates returning the output of \oleFunc to $\client$, and outputs whatever \genericAdv outputs.

		\item The simulator sends to \testAndCommit the inputs 
		$\prpOutputEmbeddingAlt$ and $\prpOutputHashAlt \assign \langle \prpOutputHashComp{1}', \dots, \prpOutputHashComp{\numberOfPointsSim}' \rangle$. 
		The functionality returns $\prfOutput$ or $\bot$.
		
		\item The simulator gets $\genericVec'$ from \genericAdv in \stepref{esp:tc:client_comb}. 
		If the output of \testAndCommit is $\bot$, then the simulator sets $\simOp{\prfOutput} = \bot$, returns $\simOp{\prfOutput}$ to the client, and outputs whatever \genericAdv outputs. Otherwise, the simulator checks whether $\genericVec'$ has been correctly computed as the sum of the values returned to $\client$ in the previous steps. If so, 
		the simulator sets $\simOp{\prfOutput} \assign \prfOutput$. Otherwise, the simulator sets $\simOp{\prfOutput} \getsr \finiteField$. The simulator sends $\simOp{\prfOutput}$ to $\client$, 
		and outputs whatever \genericAdv outputs.  
		
	\end{enumerate}

	The simulation is indistinguishable from the real execution of the protocol. In \stepref{esp:tc:rand_poly_C}, 
	$out_{\setIdx, \ptIdx}$ and $\simOp{out_{\setIdx, \ptIdx}}$ are indistinguishable. This is because in the real execution of $\protocolCommitESP$, $out_{\setIdx, \ptIdx} = \tmpVar'_{\setIdx, \ptIdx} - 
	\clientsRandomPoly[\setIdx][\evalPt[\ptIdx]] \cdot \polyFromEmbedding[\lanElmt_\setIdx][\evalPt[\ptIdx]] \getsr \finiteField$, when $\tmpVar'_{\setIdx, \ptIdx} \getsr \finiteField$, which is indistinguishable from $\simOp{out_{\setIdx, \ptIdx}} \getsr \finiteField$. In \stepref{esp:tc:rand_poly_S},, the simulator simply simulates 
	\oleFunc using \genericAdv's input and thus follows the protocol exactly.

	In \stepref{esp:tc:compute_vec},  first note that the ideal functionality outputs $\bot$ when $\symmURDist{\embed[\embedKey](\cInput)}{\lanElmt} \le \threshold$, which equivalent to checking that the degree of $\gcd(\polyFromEmbedding[\cInput], \polyFromEmbedding[\lanElmt]) \ge \degreeOfKey{1} - 
	\frac{\threshold}{2}$. On the other hand, the simulation outputs $\bot$ when the degree of $\combPoly \assign \gcd(\polyFromEmbedding[\cInput], \serversRandomPoly[\setIdx] \cdot  \polyFromEmbedding[\cInput]  + \clientsRandomPoly[\setIdx]   \cdot \polyFromEmbedding[\lanElmt]) \ge \degreeOfKey{1} - 
	\frac{\threshold}{2}$. From 
	\propref{prop:symmURFromRand}, we have 
	
	\[
	\prob{\symmURDist{\embedESP(\cInput)}{\lanElmt} \neq 
	2\cdot \degreeOfKey{1} - \degreeOfPoly(\combPoly)} \le 1/\setSize{\finiteField}
	\]
	
	Thus, 
	
	\begin{multline}
	\cprob{}{\simOp{\prfOutput} \neq \bot}{\prfOutput = \bot} \\
	= \prob{\symmURDist{\embedESP(\cInput)}{\lanElmt} \neq 
		2\cdot \degreeOfKey{1} - \degreeOfPoly(\combPoly)} \le 1/\setSize{\finiteField}
	\end{multline}
	
	Thus, with overwhelming probability, the simulation output and the real execution output are identical.
	
	Next, note that when $\genericVec'$ has been correctly computed, the simulator simply returns the output from 
	\testAndCommit. The functionality and in turn the simulator return $\simOp{\prfOutput} \assign \PRF[1][\PRFKey[1]](\authToken)$
	where 
	$\authToken \assign \PRP[\PRPKey[1]]^{-1}\left( \prpOutputEmbeddingAlt \right) + \PRP[\PRPKey[2]]^{-1}\left( \prpOutputHashAlt\right)$. 
	In the real execution, $\authToken$ is identical when the 
	value of $\genericVec'$ has been correctly computed. 
	
	Finally, if \genericAdv provides inconsistent value of $\genericVec'$, either due to the fact that $\genericVec'$ is not correctly computed as a sum of the results returned before to $\client$, or 
	because $\clientsRandomPoly[\setIdx][\evalPt[\ptIdx]]^{\ast} \neq \clientsRandomPoly[\setIdx][\evalPt[\ptIdx]]$  for some $\ptIdx, \setIdx$, 
	then the simulator returns $\simOp{\prfOutput} \getsr \finiteField$. Since,
	in this case, $\authTokenAlt \assign \left\langle \frac{1}{\randomPolyComb[\evalPt[\ptIdx]]} \left(\sum\limits_{\setIdx = 1}^{\blockListSize} \combVector_\setIdx 
	+ \genericVec' - \genericVec \right) \right\rangle_{\ptIdx = 1}^{\numberOfPointsSim} \neq \PRP[\PRPKey[1]]^{-1}\left( \prpOutputEmbeddingAlt \right) + \PRP[\PRPKey[2]]^{-1}\left( \prpOutputHashAlt\right)$, there is at least one component of $\authTokenAlt$ that differs from $\PRP[\PRPKey[1]]^{-1}\left( \prpOutputEmbeddingAlt \right) + \PRP[\PRPKey[2]]^{-1}\left( \prpOutputHashAlt\right)$. The real protocol execution will return 
	$\PRF[1][\PRFKey[1]](\authTokenAlt)$ while the simulation returns a random element in the range of $\PRF[1]$. This implies in the real execution, \genericAdv gets exactly one query to the $\PRF[1]$ oracle. Thus, \genericAdv's advantage in distinguishing between the simulation and the real protocol execution is exactly equal to \genericAdv's advantage in distinguishing between a random function and a $\PRF[1]$ with one query to $\PRF[1]$ oracle. For a secure $\PRF[1]$ construction, this is negligible and thus the simulation is indistinguishable to \genericAdv. 

\end{proof}

\begin{theorem*}
	Assuming that there is a protocol to securely realize \oleFunc,
	\protocolValidate securely realizes \authentication against a
	malicious server (threat model \ref{model:server-mal}).
\end{theorem*}

\begin{proof}
	We will show that there is a polynomial time simulator that indistinguishably simulates \server's and $\client$'s views of the protocol in the ideal world simulation.
	
	\smallskip\textbf{When \server is corrupt:}
	\server does not obtain any output from the protocol. So it is straightforward to simulate \server's view when interacting with \oleFunc. Specifically, the simulator simulates \oleFunc with \server's inputs $\genericRand[\ptIdx]$ and $\secretShare{\ptIdx}$ for $\ptIdx \in [1, \numberOfPointsSim]$ (which it gets from \server's input tape), and $\genericRand[\ptIdx]' \getsr \finiteField$ as $\client$'s input. The simulation is indistinguishable from the protocol execution if \oleFunc can be simulated since $\client$'s input in the real protocol execution $\authToken \getsr \finiteField$ is identically distributed to the simulated input 
	assuming that $\hash(\cdot)$ is a random oracle. 
	
	\textbf{When $\client$ is corrupt:}
	The simulator implements the random oracle $\hash(\cdot)$ in the simulation. When $\client$ queries $\hash(\cdot)$ with the input $(\cInput, \embedESP(\cInput), \embedKey)$ in \stepref{esp:auth:hash}, the simulator responds with   $\genericRand \in \metricSpaceESP$ and stores this locally. 
	The simulator computes $\authTokenAlt \assign  \embedESP(\cInputAlt) + \genericRand$. The simulator sends $\authTokenAlt$ to \authentication,  which either returns $\prfOutput$ or $\bot$. 
	If the output from \authentication is $\bot$, then the simulator sets $\simOp{\prfOutput} \getsr \finiteField$. Otherwise, the simulator sets $\simOp{\prfOutput} \assign \prfOutput$. The simulator secret shares $\simOp{\prfOutput}$ into $\numberOfPointsSim$ shares $\simOp{\secretShare{1}}, \dots, \simOp{\secretShare{\numberOfPointsSim}}$.

	in \stepref{esp:auth:ole}, the simulator gets inputs to \oleFunc  from \genericAdv. For $\ptIdx \in [1, \numberOfPointsSim]$, the simulator simulates returning $\simOp{\secretShare{\ptIdx}}$ to $\client$. 
	The simulation is indistinguishable from the real protocol execution. Consider first the case when $\client$'s input $\authTokenAlt \neq \authToken$. In this case, the real protocol execution returns for $\ptIdx \in [1, \numberOfPointsSim]$, $\genericRand[\ptIdx] (\authTokenComp{\ptIdx} - \authTokenAltComp{\ptIdx}) + \secretShare{\ptIdx}$. Since $\authToken \neq \authTokenAlt$, there is some $\rootIdx \in [1, \numberOfPointsSim]$ where $\authTokenComp{\rootIdx} \neq \authTokenAltComp{\rootIdx}$ and thus $\genericRand[\rootIdx] (\authTokenComp{\rootIdx} - \authTokenAltComp{\rootIdx}) + \secretShare{\rootIdx} \getsr \finiteField$. Consequently, in the real execution, $\client$ does not get all the correct $\numberOfPointsSim$ secret shares of $\prfOutput$, i.e., at least one of the secret shares is replaced with a random element. Thus, for an information theoretically secure secret sharing scheme, $\client$ obtains a random element upon recombining the shares. 
	In the ideal execution, the simulator secret shares a random element and provides the shares to $\client$. Combining the shares, $\client$ obtains a random element. The outputs of $\client$ are identically distributed in the real world execution and the ideal world simulation. 
	
	For the case when $\authToken = \authTokenAlt$, in the real execution $\forall \ptIdx \in [1, \numberOfPointsSim]$, $\genericRand[\ptIdx] (\authTokenComp{\ptIdx} - \authTokenAltComp{\ptIdx}) + \secretShare{\ptIdx} = \secretShare{\ptIdx}$. In the simulation, the simulator secret shares $\prfOutput$ into $\simOp{\secretShare{1}}, \dots, \simOp{\secretShare{\numberOfPointsSim}}$ and provides these to $\client$. Thus, $\client$'s output is identical in both cases, and the simulation is indistinguishable from the real protocol execution.

\end{proof}

\end{document}